\documentclass{aa}

\usepackage{amsmath,amssymb,textcomp}
\usepackage{graphicx}
\usepackage[varg]{txfonts}
\usepackage{enumitem}

\usepackage{natbib}
\usepackage{import}
\usepackage{multirow}
\makeatletter
\def\hlinewd#1{%
\noalign{\ifnum0=`}\fi\hrule \@height #1 %
\futurelet\reserved@a\@xhline}
\makeatother
\newcommand{\dif}{\mathrm{d}}
\newcommand{\idem}{~~\textacutedbl}
\newcommand{\lec}{\mathrm{e}}
\newcommand{\ionperso}{\mathrm{i}}
\newcommand{\therm}{\mathrm{th}}
\newcommand*{\f}[4]{#1_{#2\,|\,#3}^{#4}}
\def\dintcl{\int\limits_{\mathrm{cell}}\kern-0.7em\int\kern-1.4em\bigcirc\kern.7em}
\usepackage{color}
\definecolor{orange}{rgb}{0.6,0.1,0}

\renewcommand{\note}[1]{}
\begin{document}

\title{Apar-T: code, validation, and physical interpretation\\ of particle-in-cell results}

\author{Micka\"el Melzani\inst{1}         \and
        Christophe Winisdoerffer\inst{1}  \and
        Rolf Walder\inst{1}               \and
        Doris Folini\inst{1}              \and 
        Jean M. Favre\inst{2}             \and
        Stefan Krastanov\inst{1}          \and
        Peter Messmer\inst{3}                }

\institute{\'{E}cole Normale Sup\'{e}rieure, Lyon, CRAL, UMR CNRS 5574, 
           Universit\'{e} de Lyon, France\\
           \hspace{0.3cm} E-mail: mickael.melzani@ens-lyon.fr
           \and
           CSCS Lugano, Switzerland
           \and
           NVIDIA Switzerland, Technoparkstr 1, 8005 Z\"urich, Switzerland
}

\offprints{M. Melzani}

\date{Received ... ; accepted ...}

\authorrunning{M. Melzani et al.}
\titlerunning{Apar-T: code, validation, and physical interpretation of particle-in-cell results}

%%%%%%%%%%%%%%%%%%%%%%%%%%%%%%%%%%%%%%%%%%%%%%%%%%%%%%%%%%%%%%%%%%%%%%%%%%%%%%%%%%%%%%%%%%%%%%%%%%%%%%%%%%%%%%%%%%%%%%%%%%%%%%%%%%%%%%%%%%%%%%%%
%%%%%%%%%%%%%%%%%%%%%%%%%%%%%%%%%%%%%%%%%%%%%%%%%%%%%%%%%%%%%%%%%%%%%%%%%%%%%%%%%%%%%%%%%%%%%%%%%%%%%%%%%%%%%%%%%%%%%%%%%%%%%%%%%%%%%%%%%%%%%%%%
%%%%%%%%%%%%%%%%%%%%%%%%%%%%%                             Abstract                                                %%%%%%%%%%%%%%%%%%%%%%%%%%%%%%
%%%%%%%%%%%%%%%%%%%%%%%%%%%%%%%%%%%%%%%%%%%%%%%%%%%%%%%%%%%%%%%%%%%%%%%%%%%%%%%%%%%%%%%%%%%%%%%%%%%%%%%%%%%%%%%%%%%%%%%%%%%%%%%%%%%%%%%%%%%%%%%%
%%%%%%%%%%%%%%%%%%%%%%%%%%%%%%%%%%%%%%%%%%%%%%%%%%%%%%%%%%%%%%%%%%%%%%%%%%%%%%%%%%%%%%%%%%%%%%%%%%%%%%%%%%%%%%%%%%%%%%%%%%%%%%%%%%%%%%%%%%%%%%%%
\abstract{%
We present the parallel particle-in-cell (PIC) code Apar-T 
and, more importantly, address the fundamental question of the relations between the PIC model, the Vlasov-Maxwell theory, and real plasmas.
\newline
First, we present four validation tests: spectra from simulations of thermal plasmas,
linear growth rates of the relativistic tearing instability and of the filamentation instability, 
and non-linear filamentation merging phase.
For the filamentation instability we show that
the effective growth rates measured on the total energy can differ by more than 50\% from the linear cold predictions
and from the fastest modes of the simulation.
We link these discrepancies to the superparticle number per cell and to the level of field fluctuations.
\newline
Second, we detail a new method for initial loading of Maxwell-J\"uttner particle distributions
with relativistic bulk velocity and {relativistic} temperature, and explain why the
traditional method with individual particle boosting fails. 
The formulation of the relativistic Harris {equilibrium} is generalized to
arbitrary temperature and mass ratios. Both are required for the tearing instability setup.
\newline
Third, we turn to the key point of this paper and scrutinize the question of 
what description of (weakly coupled) physical plasmas is obtained by PIC models.
These models rely on two building blocks: coarse-graining, i.e., grouping of the order of $p\sim 10^{10}$ real particles into a single computer superparticle, 
and field storage on a grid with its subsequent finite superparticle size.
We introduce the notion of coarse-graining dependent quantities, i.e., quantities depending on $p$.
They derive from the PIC plasma parameter $\Lambda^\mathrm{PIC}$, which we show to behave as $\Lambda^\mathrm{PIC} \propto 1/p$.
We explore two important implications.
One is that PIC collision- and fluctuation-induced thermalization times are expected to scale 
with the number of superparticles per grid cell, and thus to be a factor $p \sim 10^{10}$ smaller than in real plasmas,
a fact that we confirm with simulations.
The other is that the level of electric field fluctuations scales as $1/\Lambda^\mathrm{PIC} \propto p$.
We provide a corresponding exact expression, taking into account the finite superparticle size.
We confirm both expectations with simulations.
\newline
Fourth, we compare the Vlasov-Maxwell theory, often used for code benchmarking, to the PIC model.
The former describes a phase-space fluid with $\Lambda=+\infty$ and no correlations, while the PIC plasma 
features a small $\Lambda$ and a high level of correlations when compared to a real plasma.
These differences have to be kept in mind when interpreting and validating PIC results against Vlasov-Maxwell theory
and when modeling real physical plasmas.
}


\maketitle

%%%%%%%%%%%%%%%%%%%%%%%%%%%%%%%%%%%%%%%%%%%%%%%%%%%%%%%%%%%%%%%%%%%%%%%%%%%%%%%%%%%%%%%%%%%%%%%%%%%%%%%%%%%%%%%%%%%%%%%%%%%%%%%%%%%%%%%%%%%%%%%%
%%%%%%%%%%%%%%%%%%%%%%%%%%%%%%%%%%%%%%%%%%%%%%%%%%%%%%%%%%%%%%%%%%%%%%%%%%%%%%%%%%%%%%%%%%%%%%%%%%%%%%%%%%%%%%%%%%%%%%%%%%%%%%%%%%%%%%%%%%%%%%%%
\section{Introduction}\label{sec:intro}
%%%%%%%%%%%%%%%%%%%%%%%%%%%%%%%%%%%%%%%%%%%%%%%%%%%%%%%%%%%%%%%%%%%%%%%%%%%%%%%%%%%%%%%%%%%%%%%%%%%%%%%%%%%%%%%%%%%%%%%%%%%%%%%%%%%%%%%%%%%%%%%%
%%%%%%%%%%%%%%%%%%%%%%%%%%%%%%%%%%%%%%%%%%%%%%%%%%%%%%%%%%%%%%%%%%%%%%%%%%%%%%%%%%%%%%%%%%%%%%%%%%%%%%%%%%%%%%%%%%%%%%%%%%%%%%%%%%%%%%%%%%%%%%%%

Full particle codes are now used by a large number of groups worldwide to study plasmas out of equilibrium
in a large variety of environments,
including electronic devices, inertial fusion \citep{Dieckmann2006, Bret2010b},
tokamaks, Earth and solar magnetospheres \citep{Hesse2001, Daughton2006, Drake2006, Klimas2008, Markidis2012, Baumann2012},
or high-energy astrophysics \citep{Silva2003, Petri2007b, Zenitani2008b, Cerutti2012b, Jaroschek2005, Nishikawa2008, Sironi2011}.

Particle simulations actually appeared along with the first computers at universities and in industry around 1950.
They first concerned electron beams in vacuum tubes, a device extensively used in the computers themselves.
The beams were cold \citep{Hartree1950} or later hot \citep{Tien1956}, and consisted of roughly
300 electron slabs moving in one dimension.

The step to plasma simulations was taken by \citet{Buneman1959}.
He simulated an electrostatic plasma of 512 ions and electrons in one dimension, and showed that particle
codes could be used to study the linear, non-linear, and saturation phases of
instabilities. At the time, the relevance of simulations with so few particles per Debye sphere was not 
clear and in 1962 \citet{Dawson1962} and \citet{Eldridge1962} made an important contribution by
showing that correct thermal behavior was produced.

All these algorithms used particle-particle interactions, and the 
first particle-mesh codes to introduce a grid appeared only later \citep{Burger1965, Hockney1966, Yu1965}.
A great deal of literature on the drawbacks and benefits of the grid then appeared, and is now mostly concentrated in
the two reference books of \citet{Birsdall1985} and \citet{Hockney1988}. Refinements of the algorithms quickly appeared
(higher order grid interpolation, quiet codes, etc.), as well as code optimizations, at a time when programs were written
in assembly language and depended heavily on machine architecture.
Fully electromagnetic, relativistic, and 3D codes appeared with the studies of laser
induced fusion \citep{Buneman1976}, and closely resemble today's codes.

In 1967, it was possible for \citet{Birsdall1967} to list the papers concerning simulations, and he
noticed that they had grown exponentially since 1956.
In 1956, ``many particles'' meant 300; in 1985, $10^6$; in 2008, $10^9$; and in 2012, $10^{12}$.
For more historical details on PIC simulations, 
one may consult \citet{Birsdall1967, Birsdall1999}, \citet[Sect.~9.1]{Hockney1988},
or the introduction of \citet{Birsdall1985}.

Today, the latest generation of PIC algorithms consists of large versatile codes, featuring high order integration schemes and
efficient parallelization. We can quote codes such as TRISTAN-MP \citep{Spitkovsky2005}, OSIRIS \citep{Fonseca2002, Fonseca2008}, VORPAL \citep{Nieter2004},
WARP \citep{Grote2005}, ALaDyn \citep{Benedetti2008}, iPIC3D \citep{Markidis2010}, Photon-Plasma \citep{Haugboelle2012}, or Zeltron \citep{Cerutti2013}.
They employ various simulation methods. 
The equations solved can differ: electrostatic codes, electromagnetic codes, Darwin approximation \citep{Huang2008}.
The integration scheme can also vary: integration directly on the fields staggered on a grid,
either with a charge conserving scheme or via solution of Poisson equation \citep{Cerutti2012b};
or integration of the potential and correction of the discrepancies to charge conservation \citep{Daughton2006}.
The time integration can be explicit or implicit \citep{Markidis2012}.
Special parts of the numerical scheme can also differ; for example, the order 
of the field integration or the use of Fourier transforms. The interpolation of particle quantities to grid 
points and reciprocally can be done by a nearest grid point method (NGP), by a linear weighting
(cloud in cell (CIC), or the PIC algorithm in the old terminology), or by a smoother shape \citep[spline interpolation,][]{Esirkepov2001}.

This article documents a new PIC code, Apar-T. Its first version, Tristan, was
written by O. Buneman in 1990 \citep{Matsumoto1993}. It was then made parallel by \citet{Messmer1998},
and used in \citet{2002A&A...382..301M} or \citet{2005PPCF...47.1925P}.
Given the large quantity of simulation methods, it is useful to first detail our algorithm.
This is done in Sect.~\ref{Sec:Problem_solved} and Appendix~\ref{Sec:Numerical_implementation}.

In Sect.~\ref{Sec:Example_validation}, we present a set of test problems and the results obtained with Apar-T.
The first test is the study of the fluctuation spectra of a thermal plasma. 
The second and third tests explore the linear and non-linear stages of the filamentation instability.
A last test is the computation of the linear growth rates of the relativistic tearing instability,
for which we give the general equilibrium relations for the relativistic 
Harris configuration with arbitrary temperature and mass ratio between the two species
(Appendix~\ref{Sec:Harris_details}).
In Sect.~\ref{Sec:Load_particles} we describe a method for loading a Maxwell-J\"uttner momentum distribution 
with an arbitrary bulk velocity and temperature. 
The naive method, which initializes the comobile distribution and then boosts particles individually,
is shown to be incorrect, mainly because space contraction is absent from the PIC code. 

In Sect.~\ref{Sec:computer_vs_real_plasma}, motivated by the above tests, we explore
the physical implications of two main differences between PIC plasmas and real plasmas:
coarse-graining, i.e., a reduction in the number of particles by a factor reaching $p\sim10^{10}$,
and the representation of the fields on a grid. 
We introduce the notion of coarse-graining dependent
quantities, i.e., physical parameters depending on the number of real particles 
represented by a single computer superparticle. 
The prototype of these quantities is the plasma parameter $\Lambda$, which scales for a PIC plasma as $1/p$.
The dependence on coarse-graining has important consequences for key physical quantities such as the thermalization time, 
the collision time, or the level of field fluctuations.

We point out more generally that a PIC code simulates a microstate constituted by
a restricted number of finite-sized particles 
each representing up to $10^{10}$ real plasma particles, while the Vlasov-Maxwell system models 
a plasma macrostate described by a continuous fluid in six-dimensional phase-space.
These two descriptions are not equivalent and in particular PIC systems, with their small numbers of 
particles per Debye sphere, suffer from abnormally high noise levels 
and include to an unknown degree particle correlations absent from Vlasov-Maxwell equations.

We conclude in Sect.~\ref{Sec:Discussion}, stressing that validation and interpretation of PIC simulations 
requires detailed knowledge of the code as presented here as well as awareness of the above differences between
the PIC model, the Vlasov-Maxwell model, and a real plasma.

%%%%%%%%%%%%%%%%%%%%%%%%%%%%%%%%%%%%%%%%%%%%%%%%%%%%%%%%%%%%%%%%%%%%%%%%%%%%%%%%%%%%%%%%%%%%%%%%%%%%%%%%%%%%%%%%%%%%%%%%%%%%%%%%%%%%%%%%%%%%%%%%
%%%%%%%%%%%%%%%%%%%%%%%%%%%%%%%%%%%%%%%%%%%%%%%%%%%%%%%%%%%%%%%%%%%%%%%%%%%%%%%%%%%%%%%%%%%%%%%%%%%%%%%%%%%%%%%%%%%%%%%%%%%%%%%%%%%%%%%%%%%%%%%%
\section{Physical model and numerical implementation}
%%%%%%%%%%%%%%%%%%%%%%%%%%%%%%%%%%%%%%%%%%%%%%%%%%%%%%%%%%%%%%%%%%%%%%%%%%%%%%%%%%%%%%%%%%%%%%%%%%%%%%%%%%%%%%%%%%%%%%%%%%%%%%%%%%%%%%%%%%%%%%%%
%%%%%%%%%%%%%%%%%%%%%%%%%%%%%%%%%%%%%%%%%%%%%%%%%%%%%%%%%%%%%%%%%%%%%%%%%%%%%%%%%%%%%%%%%%%%%%%%%%%%%%%%%%%%%%%%%%%%%%%%%%%%%%%%%%%%%%%%%%%%%%%%
\label{Sec:Problem_solved}

This section presents the numerical scheme used in Apar-T.
Broadly speaking, Apar-T is a parallel electromagnetic relativistic 
3D PIC code with a staggered grid, where the fields are
integrated via Faraday and Maxwell-Amp\`{e}re equations, 
currents are computed by charge conserving volume weighting (CIC),
and fields are interpolated {with the same CIC volume weighting method}.

%%%%%%%%%%%%%%%%%%%%%%%%%%%%%%%
\subsection{The PIC plasma}
%%%%%%%%%%%%%%%%%%%%%%%%%%%%%%%
\label{sec:problem_solved}
The code simulates the time-evolution of charged particles under the action of the electromagnetic fields that they generate,
and the evolution of these fields.

Plasmas in nature contain millions to tens of billions of particles per Debye sphere, and relevant microphysical phenomena spread over 
numerous Debye lengths. It is impossible to track these particles one by one. 
Rather, the numerical particles represent numerous real particles, and are consequently called \textit{superparticles}.

A superparticle represents either $p$ real ions (having then a rest mass $m_\mathrm{sp}=p\times m_\mathrm{i}$
and a charge $q_\mathrm{sp}=p\times q_\mathrm{i}$), or $p$ real electrons (having then a rest mass  
$m_\mathrm{sp}=p\times m_\mathrm{e}$ and a charge $q_\mathrm{sp}=p\times q_\mathrm{e}$).
The ratio of ion to electron charge is always $q_\mathrm{i}/q_\mathrm{e}=-1$, 
while that of their rest masses $m_\mathrm{i}/m_\mathrm{e}$
can be freely specified. Pair plasmas can thus be simulated.
In Apar-T the number of real particles per superparticles $p$ is the same for all superparticles
at all times, but other codes can introduce superparticle splitting {\citep{Fujimoto2006b,Haugboelle2012,Cerutti2013}}.

We denote the physical size of a grid cell by $X_0$, a reference number of superparticles per cell by $\rho_\mathrm{sp}^0$
(including both ion superparticles and electron superparticles),
and its associated number density of electrons by $n_\lec^0$.
Initially, the plasma is assumed to be quasi-neutral, 
in the sense that we load the same number of ion superparticles and electron superparticles in each cell.
We have the relation
\begin{equation}\label{equ:neandrho}
 2n_\lec^0\times X_0^3 = \mathrm{number\,of\,real\,particles\,in\,one\,cell} = p\times \rho_\mathrm{sp}^0.
\end{equation}

The equations governing the superparticle plasma are the equation of motion with the Lorentz force for each superparticle, 
and Maxwell equations coupled to the superparticle motions by the current:
\begin{subequations}
 \label{equ:dynamic}
 \begin{align}
  &\frac{\dif}{\dif t}(\gamma_\mathrm{sp} \textbf{v}_\mathrm{sp}) = \frac{q_\mathrm{sp}}{m_\mathrm{sp}} \left(\textbf{e} + \frac{\textbf{v}_\mathrm{sp}}{c}\wedge \textbf{b}\right), \label{equ:dyn1} \\
  &\frac{\dif}{\dif t}\textbf{x}_\mathrm{sp} = \textbf{v}_\mathrm{sp}, \label{equ:dyn2} \\
  &\frac{\partial\textbf{b}}{\partial t} = -c\,\nabla\wedge\textbf{e}, \label{equ:dyn3} \\
  &\frac{\partial\textbf{e}}{\partial t} = c\,\nabla\wedge\textbf{b} - \frac{1}{\epsilon_0}\textbf{j}, \label{equ:dyn4} \\
  &\textbf{j}                            = \sum_\mathrm{sp} q_\mathrm{sp} \textbf{v}_\mathrm{sp} S(\textbf{x}-\textbf{x}_\mathrm{sp}). \label{equ:dyn5}
 \end{align}
\end{subequations}
Here, $c$ is the speed of light, $\textbf{e}$ the electric field, 
$\textbf{b}=c\textbf{B}$ is $c$ times the magnetic field;
$q_\mathrm{sp}$, $m_\mathrm{sp}$, $\gamma_\mathrm{sp}$, $\textbf{v}_\mathrm{sp}$, and $\textbf{x}_\mathrm{sp}$ are the charge, mass, Lorentz factor, 
velocity, and position of the superparticle number $\mathrm{sp}$.
The fields are stored on a grid, and a consequence is that the superparticles are seen by the grid as having a finite shape $S$, linked to 
the interpolation scheme used in the code.

In Apar-T we do not integrate the two other Maxwell equations
because if they hold initially, 
then fulfilling the equation of conservation of charge and
Eqs.~\ref{equ:dyn3} and \ref{equ:dyn4} at all times insures that they remain correct to round-off errors.
That the current is indeed computed in a charge conserving way is detailed in Appendix~\ref{sec:current_simple}.
Initially, the fields and the charge density are correctly built by setting 
a magnetic field satisfying $\nabla\wedge\textbf{b} = \mu_0\textbf{j}$,
a null electric field, and by placing the superparticles by pairs with one ion superparticle 
and one electron superparticle on top of each other so that the charge density is zero.

%%%%%%%%%%%%%%%%%%%%%%%%%%%%%%%
\subsection{Numerical implementation of Apar-T}
%%%%%%%%%%%%%%%%%%%%%%%%%%%%%%%
The code is based on the PIC program Tristan written by O.~Buneman \citep{Matsumoto1993}, parallelized by \citet{Messmer1998}.
We largely modified its structure, that now uses Fortran modules and allows for more flexibility, 
including the possibility of switching between different initial conditions or 
boundaries by changing entries in a configuration textfile.

The need for a deep understanding of simulation methods to interpret their results,
as is the case for the test problems of Sect.~\ref{Sec:Example_validation},
motivated a detailed description of Apar-T. 
This section presents important points. More details (numerical scheme and parallelization efficiency)
can be found in Appendix~\ref{Sec:Numerical_implementation}.

Briefly, the global integration scheme of Eqs.~\ref{equ:dyn1}-\ref{equ:dyn5} is a time-centered and time-reversible leap-frog scheme, 
and is second-order accurate in time and space. 
The electric and magnetic fields are stored on a staggered Yee lattice, 
which allows for a simple integration explicit in time of Eqs.~\ref{equ:dyn3} and \ref{equ:dyn4} (without the current). 
The current is computed with the volume change of the superparticles in the grid cells, 
and added in a time explicit way to the integration of the electric field (Eq.~\ref{equ:dyn4}).

%%%%%%%%%%%%%%%%%%%%%%%%%%%%%%%
\subsubsection{Temporal and spatial discretization, normalization}

\label{sec:steps}
The spatial discretization of the code is $X_0 = d_\lec^0/n_x$, a fraction $n_x$ of 
the electron skin depth $d_\lec^0=c/\omega_\mathrm{pe}^0$,
where the electron plasma pulsation is 
$\omega_\mathrm{pe}^0 = \sqrt{n_\lec^0e^2/(\epsilon_0 m_\lec)}$,
with $-e$ and $m_\lec$ the electron charge and rest mass.
The timestep $\Delta t$ is a fraction $n_t$ of the electron plasma period: $\Delta t = T^0_\mathrm{pe}/n_t$,
with $T_\mathrm{pe}^0 = 2\pi/\omega_\mathrm{pe}^0$.
We stress that the superscript ${}^0$ is used for quantities based on the reference density $n_\lec^0$.

Spatial quantities are normalized by the cell length $X_0$, and normalized quantities are then denoted with a tilde.
For example, the electron Debye length $\lambda_\mathrm{De}^0 = \sqrt{\epsilon_0 T_\lec/(n_\lec^0e^2)} = v_{\therm,\lec}/\omega_\mathrm{pe}^0$
(with $v_{\therm,\lec} = \sqrt{T_\lec/m_\lec}$) has for normalized 
counterpart $\tilde{\lambda}_\mathrm{De}^0 = \lambda_\mathrm{De}^0/X_0 = n_x v_{\therm,\lec}/c$.

%%%%%%%%%%%%%%%%%%%%%%%%%%%%%%%
\subsubsection{Superparticle volume}

\label{sec:volume}
The use of a grid for PIC algorithms implies that the fields are
known at grid nodes, and that information relative to the superparticles (charge and current) need to be interpolated
on the grid. This interpolation is equivalent to considering the superparticles as clouds of charge of finite
extension \citep{Langdon1970}. The shape of the cloud then determines the interpolation formula.

In our case, a superparticle is assumed to be a cube of volume $V_\mathrm{sp} = X_0^3$,
and the current it produces is calculated by the change of the volume of the superparticle in the cell
containing its center. This interpolation scheme is equivalent to linear weighting (CIC) and is exactly charge conserving.
Details of the numerical implementation of the current computation can be found in Appendix~\ref{Sec:Numerical_implementation};
in Sect.~\ref{Sec:computer_vs_real_plasma} the implications of the superparticle finite sizes are discussed.

%%%%%%%%%%%%%%%%%%%%%%%%%%%%%%%
\subsubsection{Particle initialization in momentum space}

Several momentum distributions can be loaded: Maxwell-Boltzmann distribution with anisotropic temperature, boosted Maxwell-Boltzmann distribution, 
waterbag distribution, or Maxwell-J\"uttner distribution. 
We have not found in the literature a method for initializing the Maxwell-J\"uttner distribution 
when both the bulk velocity and the temperature are relativistic,
so we present one in Sect.~\ref{Sec:Load_particles}.

%%%%%%%%%%%%%%%%%%%%%%%%%%%%%%%
\subsubsection{Input, output, and data analysis}
\label{sec:dataanalysis}

Data output and data analysis can be very time consuming for large scale simulations.
To reduce data storage and writing time, 
we implemented parallel output in HDF5 format\footnote{\url{http://www.hdfgroup.org/HDF5/}}.
Files are written according to the .h5part format, and can be read
by the advanced visualization and data-analysis software VisIt \citep{HPV:VisIt},
which is fully parallel, but also by a reader in Python.
The .h5part files can either contain the whole simulation data, and then be used to 
restart the simulation, or be lighter with only a fraction of the particles.
These lighter files then include cell-averaged quantities related to the particles, 
such as the mean particle velocity or number, temperature, highest energy, etc.

We implemented the VisIt Libsim in situ
library \citep{visit_in_situ} into Apar-T. In this way, VisIt is
able to connect to the simulation while it is running and to
access the solver's data at the current timestep. 
It can then perform data visualization and data analysis without the need
to write data on the hard drive.
This feature is fully parallelized, by exploiting the data-distribution model of the code, and as such is
not restricted to the current parallelization model.
It allows the data-IO from memory to hard drive 
-- which is a major reason for slow-down of simulations using big data-sets --
to be significantly reduced.
For example, a volume-rendering of 3D data performed in situ takes less time than dumping 10GB of data to the disk.
VisIt in situ is also well suited to monitor ongoing simulations 
and to single-step through the execution
and is, in this way, of great help for debugging.

Finally, a set of test problems has been
implemented. Based on a Python script, these problems can be
automatically run to check code sanity after modifications.

%%%%%%%%%%%%%%%%%%%%%%%%%%%%%%%%%%%%%%%%%%%%%%%%%%%%%%%%%%%%%%%%%%%%%%%%%%%%%%%%%%%%%%%%%%%%%%%%%%%%%%%%%%%%%%%%%%%%%%%%%%%%%%%%%%%%%%%%%%%%%%%%
%%%%%%%%%%%%%%%%%%%%%%%%%%%%%%%%%%%%%%%%%%%%%%%%%%%%%%%%%%%%%%%%%%%%%%%%%%%%%%%%%%%%%%%%%%%%%%%%%%%%%%%%%%%%%%%%%%%%%%%%%%%%%%%%%%%%%%%%%%%%%%%%
\section{Examples and code validation}
%%%%%%%%%%%%%%%%%%%%%%%%%%%%%%%%%%%%%%%%%%%%%%%%%%%%%%%%%%%%%%%%%%%%%%%%%%%%%%%%%%%%%%%%%%%%%%%%%%%%%%%%%%%%%%%%%%%%%%%%%%%%%%%%%%%%%%%%%%%%%%%%
%%%%%%%%%%%%%%%%%%%%%%%%%%%%%%%%%%%%%%%%%%%%%%%%%%%%%%%%%%%%%%%%%%%%%%%%%%%%%%%%%%%%%%%%%%%%%%%%%%%%%%%%%%%%%%%%%%%%%%%%%%%%%%%%%%%%%%%%%%%%%%%%
\label{Sec:Example_validation}

%%%%%%%%%%%%%%%%%%%%%%%%%%%%%%%
\subsection{Cold plasma modes}
%%%%%%%%%%%%%%%%%%%%%%%%%%%%%%%
\label{Sec:cold_modes}

\begin{table*}[!ht]
\caption{\label{tab:results_cold_modes}Theoretical versus experimental pulsations for a magnetized cold plasma. 
$\omega_\perp^\mathrm{simu}$ and $\omega_{//}^\mathrm{simu}$ are the peak in the spectra of the parallel and perpendicular
total particle momentum from the simulations. The right columns give the theoretical pulsations 
for a cold plasma. See the main text for the formula.
Here $m_\ionperso/m_\lec=49$, for the expression of the plasma pulsation $\omega_\mathrm{P}$, and $m_\ionperso/m_\lec=\infty$ 
for the expressions of the right and left cut-off pulsations ($\omega_\mathrm{R}$ and $\omega_\mathrm{L}$).
The duration of each simulation is $100\,T_\mathrm{pe}$, implying a spectral precision 
of $\Delta\omega = 0.01\omega_\mathrm{pe}$. Ion and electron temperatures are equal.
{$\tilde{\lambda}_\mathrm{De}$ and $\tilde{r}_\mathrm{ce}$ are the electron Debye length and Larmor radius in units of cell number.
All pulsations are in units of $\omega_\mathrm{pe}$.}}
\centering
% NB: same order as in the text file, except for the line marked 'this one'.
\begin{tabular}{l|lllllllll|lll}
$\omega_\mathrm{ce}$ & $n_t$ & $n_x$ & $\rho_\mathrm{sp}$ & $v_{\therm,\lec}/c$ & $\tilde{\lambda}_\mathrm{De}$ & $\tilde{r}_\mathrm{ce}$ & $\omega_\mathrm{ce}\Delta t$ & $\omega_{//}^\mathrm{simu}$ & $\omega_\perp^\mathrm{simu}$ & $\omega_\mathrm{P}^\mathrm{theory}$ & $\omega_\mathrm{L}^\mathrm{theory}$ & $\omega_\mathrm{R}^\mathrm{theory}$ \\
\hline
    0.5 & 4000  &   32  & 4  & 0.04  & 1.28  & 2.56  & $7.9\cdot10^{-4}$ & 1.01 & 0.79, 1.28 & 1.010 & 0.78  & 1.28  \\
  \idem & \idem & \idem & 16 & \idem & \idem & \idem & \idem           & 1.01 & 0.79, 1.28 & \idem & \idem & \idem  \\
\hline
  1     & 1000  & 25    & 16 & \idem & 1     & 1     & $6.3\cdot10^{-3}$ & 1.01 & 0.64, 1.62 & 1.010 & 0.62  & 1.62~~~~Fig.~\ref{fig:PDF_B} \\
  \idem & 4000  & 32    & 4  & \idem & 1.28  & 1.28  & $1.6\cdot10^{-3}$ & 1.01 & 0.64, 1.62 & \idem & \idem & \idem \\
\hline
  2     & 1000  & 16    &  4 & \idem & 0.64  & 0.32  & $1.3\cdot10^{-2}$ & 1.01 & 0.43, 2.64 & 1.010 & 0.41  & 2.41 \\  % this one  
  \idem & 1000  & 25    & 16 & \idem & 1     & 0.5   & \idem             & 1.01 & 0.46, 2.40 & \idem & \idem & \idem \\
  \idem & 4000  & 43    &  4 & \idem & 1.72  & 0.86  & $3.1\cdot10^{-3}$ & 1.01 & 0.46, 2.40 & \idem & \idem & \idem \\
  \idem & 4000  & 64    &  4 & \idem & 2.56  & 1.28  & \idem             & 1.01 & 0.46, 2.40 & \idem & \idem & \idem \\
\hline
  4     & 1000  & 25    &  4 & \idem & 1     & 0.25  & $2.5\cdot10^{-2}$ & 1.01 & 0.32, 4.22 & 1.010 & 0.24  & 4.24 \\
  \idem & 2000  & 128   &  4 & \idem & 5.12  & 1.28  & $1.3\cdot10^{-2}$ & 1.01 & 0.32, 4.21 & \idem & \idem & \idem \\
\end{tabular}
\end{table*}

A first test is to simulate a thermal plasma at rest and to observe its modes of oscillation. 
It is an easy test if we focus on the pulsations 
for modes of zero wavevector, $\textbf{k}=0$.
To do so, we compute at each timestep $t_j$ the sum of the momentum of the particles along a given
direction, for example the $x$-direction, $\sum_\mathrm{sp} \gamma_\mathrm{sp}(t_j)\,v_{\mathrm{sp},x}(t_j)$,
where the summation runs over all the electron superparticles in the simulation 
(of Lorentz factor $\gamma_\mathrm{sp}$ and velocity $v_{\mathrm{sp}}$).
This sum is equivalent to the volume integral of the momentum, 
and is thus equal to the $\textbf{k}=0$ Fourier mode of the electron momentum,
with a spectral resolution of $2\pi/[\mathrm{box\,size\,in\,units\,of\,electron\,skin\,depth}]$. 
We then perform a Fourier transform in time to extract the pulsations of oscillation,
\begin{equation}\label{equ:Fourier}
 F_x(\omega) = \sum_{j}\,\exp(i\omega t_j)\times \sum_\mathrm{sp} \gamma_\mathrm{sp}(t_j)\,v_{\mathrm{sp},x}(t_j),
\end{equation}
{and similarly along $y$ and $z$.}

For all the simulations of this section, the initial state consists of a homogeneous plasma at rest, 
with superparticles loaded according to a classical Maxwell-Boltzmann distribution.
A uniform background magnetic field along $\textbf{z}$ is set up for the magnetized plasma case.
Periodic boundaries are used.

When the background magnetic field $\textbf{B}_0$ is strong (electron cyclotron pulsation larger 
than electron plasma pulsation, $\omega_\mathrm{ce} \gg \omega_\mathrm{pe}$), 
the particle trajectories are Larmor gyrations in $\textbf{B}_0$, unperturbed by collective effects 
such as Langmuir oscillations. This example then probes the accuracy of the particle motion integrator. 
When the background magnetic field is weaker ($\omega_\mathrm{ce} \lesssim \omega_\mathrm{pe}$), 
the dynamics is set by Langmuir oscillations possibly modified by $\textbf{B}_0$. 
These oscillations involve the creation of electric fields by local charge imbalance. 
The fields set the particles into motion, and the particles then oscillate because of their finite inertia. 
Several parts of the algorithm are thus involved: electric field production and propagation, as well as particle motion.

%%%%%%%%%%%%%%%%%%%%%%%%%%%%%%%
\subsubsection{No background magnetic field}
With no background magnetic field, the only cold modes are the
electromagnetic transverse wave of dispersion relation $\omega_\mathrm{Tr}^2 = \omega_\mathrm{P}^2  [1 + (kc/\omega_\mathrm{P})^2]$
and the electrostatic Langmuir oscillation at the plasma frequency $\omega_\mathrm{P} = \omega_\mathrm{pe} (1 + m_\lec/m_\ionperso)^{1/2}$.
The latter is modified by thermal effects to a wave of dispersion relation 
$\omega_\mathrm{La}^2 = \omega_\mathrm{P}^2  (1 + 3k^2\lambda_\mathrm{D}^2)$.
Consequently, we expect $F_a(\omega)$ to peak at $\omega_\mathrm{La}(\textbf{k}=0)=\omega_\mathrm{Tr}(\textbf{k}=0)=\omega_\mathrm{P}$.

\begin{figure}
 \centering
 \def\svgwidth{\columnwidth}
 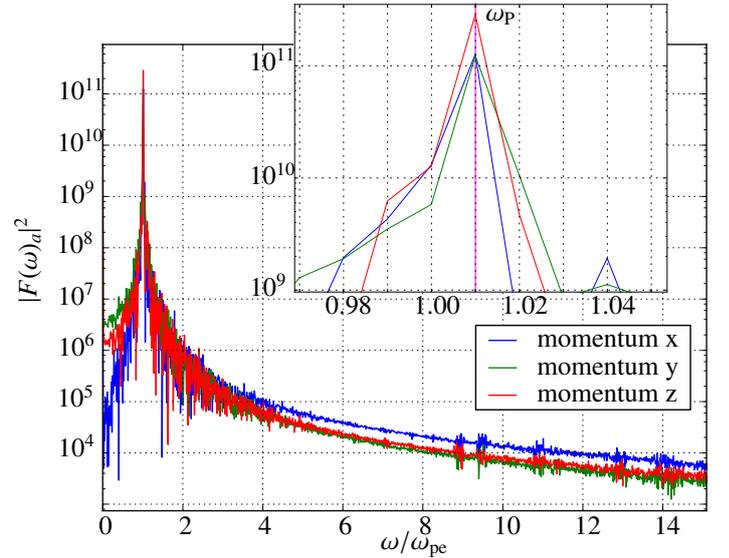
 \caption{\label{fig:PDF_B_zero}Power spectra of the total momentum of the electrons $|F_a|^2$ (Eq.~\ref{equ:Fourier}) with $a=x$, $y$, or $z$
          in the unmagnetized case.
          The inset is a zoom around the peak.
          Here $n_x=25$, $n_t=500$, $\rho_\mathrm{sp}=16$, $v_{\therm,\lec}=0.04c$, $T_\ionperso=T_\lec$, 
          box size of $25\times25\times25$ cells, duration of 100\,$T_\mathrm{pe}$.}
\end{figure}

Our simulations span a large range of parameters:
$n_x$ and $n_t$ (spatial and temporal resolution, 
see Sect.~\ref{sec:steps}) vary between 
$5$ and $50$ and between $300$ and $2000$, respectively;
$\rho_\mathrm{sp}$ (the number of superparticles per cell) varies between 
$4$ and $32$; and the thermal velocity $v_{\therm,\lec}=\sqrt{T_\lec/m_\lec}$ of the electrons between
$0.04c$ and $0.1c$ (the ions have the same temperature as the electrons).
This results in Debye lengths between 0.2 and 2 cells.
We checked that the simulation box size, comprised between 10 and 30 Debye lengths, does not influence the results.

We use a mass ratio $m_\ionperso/m_\lec = 49$, which results in $\omega_\mathrm{P} = 1.010\,\omega_\mathrm{pe}$.
Our simulations last $100\,T_\mathrm{pe}$, so that the frequency resolution is $\Delta\omega = 2\pi/(100T_\mathrm{pe}) = 0.01\omega_\mathrm{pe}$.

For all these parameters, we find that the position and width of the frequency peak 
are always the same as in Fig.~\ref{fig:PDF_B_zero}. It coincides with the theoretical plasma pulsation,
which is expected because our temporal spectra are for the wavenumber $\textbf{k}=0$, and because
$\omega(\textbf{k}=0)=\omega_\mathrm{P}$ for the two modes present in this situation.

This is the case even for simulations where the Debye length is not resolved. However, an under-resolved Debye length leads to more numerical heating
and can trigger instabilities in situations less trivial than a thermal plasma at rest (see Appendix~\ref{Sec:numerical_stability}), 
so that we have not pushed our investigations too far in this direction.

The main difference between the simulations is that less resolved ones
present noisier spectra, and thus more fluctuations. The increase of fluctuation level with decreasing resolution
is a universal feature of PIC simulations and is explored in more detail in Sect.~\ref{Sec:computer_vs_real_plasma}.

%%%%%%%%%%%%%%%%%%%%%%%%%%%%%%%
\subsubsection{With a background magnetic field}
\label{sec:fluctuation_with_a_back_mag_field}
In a uniform and cold magnetized plasma, the plasma modes depend solely on
the ratio of the electron cyclotron pulsation $\omega_\mathrm{ce}=eB/m_\lec$ 
to the electron plasma pulsation $\omega_\mathrm{pe}$ (for a fixed $m_\ionperso/m_\lec$).
This ratio sets the relative importance of individual particle motion ($\omega_\mathrm{ce}$) against
collective effects ($\omega_\mathrm{pe}$). 
In addition, the background magnetic field favors a direction, thus making the mode spectrum anisotropic
(however, the mode pulsations for $\textbf{k}=0$ remain independent of the direction of the wavevector).

\begin{figure}
 \centering
 \def\svgwidth{\columnwidth}
 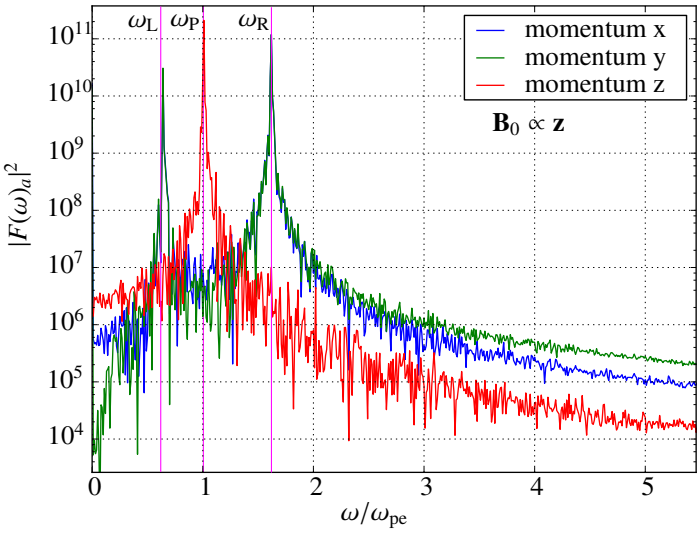
 \caption{\label{fig:PDF_B}Power spectra of the total momentum of the electrons $|F_a|^2$ (Eq.~\ref{equ:Fourier}) with $a=x$, $y$, or $z$
          in the magnetized case.
          Parameters are given in Table~\ref{tab:results_cold_modes} (case labeled Fig.~\ref{fig:PDF_B}).
          The purple lines are the pulsations $\omega_\mathrm{L}^\mathrm{theory}$, $\omega_\mathrm{P}^\mathrm{theory}$, 
          and $\omega_\mathrm{R}^\mathrm{theory}$.}
\end{figure}

For wavevectors parallel to the magnetic field, the Langmuir oscillation 
remains unchanged because the oscillations of the particles are longitudinal,
and thus along $\textbf{B}_0$ and unaffected by the magnetic field.
On the other hand, the electromagnetic wave with $\textbf{k}$ along $\textbf{B}_0$ separates into two branches, one starting at
$\omega_{\textbf{k}=0}=\omega_\mathrm{L} =  0.5[(\omega_\mathrm{ce}^2+4\omega_\mathrm{pe}^2)^{1/2}-\omega_\mathrm{ce}]$
and the other at
$\omega_{\textbf{k}=0} = \omega_\mathrm{R} = 0.5[(\omega_\mathrm{ce}^2+4\omega_\mathrm{pe}^2)^{1/2}+\omega_\mathrm{ce}]$
\citep[see, e.g.,][chap.~4]{Fitzpatrick},
in both cases with particles oscillating in the transverse plane, i.e., perpendicular to $\textbf{B}_0$. 
Two other branches appear, but they start at $\omega_{\textbf{k}=0} = 0$ and will thus not appear in $F_a(\omega)$.

For wavevectors perpendicular to the magnetic field,
the presence of the background magnetic field
deforms the Langmuir oscillation $\omega(k) = \omega_\mathrm{P}$ into
a branch starting from
$\omega_{\textbf{k}=0}=\omega_\mathrm{L}$, with particle oscillations in the plane perpendicular to $\textbf{B}_0$.
The transverse electromagnetic wave still exists with particle oscillations along $\textbf{B}_0$, unaffected by the magnetic field.
Another branch appears, which is a deformation of the transverse electromagnetic wave for particle oscillations 
not along $\textbf{B}_0$, and starts at $\omega_{\textbf{k}=0}=\omega_\mathrm{R}$ with oscillations in the plane perpendicular to $\textbf{B}_0$.
Another branch appears, starting at $\omega_{\textbf{k}=0} = 0$.

All in all, we expect to find a peak at $\omega=\omega_\mathrm{P}$ for the component of the momentum parallel to $\textbf{B}_0$,
and two peaks at $\omega=\omega_\mathrm{L}$ and $\omega_\mathrm{R}$ for the component of the momentum perpendicular to $\textbf{B}_0$.
We ran the set of simulations described in Table~\ref{tab:results_cold_modes},
with a ratio $\omega_\mathrm{ce} / \omega_\mathrm{pe}$ ranging from 0.5 to 4, 
$n_x$ from 16 to 128, $n_t$ from 1000 to 4000, and $\rho_\mathrm{sp}$ from 4 to 16,
and we did find the required pulsation peaks for $F_a(\omega)$ (see Fig.~\ref{fig:PDF_B} for a sample spectrum).
The positions and widths of these three peaks are almost constant within our parameter range.

We note that the peak positions and widths are not changed even for cases where the thermal Larmor radius
$\tilde{r}_\mathrm{ce} = r_\mathrm{ce}/X_0 = v_{\therm,\lec}/(X_0\omega_\mathrm{ce}) = \tilde{\lambda}_\mathrm{De}/(\omega_\mathrm{ce}/\omega_\mathrm{pe})$ 
is not resolved. 
It is expected that the resolution of the Larmor radius by the grid is of no importance 
to describe particle trajectories in constant fields, because
the interpolation of these fields from grid points to superparticle
position gives the same result regardless of the grid size if the fields are constant. 
The relevant constraint is instead that $r_\mathrm{ce}$ should be resolved along the trajectory,
$v_\mathrm{sp}\Delta t < r_\mathrm{ce}$, with $v_\mathrm{sp}$ the superparticle velocity.
This relation is equivalent to $\Delta t < \omega_\mathrm{ce}^{-1}$.

However, for simulations with under-resolved Larmor radii the electric field energy starts to behave abnormally
after some tens of plasma pulsations, and 
energy conservation curves present an exponential heating (see Sect.~\ref{sec:energy_conservation})
that can lead to dramatic consequences.
This parameter range must be avoided.

%%%%%%%%%%%%%%%%%%%%%%%%%%%%%%%
\subsubsection{Energy conservation}
\label{sec:energy_conservation}

Independent of the strength of the background magnetic field, we observe a linear
increase of {the total} energy with time that is due to interactions of 
particles with the grid (see Appendix~\ref{sec:globaleffects}).
The growth rate is independent of the size of the timestep
(from $n_t=2000$ down to Courant condition $\Delta t\sim X_0/c$, equivalent to $n_t=2\pi n_x$). 
Its dependence on the number of superparticles per
cell is quite precisely given by $1/\rho_\mathrm{sp}$. However, its dependence on the spatial
resolution $n_x$ and thermal spread $v_{\therm,\lec}$ is less clear. In particular, it does not depend only 
on the product $n_xv_{\therm,\lec}/c = \tilde{\lambda}_\mathrm{De}$. 
The rate increases with increasing $v_{\therm,\lec}$, and decreases with increasing $n_x$. 
Some examples are given for reference in Table~\ref{tab:results_energy_conservation}.

\begin{table}[ht]
\caption{\label{tab:results_energy_conservation}Energy conservation for simulations of a thermal plasma with 
no background magnetic field. The energy increase rate is measured on the total energy normalized by the total initial
energy, while time is again normalized with the electron plasma pulsation $T_\mathrm{pe}$.
Here, $m_\ionperso/m_\lec=49$ and $T_\ionperso=T_\lec$.}
\centering
\begin{tabular}{llll|ll}
$n_t$ & $n_x$ & $\rho_\mathrm{sp}$ & $v_{\therm,\lec}/c$ & $\tilde{\lambda}_\mathrm{De}$ & $T_\mathrm{pe}\frac{\mathrm{d}}{\mathrm{d}t}\frac{E_\mathrm{tot}(t)}{E_\mathrm{tot}(0)}$ \\
\hline
500 to 2000 & 10 & 16 & 0.04  &  0.4  &  $1.6\cdot10^{-4}$ \\
\idem       & 10 & 16 & 0.10  &  1    &  $5.8\cdot10^{-5}$ \\
\idem       & 14 & 16 & 0.07  &  1    &  $3.5\cdot10^{-5}$ \\
\idem       & 18 & 16 & 0.04  &  0.7  &  $5.3\cdot10^{-5}$ \\
300 to 2000 & 25 &  4 & 0.04  &  1    &  $1.0\cdot10^{-4}$ \\
\idem       & 25 & 16 & 0.04  &  1    &  $2.5\cdot10^{-5}$ \\
\idem       & 25 & 32 & 0.04  &  1    &  $1.3\cdot10^{-5}$ \\
\end{tabular}
\end{table}

Simulations with under-resolved Larmor radii $\tilde{r}_\mathrm{ce} < 1$ show an exponential 
(instead of linear) increase of the total energy starting after roughly $40\,T_\mathrm{pe}$.
This numerical instability is believed to arise because field perturbation at wavelength $\lambda=2r_\mathrm{ce}$
and their aliases ($\pm\lambda+n\times2\pi/X_0$, with $n$ an integer and $X_0$ the grid size) 
are allowed to couple when $2X_0 > \lambda$ (see Appendix~\ref{sec:globaleffects}).
An inspection of the energy curves shows that the energy gain is for the kinetic energy.
We note that it did not disturbed the spectra of Sect.~\ref{sec:fluctuation_with_a_back_mag_field}
because they were computed before the heating reached a significant level.
It is interesting to note that this numerical instability is develops slowly, 
so that particles with under-resolved Larmor radii in constant fields can be included in 
simulations if they spend a small amount of time before being heated or before reaching areas 
with smaller magnetic fields.

%%%%%%%%%%%%%%%%%%%%%%%%%%%%%%%
\subsection{Linear growth rates of the counter-streaming instability}
%%%%%%%%%%%%%%%%%%%%%%%%%%%%%%%
\label{Sec:two-stream_linear}

\begin{table*}[ht]
\caption{\label{tab:results_two_stream}{Theoretical versus experimental values of the filamentation growth rate $\tau$. 
Numbers in parenthesis are the discrepancy with respect to $\tau^\mathrm{theory}_\mathrm{max}$, i.e.,
$[\tau^\mathrm{simu}_\mathrm{tot\,en}-\tau^\mathrm{theory}_\mathrm{max}]/\tau^\mathrm{theory}_\mathrm{max}$ and 
$[\tau^\mathrm{simu}_\mathrm{fast\,mode}-\tau^\mathrm{theory}_\mathrm{max}]/\tau^\mathrm{theory}_\mathrm{max}$.}
$T_\mathrm{pe}$ is the plasma period comprising all electrons, and $d_\lec$ the plasma skin depth based on this period.
We recall that $d_\lec$ corresponds to $n_x$ cells. We also give the PIC plasma parameter 
$\Lambda_p = n_\mathrm{e,sp}\lambda_\mathrm{De}^3 = \rho_\mathrm{sp}(n_xv_\mathrm{th}/c)^3$ (see Sect.~\ref{Sec:plasma_behavior}).
}
\centering
\begin{tabular}{l|lllll|lll|l}
$\beta_0$ & $n_t$ & $n_x$ & $\rho_\mathrm{sp}$ & $v_{\therm}/c$ & Box size in $d_\lec$ & $\Lambda_p$ & $\tau^\mathrm{simu}_\mathrm{tot\,en}/T_\mathrm{pe}$ & $\tau^\mathrm{simu}_\mathrm{fast\,mode}/T_\mathrm{pe}$ & $\tau^\mathrm{theory}_\mathrm{max}/T_\mathrm{pe}$ \\
\hline
0.95  & 1000 & 20    & 16    & 0.1   & $3\times3\times30$ &     & 0.29~~(38\%) &              & 0.21  \\
\idem & \idem& \idem & 100   & \idem & \idem              &     & 0.27~~(29\%) &              & \idem \\
\idem &  250 & 20    & 16    & 0.1   & $9\times9\times30$ &     & 0.27~~(29\%) & 0.24~~(14\%) & \idem \\
\hline
% nx=20, nt=1000
0.995 & 1000 & 20    &  4    & 0.1   & $3\times3\times30$ & 32  & 0.52~~(44\%) &              & 0.36  \\
\idem & \idem& \idem & 16    & \idem & \idem              & 128 & 0.45~~(25\%) &              & \idem \\
\idem & \idem& \idem & 80    & \idem & \idem              & 640 & 0.43~~(19\%) &              & \idem \\%~~~~~~~~~~~Fig.\,\ref{fig:twostream1} \\
\idem & \idem& \idem & 160   & \idem & \idem              & 1280& 0.43~~(19\%) &              & \idem \\
\idem & \idem& \idem & 400   & \idem & \idem              & 3200& 0.41~~(14\%) &              & \idem \\
\idem & \idem& \idem & 560   & \idem & \idem              & 4480& 0.42~~(17\%) &              & \idem \\
% nx=20, nt=500
\idem & 500  & 20    & 4     & 0.1   & \idem              &  32 & 0.51~~(42\%) &              & \idem \\
\idem & \idem& \idem & 80    & \idem & \idem              & 640 & 0.41~~(14\%) &              & \idem \\
\idem & \idem& \idem & 128   & \idem & \idem              & 1024& 0.43~~(19\%) &              & \idem \\
\idem & \idem& \idem & 200   & \idem & \idem              & 1600& 0.42~~(17\%) &              & \idem \\
\idem & \idem& \idem & 280   & \idem & \idem              & 2240& 0.42~~(17\%) &              & \idem \\
% nx=20, nt=250
\idem & 250  & 20    & 4     & 0.1   & \idem              &  32 & 0.51~~(42\%) &              & \idem \\
\idem & \idem& \idem & 80    & \idem & \idem              & 640 & 0.41~~(14\%) &              & \idem \\
\idem & \idem& \idem & 128   & \idem & \idem              & 1024& 0.43~~(19\%) &              & \idem \\
\idem & \idem& \idem & 200   & \idem & \idem              & 1600& 0.42~~(17\%) &              & \idem \\
\idem & \idem& \idem & 280   & \idem & \idem              & 2240& 0.42~~(17\%) &              & \idem \\
% nx=40, nt=1000
\idem & 1000 & 40    & 80    & 0.1   & \idem              & 5120& 0.46~~(27\%) &              & \idem \\
% nx=40, nt=2000
\idem & 2000 & 40    &  4    & 0.1   & \idem              &  256& 0.60~~(67\%) &              & \idem \\
\idem & \idem& \idem & 16    & \idem & \idem              & 1024& 0.50~~(39\%) &              & \idem \\
\idem & \idem& \idem & 48    & \idem & \idem              & 3072& 0.46~~(27\%) &              & \idem \\
\idem & \idem& \idem & 74    & \idem & \idem              & 4736& 0.47~~(31\%) &              & \idem \\
\idem & \idem& \idem & 160   & \idem & \idem              & 10024&0.43~~(19\%) &              & \idem \\
\idem & 2000 & 40    & 16    & 0.075 & \idem              & 432 & 0.52~~(44\%) &              & \idem \\
\idem & 2000 & 40    & 16    & 0.05  & \idem              & 128 & 0.58~~(61\%) &              & \idem \\
\idem & 250  & 20    & 80    & 0.1 &$4.5\times4.5\times15$& 128 & 0.41~~(14\%) & 0.38~~(6\%)  & \idem \\
%\idem & \idem& \idem & 16    & \idem & \idem            2 & 640 & 0.41~~(\%) & 0.38~~(6\%)  & \idem \\
\idem & 250  & 20    & 4     & 0.1   & $9\times9\times30$ & 32  & 0.50~~(39\%) & 0.38~~(6\%)  & \idem \\
\idem & \idem& \idem & 16    & \idem & \idem              & 128 & 0.48~~(33\%) & 0.38~~(6\%)  & \idem ~~~~~~Figs.\,\ref{fig:twostream2},~\ref{fig:twostream_Fourier_modes},~\ref{fig:twostream_visit} \\
\idem & \idem& \idem & 80    & \idem & \idem              & 640 & 0.43~~(19\%) & 0.39~~(8\%)  & \idem \\
\idem & 125  & 10    & 128   & 0.1   & \idem              & 128 & 0.42~~(17\%) & 0.39~~(8\%)  & \idem \\
\idem & 1000 & 20    & 16    & 0.1   & \idem              & 128 & 0.46~~(27\%) & 0.38~~(6\%)  & \idem \\
\idem & 500  & 20    & 16    & 0.1   & \idem              & 128 & 0.46~~(27\%) & 0.38~~(6\%)  & \idem \\
\hline
0.999 & 1000 & 20    & 100   & 0.1   & $3\times3\times30$ &     & 0.61~~(15\%) &              & 0.53  \\
\idem &  250 & 20    & 16    & 0.1   & $9\times9\times30$ &     & 0.70~~(32\%) & 0.56~~(6\%)  & \idem \\
\end{tabular}
\end{table*}

\begin{figure}
 \centering
 \def\svgwidth{0.8\columnwidth}
 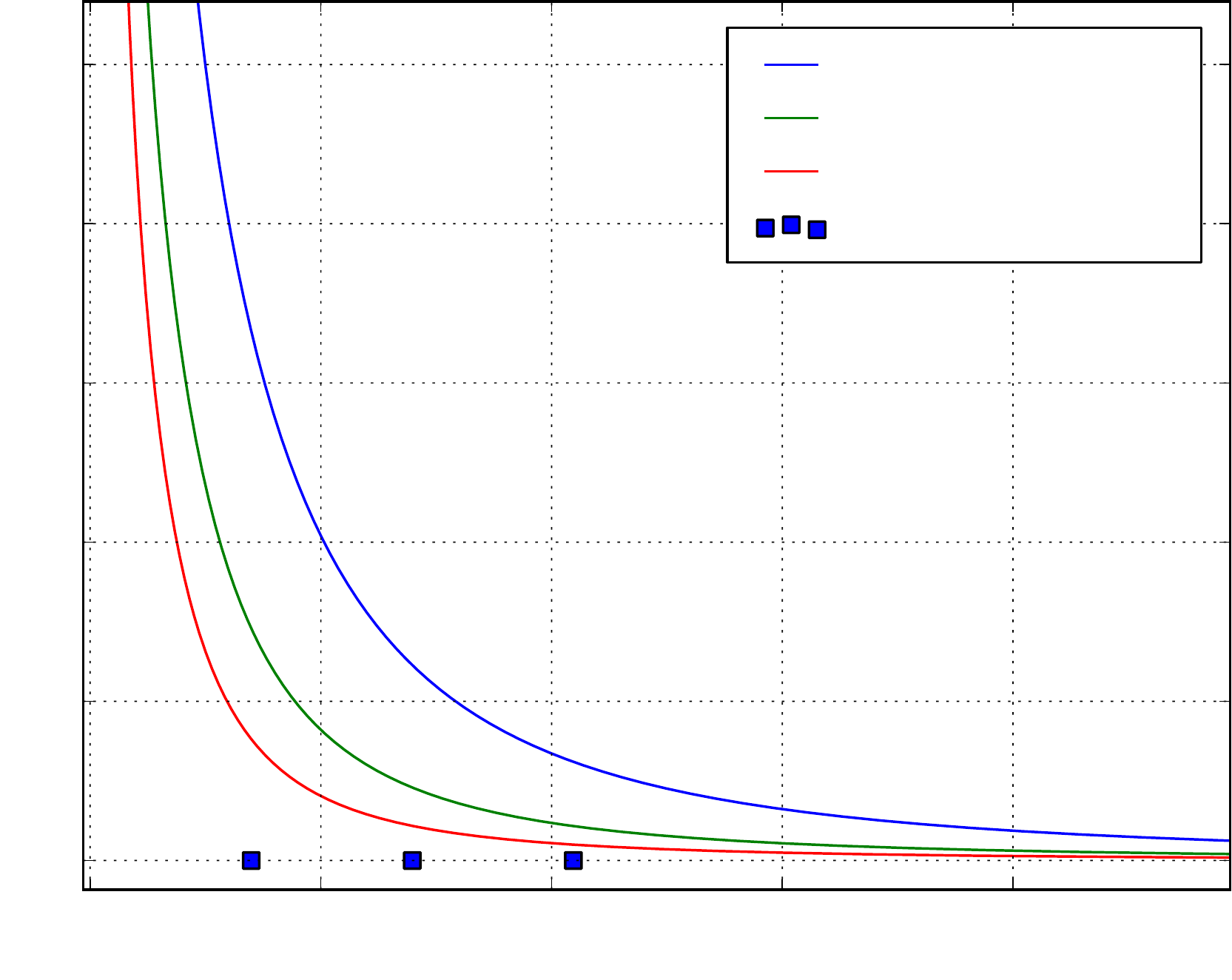
 \caption{\label{fig:theory_growth_rate}{Wavenumber dependance of the growth rate $\tau^\mathrm{theory}(k_\perp)$, 
          represented here as the deviation $[\tau^\mathrm{theory}(k_\perp)-\tau^\mathrm{theory}_\mathrm{max}]/\tau^\mathrm{theory}_\mathrm{max}$.
          We recall that $\tau^\mathrm{theory}_\mathrm{max}=\tau^\mathrm{theory}(k_\perp=+\infty)$.
          From Eq.~69 of \citet{Michno2010}. 
          The squares are the resolution in $k_\perp$ for a box of transverse size 9, 4.5, and 3 $d_\lec$.}}
\end{figure}

\begin{figure}
 \centering
 \def\svgwidth{\columnwidth}
 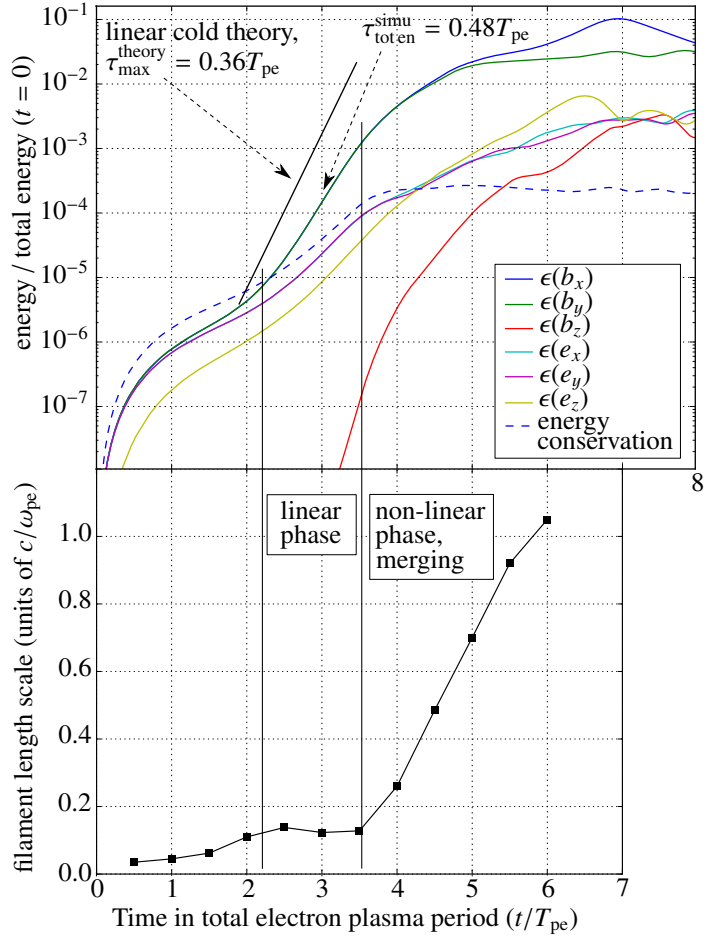
 \caption{\label{fig:twostream2}\textbf{Top}: Energy curves for the filamentation instability 
                                (case labeled Fig.~\ref{fig:twostream2} in Table~\ref{tab:results_two_stream}).
                                They are normalized by the total initial energy $E_\mathrm{tot}^0$, which is mostly the kinetic energy of the particles.
                                {For example $\epsilon(b_x) = \int\!\dif V B_x(t)^2/(2\mu_0) / E_\mathrm{tot}^0$}. 
                                The curve ``energy conservation'' is $(E_\mathrm{tot}^0 - \mathrm{total\,energy}(t))/E_\mathrm{tot}^0$.
                                After $8T_\mathrm{pe}$, the situation is more or less steady.\newline
                                \textbf{Bottom}: Autocorrelation scale of the current amplitude. 
                                }
\end{figure}

\begin{figure}
 \centering
 \def\svgwidth{\columnwidth}
 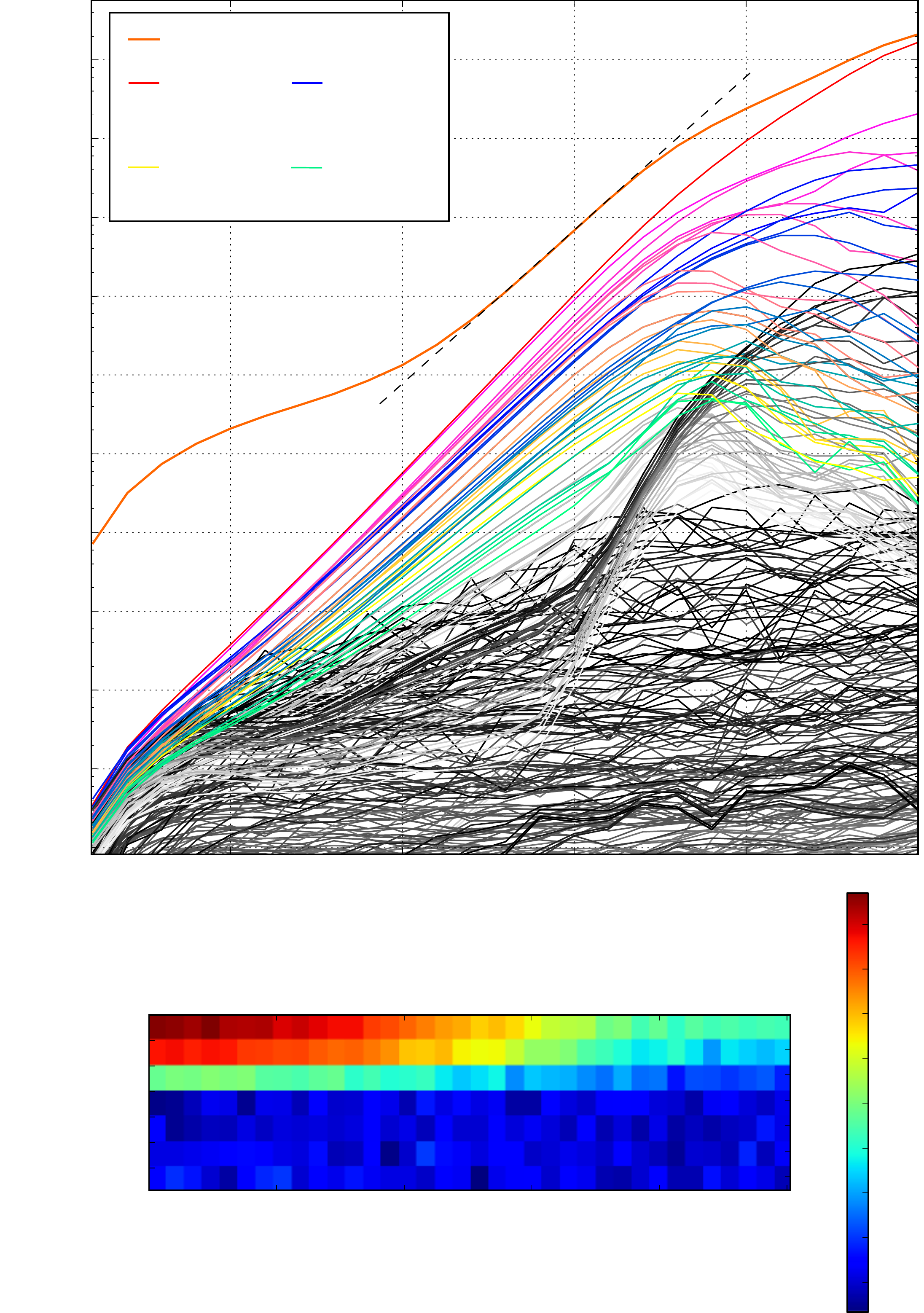
 \caption{\label{fig:twostream_Fourier_modes}\textbf{Top}: Growth of individual Fourier modes for $b_x$
                                (for the run labeled Fig.~\ref{fig:twostream2} in Table~\ref{tab:results_two_stream}). 
                                The modes shown are those for $i=0,1,2$, and $j=0,1,...,25$, 
                                and one mode every 100 modes for the remaining. 
                                We note that the graphic has been cut and that the weak modes 
                                actually fill a continuum down to an energy of $10^{-7}$.
                                The sum of all $320\times 90$ modes is shown in orange. 
                                We recall that mode $(i,j)$ corresponds to $(k_z d_\lec,k_x d_\lec)=2\pi\times 20(i/640,j/180)$.\newline
                                \textbf{Bottom}: Growth map of the Fourier modes, in units of $T_\mathrm{pe}$.}
\end{figure}

Another standard test is to study the linear phase of the counter-streaming instability.
We use relativistic streaming velocities to validate the behavior of the algorithm
for relativistic particle motions. Moreover, since magnetic fields are generated for this range of parameters,
this test also probes the integration of $\textbf{b}$.

The initial setup consists of two unmagnetized and cold counter-streaming electron-positron beams, 
with velocity $\pm\beta_0\hat{\textbf{z}}$ and associated Lorentz factor $\Gamma_0$. 
There is no background magnetic field, and
the particles are loaded according to a drifting Maxwell-Boltzmann distribution.
This situation is unstable, and the kinetic energy of the beam is converted 
into particle thermal kinetic energy and electromagnetic field energy,
the initial perturbation coming from fluctuations due to the finite superparticle number.
The linear instability spectrum is described by a branch comprising an electrostatic longitudinal two-stream mode and 
a transverse electromagnetic filamentation mode, the general case being an oblique mixed mode \citep{Bret2010b}.

\subsubsection{Theoretical model}

We take the growth rates derived analytically by \citet{Michno2010} on the basis of 
a cold two-fluid model and denote this result as theoretical. 
Our thermal velocity $v_\therm$, identical for both species, is low enough to insure 
that thermal effects are negligible \citep[Eq.\,28]{Bret2010b},
but high enough to have a resolved Debye length and to avoid numerical instabilities (Appendix~\ref{Sec:numerical_stability}).
Our parameters are chosen such that the {transverse} filamentation mode always dominates.
The fastest growing modes are those at large wavenumbers perpendicular to the beams, i.e., $k_\perp d_\lec \gg \sqrt{2}/\Gamma_0^{3/2}$
(with $d_\lec=c/\omega_\mathrm{pe}$),
and that grow according to $b_x,\,b_y \propto \exp\{{t/\tau^\mathrm{theory}_\mathrm{max}}\}$ with 
\begin{equation}
 \tau^\mathrm{theory}_\mathrm{max} = \frac{1}{2\pi} \sqrt{\frac{\Gamma_0}{2}}\beta_0^{-1}\,T_\mathrm{pe}.
\end{equation}
The $k_\perp$-dependance of the growth rate is plotted in Fig.~\ref{fig:theory_growth_rate}. 
We see that all modes above a few $d_\lec^{-1}$ quickly reach 
the maximum growth rate $\tau^\mathrm{theory}_\mathrm{max}=\tau^\mathrm{theory}(k_\perp=+\infty)$.

\subsubsection{Method of measurement}

We measure the growth rates of the magnetic fields $b_x$ and $b_y$ with two methods. 
The first is a direct measure on the total energy curve, e.g., $\int\!\dif V\, b_x^2 \propto \exp(2t/\tau)$
(see Fig.~\ref{fig:twostream2}, top, for an illustration). It gives an effective growth rate that we denote by 
$\tau^\mathrm{simu}_\mathrm{tot\,en}$, equal to $0.48T_\mathrm{pe}$ in this case.

The second consists in following the time evolution of the Fourier modes of the fields.
At a fixed time $t_0$, we compute the 2D Fourier transform of the fields in a plane 
$(x,z)$ with a fixed $y$, that we denote by $\mathrm{FT}_{y=y_0}(t_0,k_x,k_z)$. 
We then average the power spectrum over all the planes $y=\mathrm{cst}$ to obtain the power spectrum
$\mathrm{PS}(t_0,k_x,k_z) = \sum_{y_0} |\mathrm{FT}_{y=y_0}(t_0,k_x,k_z)|^2$. We then repeat this procedure for 
several $t_0$. 
The discrete mode spectrum is sampled with $(k_z d_\lec,k_x d_\lec) = 2\pi n_x(i/N_z,j/N_x)$, where 
$N_z$ and $N_x$ are the total number of cells in the $z$ and $x$ directions, and $i=0..N_z/2$, $j=0..N_x/2$. 
The spectral resolution in the direction perpendicular to the beam is thus 
$\Delta k_\perp d_\lec = 2\pi n_x/N_x = 2\pi/(\mathrm{box~width~in~}d_\lec)$.
The squares in Fig.~\ref{fig:theory_growth_rate} represent this spectral resolution for the different box sizes that we use. 

\subsubsection{Results}

Figure~\ref{fig:twostream_Fourier_modes} is an example of the temporal evolution 
of the modes of $b_x$ for the same simulation as in Fig.~\ref{fig:twostream2}.
The sum of all modes grows at the same effective growth rate as the total energy 
in $b_x$ (to within $\pm 1\%$), $\tau^\mathrm{simu}_\mathrm{tot\,en} = 0.48 T_\mathrm{pe}$.
However, the fastest growing modes are those for $k_z = 0$ and $0 \leq k_xd_\lec \leq 5$, 
with $\tau^\mathrm{simu}_\mathrm{fast\,mode} = 0.38T_\mathrm{pe}$, which is 
close to the cold-fluid result $\tau^\mathrm{theory}_\mathrm{max} = 0.36T_\mathrm{pe}$.
It is seen from Fig.~\ref{fig:twostream_Fourier_modes} that the 
large difference between the effective growth rate $\tau^\mathrm{simu}_\mathrm{tot\,en}$ 
and the growth rate of the fastest modes $\tau^\mathrm{simu}_\mathrm{fast\,mode}$ 
is due to a significant contribution of all the modes during 
the whole linear phase. The fastest mode thus never dominates the total energy in the linear phase. 
We suspect that this is due to the large noise level present in PIC simulations.

\begin{figure}
 \centering
 \def\svgwidth{0.99\columnwidth}
 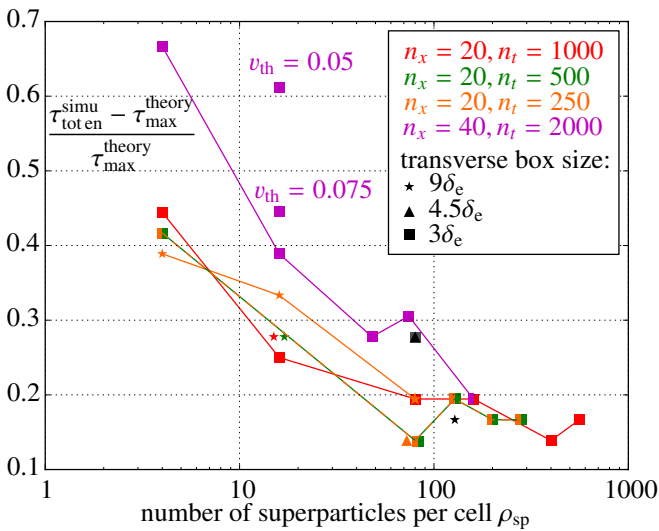
 \caption{\label{fig:error_toten_theory}Difference between the growth rate measured on 
          the total energy curve $\tau^\mathrm{simu}_\mathrm{tot\,en}$ and the theoretical growth rate of the fastest modes
          $\tau^\mathrm{theory}_\mathrm{max}$. 
          Except when labeled otherwise, $v_\mathrm{th}=0.1c$. Each symbol corresponds to a transverse box size,
          and each color to fixed $(n_x,n_t)$.
                                }
\end{figure}

These results hold for all the test simulations that we conducted, which are summarized in Table~\ref{tab:results_two_stream}.
The effective growth rates $\tau^\mathrm{simu}_\mathrm{tot\,en}$ measured on 
the total energy present various levels of discrepancies with $\tau^\mathrm{theory}_\mathrm{max}$, 
between 14\% and 67\%. Figure~\ref{fig:error_toten_theory} shows the dependance of these discrepancies.
There is a small sensitivity with respect to the timestep ($n_t$) and the box size, and an important influence of the spatial resolution ($n_x$).
There is a systematic decrease in the difference when the superparticle number per cell $\rho_\mathrm{sp}$
is increased (all other parameters are kept constant).
Since the fluctuation level in the PIC plasma decreases with increasing $\rho_\mathrm{sp}$,
this indicates that the high fluctuation level excites all the modes and prevents the fastest ones from
dominating the energy.

On the other hand, the growth rates $\tau^\mathrm{simu}_\mathrm{fast\,mode}$ measured on the fastest modes 
differ from $\tau^\mathrm{theory}_\mathrm{max}$ by a more systematic factor, 14\% for $\beta_0=0.95$, 
$7\pm1\%$ for $\beta_0=0.995$, and 6\% for $\beta_0=0.999$.
These systematic differences can be explained by looking at the mode spectrum. 
In all the simulations, the fastest modes 
are for $k_z = 0$ and $0 \leq k_xd_\lec \leq 5-15$
(see, e.g., Fig.~\ref{fig:twostream_Fourier_modes}, bottom).
Given the spectral resolution $\Delta k_\perp d_\lec = 2\pi/(\mathrm{box~length~in~}d_\lec)$, these modes 
actually cover a portion of $k_\perp d_\lec$ where the curves $\tau^\mathrm{theory}(k_\perp)$ 
of Fig.~\ref{fig:theory_growth_rate} vary significantly
and do not yet reach $\tau^\mathrm{theory}_\mathrm{max}$. 
It explains the sign and order of magnitude of the difference $\tau^\mathrm{simu}_\mathrm{fast\,mode} - \tau^\mathrm{theory}_\mathrm{max}$.
It also explains the increasingly better agreement when $\beta_0$ increases.

We note that $\textbf{e}$ and $b_z$ are zero in the linear two-fluid theory, 
and the fact that they are not zero in our simulation
(see Fig.~\ref{fig:twostream2}) 
reflects an early non-linear evolution or the effects of fluctuations and correlations absent from the {linear} model
but present in PIC simulations. 
These differences are discussed further in Sect.~\ref{Sec:computer_vs_real_plasma}.

%%%%%%%%%%%%%%%%%%%%%%%%%%%%%%%
\subsection{Non-linear evolution of the filamentation instability: filament merging}
%%%%%%%%%%%%%%%%%%%%%%%%%%%%%%%
\begin{figure*}
 \centering
 \includegraphics[width=\textwidth]{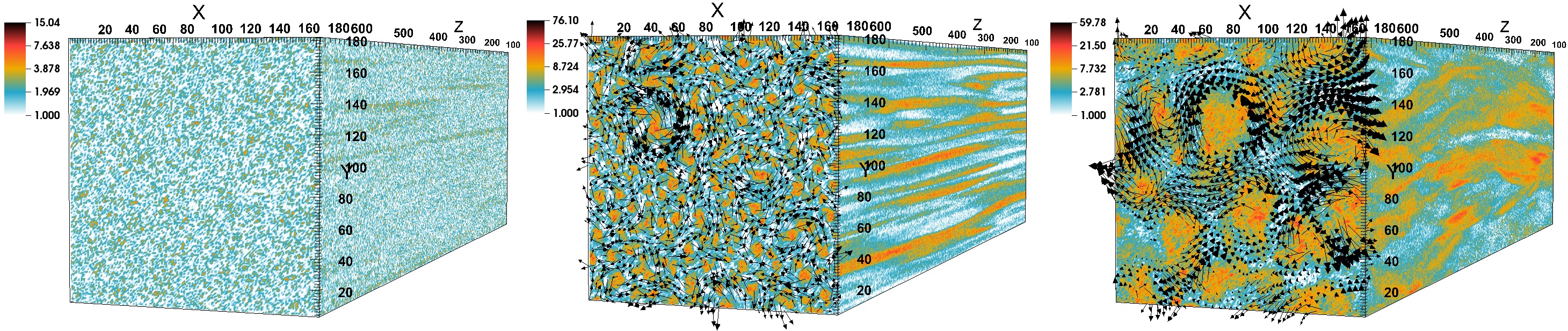}
 \caption{\label{fig:twostream_visit}Three snapshots of the filamentation instability, 
 at times $2.5$, $4$, and $5.5T_\mathrm{pe}$ (from left to right) for the simulation of Fig.~\ref{fig:twostream2}.
 Colors represent the current magnitude per cell (units are number of superparticles per cell times their mean velocity),
 while arrows are the magnetic field. {Lengths are in cell number.}}
\end{figure*}

We now consider the non-linear phase of the filamentation instability.
We study the same counter-streaming configuration as in Sect.~\ref{Sec:two-stream_linear}, with the setup 
labeled by Fig.~\ref{fig:twostream2} in Table~\ref{tab:results_two_stream}. The energy curves are shown in 
Fig.~\ref{fig:twostream2}.

As a diagnostic, we focus on the filament growing and merging processes. 
The linear phase of the filamentation instability produces current filaments. 
Since they are threaded by parallel currents, they attract each other and, starting
from the end of the linear phase, start to merge to produce larger and larger filaments \citep{Medvedev2005}.
This is clearly visible in Fig.~\ref{fig:twostream_visit}.

We measure the size of the filaments by computing the two-dimensional autocorrelation function, in the $x-y$ plane, of the 
$z$-averaged current amplitude ($z$ is the direction of the beams). This autocorrelation function is 
azimuthally averaged to obtain a radial function $\mathrm{corr}(r)$. We normalize $\mathrm{corr}(0)$ to 1.
The scale of the filaments is then taken to be five times the radius where $\mathrm{corr}(r) = 0.8$ 
(taking this scale as the radius at which $\mathrm{corr}(r)$ first vanishes yields the same results)\note{The factor 5
is here to make this correlation radius coincide with $r$ such that the tangent at $r=0$
intersects the $r$-axis, when this is applicable. Also, taking the scale as the radius at which $\mathrm{corr}(r)$ first vanishes
yields the same results.}.

The results are shown in Fig.~\ref{fig:twostream2} (bottom). We clearly see two regimes:
one during the linear growth of the filamentation instability (from $t=2.2$ to $3.6T_\mathrm{pe}$)
where the filament correlation length is set by the wavelength of the fastest growing mode and remains constant,
and one in the non-linear regime (after $t=3.6T_\mathrm{pe}$) where the filaments merge and 
thus quickly increase their size. In the second case, the growth is roughly linear with time,
which agrees with the PIC simulation results of \citet{Dieckmann2009} for a similar setup.
After $t=6.5T_\mathrm{pe}$, the filament growth stops. However, we suspect that the periodic boundaries start influencing the 
dynamics at this point.

%%%%%%%%%%%%%%%%%%%%%%%%%%%%%%%
\subsection{Linear growth rates of the relativistic tearing instability}
%%%%%%%%%%%%%%%%%%%%%%%%%%%%%%%
\label{Sec:tearing_instability}

\begin{table*}[ht]
\caption{\label{tab:results_tearing}Theoretical versus experimental values of the tearing growth rate $\tau$. 
$T_\mathrm{pe}$ is the plasma period comprising all electrons, and $d_e$ the plasma skin depth based on this period.
We recall that $d_\lec$ corresponds to $n_x$ cells.
Other simulations were performed in the case $\Theta=0.01$, but with an initial setup slightly out of equilibrium, and 
they presented variations in the growth rates of less than 3\% when $n_x$, $n_t$, and $\rho_\mathrm{sp}$
were doubled, or when $\rho_\mathrm{sp}$ was divided by two. 
         }
\centering
\begin{tabular}{l|lllllll|lll}
$\Theta$ & $\omega_\mathrm{ce}/\omega_\mathrm{pe}$ & $\Gamma_\lec U_\lec/c$ & $L/d_\lec$ & $n_t$ & $n_x$ & $\rho_\mathrm{sp}$ & Box size in $d_\lec$ & $\tau^\mathrm{simu}/T_\mathrm{pe}$ & $\tau^\mathrm{theory}/T_\mathrm{pe}$  & $(\tau^\mathrm{simu}-\tau^\mathrm{theory})/\tau^\mathrm{simu}$ \\
\hline
0.01  & 0.2   & 0.14  & 0.72  & 250   & 15    & 1000  & $53\times0.3\times42$ & 5.4 & 5     & 8\%  \\
\idem & \idem & \idem & \idem & \idem & \idem & \idem & $80\times0.3\times42$ & 5.4 & \idem & 8\%  \\
\hline
0.1   & 0.61  & 0.40  & 0.83  & 250 & 15   & 1000  & $53\times0.3\times42$  & 2.1   & 2     & 5\%~~~~Fig.~\ref{fig:tearing}  \\
\hline
1     & 1.82  & 0.68  & 1.62  & 250 & 15   & 1000  & $53\times0.3 \times42$ & 2.5   & 2.7   & 7\%  \\
%\idem & \idem & \idem & \idem & 500 & 30   & 1000  & $53\times0.16\times42$ &       &       &       \\ TODO ?
\end{tabular}
\end{table*}

We also study the linear phase of the tearing mode for a relativistic Harris sheet in a pair plasma.
Contrary to the preceding case, this example provides a test of the algorithm in a situation where 
thermal effects are essential.

The equilibrium consists of a magnetic field
\begin{equation}\label{B_field_Harris}
 \textbf{B} = \hat{\textbf{z}}\, B_0 \tanh\left( \dfrac{x}{L} \right),
\end{equation}
sustained by a population of electrons and positrons of density $\propto 1 / \cosh^2 (x/L)$ flowing
with opposite bulk velocities $U_\lec=-U_\ionperso$ in the $\pm y$ directions. We denote the associated Lorentz factor by $\Gamma_\lec$,
and the temperature of the two species by $\Theta = T/(m_\lec c^2)$.
The exact relations between the different parameters to satisfy the equilibrium are given in Appendix~\ref{Sec:Harris_details}.
In particular, one should be careful to distinguish between quantities in the frame
moving with one species (denoted with a prime) and quantities in the simulation frame. 
For example, contraction of the electron density leads to $\omega_\mathrm{pe}=\sqrt{\Gamma_\lec}\omega'_\mathrm{pe}$. 

The superparticles are loaded according to a drifting Maxwell-J\"{u}ttner distribution with the method described 
in Sect.~\ref{Sec:Load_particles}.
There is no initial perturbation, and the instability grows out of the fluctuations produced by the finite number of superparticles.

As can be seen in Table~\ref{tab:results_tearing}, the bulk velocities and the temperatures of electrons and positrons are relativistic.
Loading these distributions in a PIC code is a non-trivial task, and we have developed a special method for this. It is presented in
Sect.~\ref{Sec:Load_particles}.

The simulation domain is periodic along $z$ and $y$. Reflecting boundaries for particles and fields are present along the $x$ direction.
There are no background particles, only that of the current sheet. 

\begin{figure}
 \centering
 \def\svgwidth{\columnwidth}
 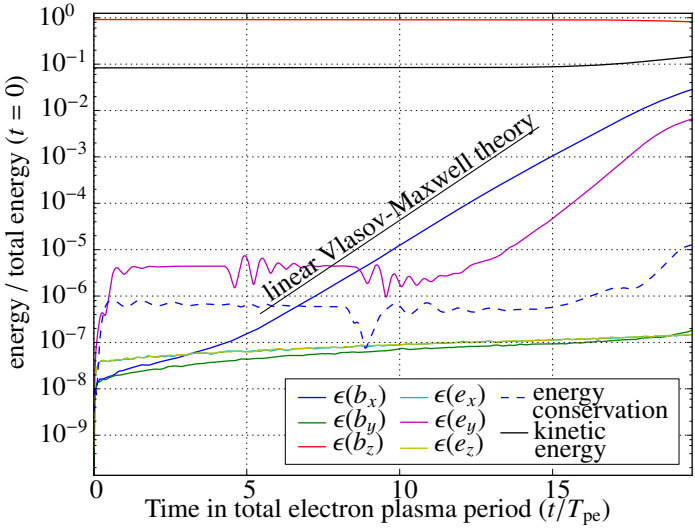
 \caption{\label{fig:tearing}Energy curves for the tearing instability (labeled Fig.~\ref{fig:tearing} in Table~\ref{tab:results_tearing}).
                             Shown are the energy curves, for example $\int\!\dif V B_x(t)^2/(2\mu_0)$. 
                             They are normalized by the total initial energy $E_\mathrm{tot}^0$ (which is mostly the energy in $b_z$).
                             The curve labeled ``energy conservation'' is $(E_\mathrm{tot}^0 - \mathrm{total\,energy}(t))/E_\mathrm{tot}^0$.}
\end{figure}

An example of the energy evolution is presented in Fig.~\ref{fig:tearing}.
After some time, the system becomes unstable and the magnetic field starts reconnecting.
As expected, $b_z$ dwindles while $b_x$ rises, which corresponds to the formation of magnetic islands.
We measure the linear growth rate on $b_x$ as $\int\!\dif V\, b_x^2 \propto \exp(2t/\tau)$. 
For comparison, we use the linear growth rates derived by \citet{Petri2007} by linearizing 
the Vlasov-Maxwell system around the drifting Maxwell-J\"uttner 
distribution \ref{equ:Jutt_Harris} and the magnetic field of Eq.~\ref{B_field_Harris}.

The results are summarized in Table~\ref{tab:results_tearing}.
Discrepancies with \citet{Petri2007} range between 5\% and 8\%. 
These growth rates vary by less than 3\% when the numerical resolution is doubled
(i.e., when $\rho_\mathrm{sp}$, $n_x$, and $n_t$ are doubled all together).
We restrict our analysis to total energy curves because contrary to the case of the filamentation instability, the linear growth of the 
field energy spans several orders of magnitude and the fastest modes have enough time to dominate the 
total energy.

We note that when the simulation is launched, an electromagnetic wave is seen to propagate from the sheet in the $\pm x$ directions.
This wave is a necessary consequence of the fact that at $t=0$ we load the superparticles by pairs of electron-positron at the same location,
and we set a zero electric field everywhere (see Sect.~\ref{sec:problem_solved}). The system then has to relax from this very peculiar state:
in less than one plasma period, charge screening is established and an electric field appears. It is this field that partly propagates outside of the sheet.
It is then reflected on the $\pm x$ boundaries and propagates back to the sheet, causing the oscillations in $e_y$ seen in Fig.~\ref{fig:tearing}.
We have checked that their incidence does not influence the linear growth by using different domain sizes. 

It is more puzzling that the sheet contracts slightly just after the beginning of the simulation and the current magnitude at its center rises by $\lesssim 10\%$. 
This may be related to the fact that our algorithm does not solve the Vlasov-Maxwell system in a strict sense
(see also Sects.~\ref{Sec:computer_vs_real_plasma} and~\ref{Sec:discussion_coarse_graining}).

%%%%%%%%%%%%%%%%%%%%%%%%%%%%%%%%%%%%%%%%%%%%%%%%%%%%%%%%%%%%%%%%%%%%%%%%%%%%%%%%%%%%%%%%%%%%%%%%%%%%%%%%%%%%%%%%%%%%%%%%%%%%%%%%%%%%%%%%%%%%%%%%
%%%%%%%%%%%%%%%%%%%%%%%%%%%%%%%%%%%%%%%%%%%%%%%%%%%%%%%%%%%%%%%%%%%%%%%%%%%%%%%%%%%%%%%%%%%%%%%%%%%%%%%%%%%%%%%%%%%%%%%%%%%%%%%%%%%%%%%%%%%%%%%%
\section{Loading a relativistic particle distribution}
%%%%%%%%%%%%%%%%%%%%%%%%%%%%%%%%%%%%%%%%%%%%%%%%%%%%%%%%%%%%%%%%%%%%%%%%%%%%%%%%%%%%%%%%%%%%%%%%%%%%%%%%%%%%%%%%%%%%%%%%%%%%%%%%%%%%%%%%%%%%%%%%
%%%%%%%%%%%%%%%%%%%%%%%%%%%%%%%%%%%%%%%%%%%%%%%%%%%%%%%%%%%%%%%%%%%%%%%%%%%%%%%%%%%%%%%%%%%%%%%%%%%%%%%%%%%%%%%%%%%%%%%%%%%%%%%%%%%%%%%%%%%%%%%%
\label{Sec:Load_particles}
This section deals with the general problem of loading particles with momenta that reproduce 
a given distribution function. 

Very common cases are waterbag and Maxwell-Boltzmann distributions with a mean bulk velocity $U_0$. 
A simple method is then to load the particles in the frame comoving with the 
plasma, which is fairly easy because in this frame the distributions are isotropic, and then 
to add to every particle the velocity $U_0$ or, if $U_0$ is close to $c$, to boost every particle 
with the Lorentz boost corresponding to $U_0$. We will see, however, that this method is no longer correct  
when both $U_0$ and the rest frame distribution are relativistic, mainly because boosting particles in a PIC code does
not boost space. We present here another method, correct in all cases\footnote{We note that after the submission of our article,
\citet{Swisdak2013} published a similar method.}.

%%%%%%%%%%%%%%%%%%%%%%%%%%%%%%%
\subsection{Transformation of the distribution function}
%%%%%%%%%%%%%%%%%%%%%%%%%%%%%%%
\label{Sec:transform_of_distrib}

We start by explaining how the particle 
distribution changes from one frame to another \citep[see e.g.][]{Mihalas1984,Pomraning1973}.

We consider a frame $\mathcal{R}$ where the plasma 
has a mean velocity $U_0$ and follows the distribution $f(\textbf{x},\textbf{p})$.
In the comoving frame or plasma rest frame $\mathcal{R}_0$, 
the plasma mean velocity is zero and follows the distribution $f_0(\textbf{x},\textbf{p})$. 
We follow a group of particles. Seen from $\mathcal{R}$, they are in a volume $\dif^3x$ 
around $\textbf{x}$ and have momentum $\textbf{p}$ with a scatter $\dif^3p$;
seen from $\mathcal{R}_0$ these quantities change respectively to $\dif^3x_0$, $\textbf{x}_0$, $\textbf{p}_0$, and $\dif^3p_0$. 
The number of particles in our group is 
\begin{equation}
 f_0(\textbf{x}_0,\textbf{p}_0)\dif^3x_0\dif^3p_0 = f(\textbf{x},\textbf{p})\dif^3x\dif^3p, 
\end{equation}
so that to find the link between $f$ and $f_0$ we have to find a relation between $\dif^3x_0\dif^3p_0$ and $\dif^3x\dif^3p$.

We start with the momentum. In the rest frame, our group of particles have momenta spanning a range $\dif^3p_0$. 
Seen in the boosted frame, their momenta transform according to the Lorentz transformation, and span a new range $\dif^3p$. 
These two volumes are thus linked by the Jacobian of the Lorentz transformation, and it can be shown that
\begin{equation}\label{equ:transf_mom}
  \dif^3p/\gamma = \dif^3p_0/\gamma_0
\end{equation}
where $\gamma$ and $\gamma_0$ are the Lorentz factors associated to $\textbf{p}$ and $\textbf{p}_0$.

We now consider the space volumes. Because of space contraction/dilatation, 
the group of particles will occupy a different volume in different frames. 
We consider the frame $\mathcal{R}'$ comoving with the group of particles. 
This is possible because the particles all move at nearly the same 
velocity $\textbf{v}_0=\textbf{p}_0/\gamma_0$. We denote by a prime all quantities seen from this frame. 
In $\mathcal{R}'$, the particles occupy a volume $\dif^3x'$. 
Since only one direction is contracted, and since $\mathcal{R}'$ moves relative to $\mathcal{R}_0$ 
with Lorentz factor $\gamma_0=\sqrt{1+p_0^2}$, we have the relation $\dif^3x_0 = \dif^3x' / \gamma_0$.
Similarly, $\mathcal{R}'$ moves relative to $\mathcal{R}$ with Lorentz factor $\gamma=\sqrt{1+p^2}$, 
and we have $\dif^3x = \dif^3x' / \gamma$. All in all:
\begin{equation}\label{equ:transf_vol}
 \gamma\, \dif^3x = \gamma_0\, \dif^3x_0.
\end{equation}

From this, we deduce that
\begin{equation}\label{equ:inv_of_f}
 f(\textbf{x},\textbf{p}) = f_0(\textbf{x}_0,\textbf{p}_0).
\end{equation}

%%%%%%%%%%%%%%%%%%%%%%%%%%%%%%%
\subsection{Why boosting particles from the rest frame is incorrect for relativistic distributions}
%%%%%%%%%%%%%%%%%%%%%%%%%%%%%%%
\label{Sec:incorrect_loading}

We now come back to PIC simulations. We assume that we load particles uniformly in space,
with momenta following $f_0(\textbf{x}_0,\textbf{p}_0)$, and that we boost each particle with a velocity $U_0$.
The momentum volume elements are then transformed according to Eq.~\ref{equ:transf_mom}, but positions are not changed. 
Equation~\ref{equ:transf_vol} does not hold, and we obtain a particle distribution
\begin{equation}\label{equ:PIC_transf}
 f_\mathrm{PIC}(\textbf{x},\textbf{p}) = f_0(\textbf{x},\textbf{p}_0)\dif^3p_0/\dif^3p = f_0(\textbf{x},\textbf{p}_0) \gamma_0/\gamma. 
\end{equation}
The volume contraction/dilatation is not performed, and the factor $\gamma_0/\gamma$ does not cancel.

Consequently, boosting each particle from the rest frame leads to the expected distribution only if
$\gamma_0/\gamma$ is independent of the particle. We can write this ratio as
\begin{equation}
 \frac{\gamma_0}{\gamma} = \frac{\gamma_0}{\Gamma_0(\gamma_0+p_{0,y}\,U_0/c)},
\end{equation}
with $p_{0,y}$ the $y$ component of $\textbf{p}_0$ and $\Gamma_0 = (1-U_0^2/c^2)^{-1/2}$, so that this is the case
only if $p_{0,y} \ll 1$ (or if the boost is non-relativistic, $U_0\ll c$).
If $p_{0,y} \ll 1$, then $\gamma_0/\gamma \sim 1/\Gamma_0$ and when it is inserted back into Eq.~\ref{equ:PIC_transf}, 
we find the usual result of density contraction.

However, when the particle distribution is relativistic in the rest frame of the plasma, 
$\gamma_0/\gamma$ is not a constant factor and Eq.~\ref{equ:PIC_transf} does not have the expected 
dependence on momentum $\textbf{p}$.

\begin{figure}
 \centering
 \def\svgwidth{0.967\columnwidth}
 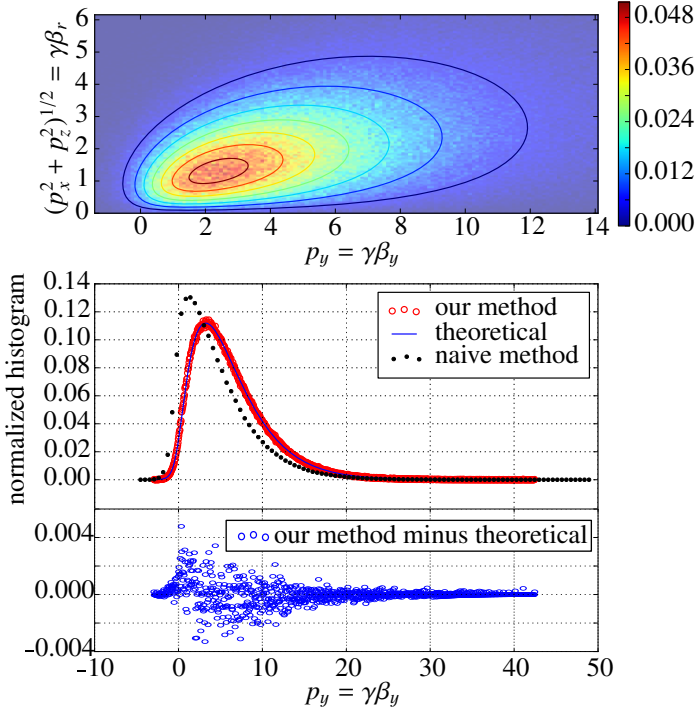
 \caption{\label{fig:testJuttner}Maxwell-J\"{u}ttner distribution for $T=mc^2$ and $\Gamma_0\beta_0=\sqrt{2}$.\newline
\textbf{Upper plot}: contours are drawn from the exact expression $2\pi r g(r,0,p_y)$, 
with $g$ from Eq.~\ref{equ:Jutt} and $(r,0,p_y)$ defined in Appendix~\ref{Sec:detail_load_Juttner}.
The background color map is the 2D histogram of $((p_x^2+p_z^2)^{1/2},p_y)$ for $10^6$ particles generated according to our method.\newline
\textbf{Middle plot}: normalized histogram of $p_y$ for the particles (red dotted),
to be compared to the exact expression in Eq.~\ref{equ:distr_y} (blue line), 
and for comparison (black dots) the histogram of $p_y$ for
particles generated the wrong way (initialization in the rest frame and boost of $\Gamma_0$).\newline
\textbf{Bottom plot}: difference between the red points and the blue curve. }
\end{figure}

%%%%%%%%%%%%%%%%%%%%%%%%%%%%%%%
\subsection{Loading a drifting Maxwell-J\"uttner distribution with arbitrary temperature and drift speed}
%%%%%%%%%%%%%%%%%%%%%%%%%%%%%%%
\label{Sec:correct_loading}

We now present a method for loading the superparticle momenta directly 
in the frame where the distribution has a bulk velocity.

In the literature, there is some agreement around the fact that the particle distribution of a relativistic plasma 
in thermodynamic equilibrium is given 
by the Maxwell-J\"{u}ttner distribution \citep{Juttner1911,Cubero2007,Chacon2010,Dunkel2009} and, even if some alternatives are also debated \citep{Treumann2011}, 
this is the distribution used in PIC simulations \citep[e.g.,][]{Petri2007b, Zenitani2008b, Jaroschek2009}.

In the plasma rest frame, the Maxwell-J\"uttner distribution is given by 
\begin{equation}\label{equ:Jutt_Rest}
 f_0(\textbf{x}_0,\textbf{p}_0) = n_0g_0(\textbf{p}_0) = n_0\frac{\mu}{4\pi K_2(\mu)} \exp\left\{ -\mu\sqrt{1+p^2_0} \right\},
\end{equation}
with $n_0$ the uniform particle number density and $g_0$ the momentum distribution, both in the restframe, 
$\mu=mc^2/T$, $\textbf{p}=\gamma\textbf{v}/c$, and $K_2$ the modified Bessel function of the second kind.
If we now denote by $f$ the distribution in the frame where the plasma moves with a bulk velocity $c\beta_0$ 
(and associated Lorentz factor $\Gamma_0$), by $n$ the particle density, and by $g$ the momentum
distribution, both in this same frame, we have $f(\textbf{x},\textbf{p})=ng(\textbf{p})$. 
Equation~\ref{equ:inv_of_f} then gives $g(\textbf{p}) = g_0(\textbf{p}_0)\times n_0/n$. Now using $n=\Gamma_0n_0$, we arrive at
\begin{equation}\label{equ:Jutt}
 g(\textbf{p}) = \frac{\mu}{4\pi K_2(\mu)\Gamma_0} \exp\left\{ -\mu\Gamma_0\left(\sqrt{1+p^2}  - \beta_0 p_y\right) \right\}.
\end{equation}
This distribution is normalized to unity: $\iiint\!\dif^3\textbf{p}\,g(\textbf{p}) = 1$.

The main difficulty with Eq.~\ref{equ:Jutt} is that the variables $p_x$, $p_y$, and $p_z$ 
are coupled and cannot be chosen independently. The solution is to compute the marginal distribution for 
the variable $p_y$. With this, one can choose the $y$-component of $\textbf{p}$ independently of the others,
and then use the distribution $g(p_\perp,p_y)$ knowing $p_y$ to choose the component normal to $\textbf{y}$.
Details are presented in Appendix~\ref{Sec:detail_load_Juttner}, where we provide the expressions for the marginal distribution 
and the conditional probability, as well as a method for generating the velocity by using the cumulative distributions.

Figure~\ref{fig:testJuttner} shows an example of the distribution generated with this method for $T=mc^2$ and $\Gamma_0\beta_0=1.41$,
and compares it to the theoretical expectation and to the the distribution obtained by boosting
individually the superparticles from the restframe. We clearly see the accuracy of our algorithm,
and the mismatch between the simpler boosting method and the expected result.
We note that this mismatch can have significant consequences. 
For example, when used for the tearing instability of Sect.~\ref{Sec:tearing_instability} it leads to large adjustments 
in the initial conditions, and for the most extreme temperatures to
a complete disruption of the current sheet.

We also provide in Table~\ref{tab:averages_Juttner} the analytical expressions of various quantities
averaged over the Maxwell-J\"uttner distribution, for example, the mean Lorentz factor, mean momentum, or the enthalpy.
The method used to derive these expressions is presented in Appendix~\ref{sec:detail_calc_averages}.
Besides being of general interest, these expressions served to further validate our generated distributions
by comparing the analytical results of Table~\ref{tab:averages_Juttner} with averages performed over 
particles generated with our method.

\begin{table}[ht]
\caption{\label{tab:averages_Juttner}{Useful averages for the Maxwell-J\"uttner distribution $f(\textbf{x},\textbf{p})=ng(\textbf{p})$
(Eq.~\ref{equ:Jutt}) of temperature $\Theta=1/\mu=T/(mc^2)$.
We define $\kappa_{ij}(\mu)=K_i(\mu)/K_j(\mu)$, with $K_n$ the modified Bessel function of the nth kind.
The Larmor radius is defined by $\langle r_\mathrm{ce}\rangle = \langle(\gamma v_{\perp})^2\rangle^{1/2}/\omega_\mathrm{ce}$,
the pressure by $P = (1/3)n\langle\textbf{v}\cdot(\gamma m\textbf{v})\rangle$,
the enthalpy by $h = (n\langle\gamma mc^2\rangle+P)/(nmc^2)$,
and the adiabatic exponent by $\hat{\gamma}-1 = P/(n\langle\gamma-1\rangle mc^2)$.
NR means non-relativistic limit ($\Theta\rightarrow0$, $\kappa_{32}(\mu)\sim 1+5\Theta/2$), 
and UR ultra-relativistic limit ($\Theta\rightarrow+\infty$, $\kappa_{32}(\mu)\sim 4\Theta$),
with in both cases no constraints on $\Gamma_0$.}
}
\centering
\begin{tabular}{llll}
\multicolumn{4}{l}{Without drift velocity: $U_0=0$} \\
 Parameter                       & Value                                                  & NR              & UR \\
\hline
 $\langle\gamma\rangle$          & $\kappa_{32}(\mu) - \mu^{-1}$                          & $1 + 3\Theta/2$ & $3\Theta$  \\
 $\langle(\gamma\beta)^2\rangle$ & $3\Theta\kappa_{32}(\mu)$                              & $3\Theta$       & $12\Theta^2$ \\
 $\langle\beta^2\rangle$         & $3\Theta\kappa_{12}(\mu)$                              & $3\Theta$       & $3/2$ \\
 $\langle\gamma\beta^2\rangle$   & $3\Theta$                                              &                 & \\
 Larmor radius                   & $(c/\omega_\mathrm{ce})\sqrt{2\Theta\kappa_{32}(\mu)}$ &                 & \\
 Pressure $P$                    & $nT$                                                   &                 & \\
 Enthalpy $h$                    & $\kappa_{32}(\mu)$                                     & $1+5\Theta/2$   & $4\Theta$ \\
 \shortstack{Adiabatic \\ exponent $\hat{\gamma}$} & $1 +\left(\mu \kappa_{32}(\mu) - \mu - 1\right)^{-1}$ & 5/3 & 4/3 \\
\vspace*{0.0cm}
\end{tabular}
\begin{tabular}{llll}
\multicolumn{4}{l}{With a drift velocity $\textbf{U}_0=U_0\hat{\textbf{y}}$} \\
 Parameter                                                 & Value                                                          & NR         & UR \\
\hline
 $\langle\textbf{v}\rangle$                                     &  $\textbf{U}_0$                                                     & & \\
 $\langle\gamma\textbf{v}\rangle$                               & $\kappa_{32}(\mu) \Gamma_0 \textbf{U}_0$                            & $\Gamma_0 \textbf{U}_0$ & $4\Theta \Gamma_0 \textbf{U}_0$\\
 $\langle \gamma \rangle$                                  & $\Gamma_0\kappa_{32}(\mu) - \frac{1}{\mu\Gamma_0}$             & $\Gamma_0$         & $\frac{4\Gamma_0}{\mu}-\frac{1}{\mu\Gamma_0}$ \\
 $\langle p_x v_x \rangle=\langle p_z v_z \rangle$         & $\Theta/\Gamma_0$                                              & & \\
 $\langle p_y v_y \rangle$                                 & $\frac{\Theta}{\Gamma_0} + \Gamma_0\beta_0^2 \kappa_{32}(\mu)$ & & \\
 $\big\langle (p_i- \langle p_i\rangle) (v_j-\langle v_j\rangle) \big\rangle$ & $\delta_{ij}\,\Theta/\Gamma_0$                                 & & \\
\end{tabular}
\end{table}

%%%%%%%%%%%%%%%%%%%%%%%%%%%%%%%%%%%%%%%%%%%%%%%%%%%%%%%%%%%%%%%%%%%%%%%%%%%%%%%%%%%%%%%%%%%%%%%%%%%%%%%%%%%%%%%%%%%%%%%%%%%%%%%%%%%%%%%%%%%%%%%%
%%%%%%%%%%%%%%%%%%%%%%%%%%%%%%%%%%%%%%%%%%%%%%%%%%%%%%%%%%%%%%%%%%%%%%%%%%%%%%%%%%%%%%%%%%%%%%%%%%%%%%%%%%%%%%%%%%%%%%%%%%%%%%%%%%%%%%%%%%%%%%%%
\section{Real, PIC, and Vlasov-Maxwell plasmas}
%%%%%%%%%%%%%%%%%%%%%%%%%%%%%%%%%%%%%%%%%%%%%%%%%%%%%%%%%%%%%%%%%%%%%%%%%%%%%%%%%%%%%%%%%%%%%%%%%%%%%%%%%%%%%%%%%%%%%%%%%%%%%%%%%%%%%%%%%%%%%%%%
%%%%%%%%%%%%%%%%%%%%%%%%%%%%%%%%%%%%%%%%%%%%%%%%%%%%%%%%%%%%%%%%%%%%%%%%%%%%%%%%%%%%%%%%%%%%%%%%%%%%%%%%%%%%%%%%%%%%%%%%%%%%%%%%%%%%%%%%%%%%%%%%
\label{Sec:computer_vs_real_plasma}

Particle-in-cell simulations have brought tremendous new
insights into astrophysical plasmas, for example through studies of
kinetic instabilities in their non-linear phase, kinetic
turbulence, particle acceleration via the Fermi-process,
or 3D magnetic reconnection and the associated
particle acceleration. However, as we will detail in this section,
there remain a number of questions with respect to the degree to which PIC
models are able to completely mirror real plasmas.

The modeling of a real plasma by a PIC plasma implies two steps
(see the right branch of Fig.~\ref{fig:plasma_models}): 
the grouping of many real particles into a single superparticle, known as coarse-graining, 
and the discretization of the equations with the presence of a grid. 
Each of these steps raises questions:
\begin{enumerate}
 \item Coarse-graining: Is plasma behavior still expected with 
       so few superparticles per Debye sphere? 
       Is the noise level too large? With this, do non-linearities appear sooner than in real plasmas?
       Does the PIC plasma remain collisionless?
       And what are we losing when we gather the particles into the superparticles?
 \item Discretization and grid: At least for explicit schemes, they bring with them numerical stability problems, 
       reviewed in Appendix~\ref{Sec:numerical_stability}.
       Moreover, the interpolation of superparticle quantities to grid points implies a finite volume for the superparticles,
       which in turn implies a vanishing two-point force at short distances and thus reduces drastically 
       the influence of collisions; it helps the PIC plasma to be collisionless, but is it enough?
       And what are the consequences of having superparticles whose sizes reach a significant fraction of the Debye length?
\end{enumerate}
We discuss some of these questions in Sects~\ref{Sec:plasma_behavior} to \ref{Sec:noise_level}.

The distinguishing feature of the Vlasov-Maxwell description of a plasma is the absence of collisions and of correlations between particles.
Given the two preceding points, we can wonder if a PIC plasma can be described by the Vlasov-Maxwell system,
or if it has too few superparticles per cell and thus correlation levels that are too high for this description 
to be accurate. The differences between PIC and Vlasov-Maxwell descriptions are examined in Sect.~\ref{sec:from_real_to_PIC}.

%%%%%%%%%%%%%%%%%%%%%%%%%%%%%%%
\subsection[The plasma parameter: a coarse-graining dependent quantity]{The plasma parameter $\Lambda$: a coarse-graining dependent quantity}
%%%%%%%%%%%%%%%%%%%%%%%%%%%%%%%
\label{Sec:plasma_behavior}
As said earlier and expressed in Eq.~\ref{equ:neandrho}, a real plasma is
represented in the computer by grouping many particles into superparticles.
This is what is called coarse-graining.
Unlike fluid equations\footnote{By a fluid model we mean any set of equations 
where the individual nature of the particles has been smoothed. 
This is the case of the MHD family, two-fluid models, or the Vlasov-Maxwell system.},
Eqs~\ref{equ:dyn1}-\ref{equ:dyn5} are not invariant under coarse-graining (because of the definition of the current, Eq.~\ref{equ:dyn5}). 
The prototype of $p$-dependent quantities is the plasma parameter\footnote{In a fully ionized plasma, 
all coarse-graining dependent quantities can 
be expressed as the product of a fluid quantity (which is coarse-graining independent)
and a parameter expressing a number of particles per fluid volume.
Examples of these parameters include $\Lambda = n\lambda_\mathrm{D}^3$, $n(c/\omega_\mathrm{pe})^3$, $n(c/\omega_\mathrm{pi})^3$,~...
% We note that the latter is used by \citet{Bret2013} with the same 
} $\Lambda = n\lambda_\mathrm{D}^3$,
which is close to the number of particles per Debye sphere (we recall that $p$ is the number of particles
per superparticles, $\lambda_\mathrm{D}$ the Debye length and $n$ the real particle number density).

The plasma parameter $\Lambda$ also expresses the ratio of the particles' kinetic energy to their electrostatic potential energy of interaction and,
as such, varies as $1/p$ because kinetic energy is proportional to the superparticles' mass $m_\mathrm{sp} \propto p$ 
while charge interaction energy involves their charge $q_\mathrm{sp}^2 \propto p^2$. 
This can be seen directly by writing $\Lambda_p = n_\mathrm{sp}\lambda_\mathrm{D}^3$ for the superparticle plasma, 
with $ n_\mathrm{sp}$ the number density of superparticles.
The Debye length, being derived from fluid theory, is invariant under coarse-graining, and
since $n = p\times n_\mathrm{sp}$, one has that 
\begin{equation}\label{equ:Lambda_equals_p_Lambdap}
 \Lambda_p = \frac{\Lambda}{p}, 
\end{equation}
with $\Lambda = \Lambda_{p=1}$ the real plasma parameter.

In a real plasma $\Lambda$ ranges from $10^4$ to $10^{20}$
(for example $\Lambda\sim10^{6}$ in solar coronal loops; $10^{12}$ in the magnetotail, 
magnetopause, or in typical Crab flares; $10^{17}$ in AGN jets),
while in computer experiments where we have to simulate thousands
to millions of Debye spheres, $\Lambda$ reaches hardly a few tens \citep[for a discussion see, e.g.,][Sect.~4]{Bykov2011}.
The corresponding number of particles per superparticles then reaches $p\sim 10^3$ to $10^{19}$.
The question of the relevance of PIC simulations for describing \textit{collisionless plasmas}
has thus been asked from the beginning, and concerns both terms, collisionless and plasma, that we now discuss.

\subsubsection{Plasma behavior}
A weakly coupled plasma is characterized by the predominance of collective effects over individual effects.
The ratio of these effects is contained in the plasma parameter $\Lambda_p$, 
which can be seen as the ratio of collective behavior (the interaction of one particle with the electromagnetic fields
collectively generated by all others, which is coarse-graining independent)
to binary effects (which are proportional to $q_\mathrm{sp}^2 / m_\mathrm{sp} \propto p$).

Since plasma behavior, with Debye screening and local charge neutrality, requires a high plasma parameter,
it is wise to ask how large it should be in a PIC plasma.
\citet[chap.~1]{Birsdall1985} and \citet{Hockney1988} have shown that 
it is not necessary for this ratio to be as high as in real plasmas, 
and that a $\Lambda_p$ of about a few suffices for correct plasma behavior.

\subsubsection{Collisionless behavior}
A plasma behaves collisionlessly if the time and length scales of interest are negligible toward 
the collision time and the mean-free path, respectively.
Since the collision time scales as $\Lambda_p$ times the plasma period, 
it is not clear whether a PIC computer plasma with a plasma parameter on the order of unity will be collisionless.

Particle-in-cell plasmas are helped by the superparticle finite sizes, which imply that the two-point force decreases to zero for separations smaller than this size.
This fact, albeit degrading the accuracy of single particle dynamics, 
greatly reduces the relative importance of binary collisions so that in order to correctly simulate a collisionless plasma
for scales accessible in simulations, one has to insure that \citep[chap.~1]{Birsdall1985,Hockney1988}
\begin{equation}
   \Lambda_p = n_\mathrm{sp} {\lambda}_\mathrm{D}^3 = \rho_\mathrm{sp} \tilde{\lambda}_\mathrm{D}^3 > \mathrm{a~few}.
\end{equation}
(Here, $\tilde{\lambda}_\mathrm{D} = \lambda_\mathrm{D}/X_0$ is the normalized Debye length. It can be expressed as 
$\tilde{\lambda}_\mathrm{D} = n_x v_{\therm}/c$ with $v_{\therm}=\sqrt{T/m}$.)

This is all the more true if $r_\mathrm{c} < X_0$, where $X_0$ is the grid size and $r_\mathrm{c}$ is the effective collision radius for Coulomb encounters,
expressed by equating the kinetic energy of the meeting particles to their potential energy of interaction:
$r_\mathrm{c}=q_\mathrm{sp}^2/(4\pi\epsilon_0\, T)=\lambda_\mathrm{D}/\Lambda_p$ ($r_\mathrm{c}$ is also $p$-dependent). 
The Debye length must be resolved for reasons of numerical stability (Sect.~\ref{Sec:cold_modes} and Appendix~\ref{Sec:numerical_stability}), 
so that we arrive at an optimal ordering $r_\mathrm{c} < X_0 < \lambda_\mathrm{D}$, which is allowed only if, 
again,  $\lambda_\mathrm{D}/r_\mathrm{c}=\Lambda_p>1$.

However, because the PIC model is a description of a plasma of cloud charges, grazing collisions will still be present 
and will lead to thermalization 
(\citet[chap.~1]{Birsdall1985}, \citet[chaps.~1 and 9]{Hockney1988}), a point discussed in the next section.

%%%%%%%%%%%%%%%%%%%%%%%%%%%%%%%
\subsection{Thermalization time: a coarse-graining dependent quantity}
%%%%%%%%%%%%%%%%%%%%%%%%%%%%%%%
\label{Sec_plasma_behavior_thermalization_time}

\begin{figure}
 \centering
 \def\svgwidth{\columnwidth}
 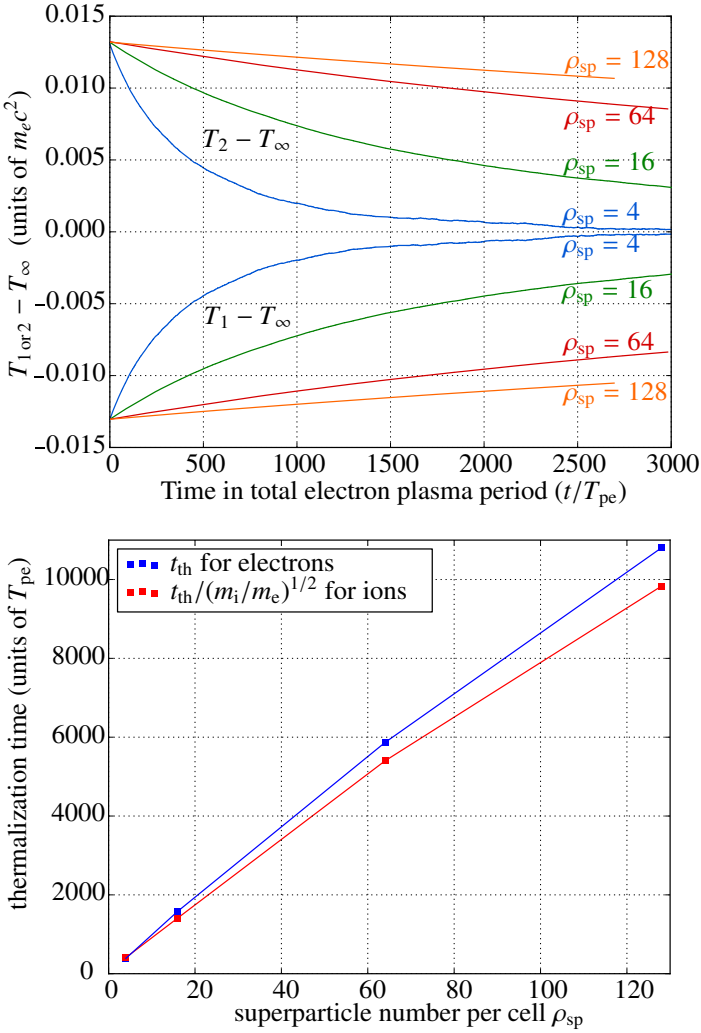
 \caption{\label{fig:thermalization}%
                 \textbf{Top}: Electron temperatures for the hot (2) and cold (1) plasmas, from four simulations with different 
                 $\rho_\mathrm{sp}$. The curves for ion temperatures are similar, except for an overall time dilatation by a factor 
                 $\sim(m_\ionperso/m_\lec)^{1/2}=5$. In this figure we use $T_\infty(t)=(T_1(t)+T_2(t))/2$. Except for $\rho_\mathrm{sp}=4$ where
                 there is significant numerical heating, $T_\infty(t)$ is constant in time.\newline
                 \textbf{Bottom}: Half-thermalization time for ions and electrons, versus number of superparticles per cell. 
                 For ions, we have plotted $t_\mathrm{th}/(m_\ionperso/m_\lec)^{1/2}$.
                 The times reported are measured as the initial slope of the temperature curves in a log-lin plot, and thus correspond to $t_\mathrm{th}/2$.
                 We see the scaling $t_\mathrm{th}\propto\rho_\mathrm{sp}$.
 }
\end{figure}

In a PIC plasma, the behavior of plasma quantities depending on $\Lambda$ can be guessed by replacing $\Lambda$
by $\Lambda_p$. 
This is the case for the thermalization time of a plasma 
by grazing Coulomb collisions \citep{Spitzer1965} or by electric field fluctuations \citep[p.\,282]{Birsdall1985},
which is on the order of $t_\mathrm{th}\sim T_\mathrm{P}\times\Lambda$ (with $T_\mathrm{P}$ the plasma period).
This has two important consequences:
\begin{itemize}
 \item We expect $t_\mathrm{th}$ to depend on resolution and coarse-graining, roughly as 
       \begin{equation}
          \frac{t_\mathrm{th}}{T_\mathrm{P}} \propto \Lambda_p = \rho_\mathrm{sp} \tilde{\lambda}_\mathrm{D}^3.
       \end{equation}
 \item Since $\Lambda_p=\Lambda/p$ is several orders of magnitude smaller than the real plasma parameter $\Lambda$,
       we expect the thermalization {by grazing collisions and fluctuations} 
       to be vastly more efficient in PIC codes than in reality.
\end{itemize}
This can have important consequences in simulations where thermalization plays a key role.
For example in real collisionless shocks, 
the mean free path for collisions $l_\mathrm{mean\,free\,path}$ is far larger than the shock thickness $\Delta_\mathrm{shock}$ 
and the thermalization processes are collisionless kinetic instabilities.
Since the mean free path $l^\mathrm{PIC}_\mathrm{mean\,free\,path}\propto\Lambda_p$ 
in a PIC plasma is smaller by a factor of $p\sim 10^{10}$
than in a real plasma, it is not obvious that the ordering 
$l^\mathrm{PIC}_\mathrm{mean\,free\,path} \gg \Delta_\mathrm{shock}$ still holds. 
To truly describe a collisionless shock with a PIC algorithm, 
one has to be careful that the unphysically fast thermalization by
collisions or fluctuations remains slower than thermalization by kinetic instabilities.

To illustrate the dependance of the collision and fluctuation induced thermalization time, 
we present simulations that initially have two thermal ion-electron plasmas.
The first is cold, with a temperature $T_{1,\lec}(0) = T_{1,\ionperso}(0) = 1.6\cdot10^{-3}m_\lec c^2$ for its electrons and ions,
while the second is hot, with $T_{2,\lec}(0) = T_{2,\ionperso}(0) = 1.8\cdot10^{-2}m_\lec c^2$. The mass ratio is $m_\ionperso/m_\lec = 25$.
The four species interact via collisions and correlations (no sign of plasma kinetic instabilities were found)
and tend to reach the same final temperature
\begin{equation}
 T_\infty = \frac{T_{1,\lec}(0)+T_{2,\lec}(0)}{2}.
\end{equation}

Particle-in-cell results are shown in Fig.~\ref{fig:thermalization} (top) for the electrons, for four simulations
with a number of superparticles per cell (including all species) $\rho_\mathrm{sp}=4$, 16, 64, or 128. 
The other parameters are kept fixed: $n_x=25$, $n_t=500$, and box size of $25^3$ cells.
It results in $\Lambda_\infty = 0.25\rho_\mathrm{sp}(n_x\sqrt{T_\infty/m_ec^2})^3 = 15$, 61, 243, or 485.
The temperatures are measured with $T = m\sum_\mathrm{sp}\textbf{v}^2_\mathrm{sp}/3$, where the sum runs over all the superparticles of a given species.
In Fig.~\ref{fig:thermalization} (top) we clearly see a slower thermalization as $\rho_\mathrm{sp}$ increases.

To evaluate the thermalization times, we use the result of \citet{Spitzer1965}:
for two species at temperature $T_1$ and $T_2$, thermalization occurs according to
\begin{equation}
\begin{split}
 \frac{dT_1}{dt} &= - \frac{dT_2}{dt} = \frac{T_2-T_1}{t_\mathrm{th}}, \\
 t_\mathrm{th}   &= \frac{3\pi}{2\sqrt{2\pi}}\frac{n\omega_{\mathrm{p}1}^{-2}\omega_{\mathrm{p}2}^{-2}}{\ln\Lambda^\mathrm{c}}\left(\frac{T_1}{m_1}+\frac{T_2}{m_2}\right)^{3/2},
\end{split}
\end{equation}
with $n = n_1 = n_2$ the particle number density,
$\omega_{\mathrm{p}i}=\sqrt{ne^2/(\epsilon_0m_i)}$, and $\ln\Lambda^\mathrm{c}$ the Coulomb logarithm, or the 
logarithm of the ratio of the largest to closest distances used in the collision integral. 
In a PIC code, $\ln\Lambda^\mathrm{c}=\ln\lambda_\mathrm{D}/X_0$. 
Clearly, $(T_1(t)+T_2(t))/2$ is constant and equal to $T_\infty$. It follows that if mass $m_1$ and $m_2$ are equal, 
the thermalization time is also constant and can be written
\begin{equation}
 t_\mathrm{th} = \frac{3}{2\sqrt{\pi}} \frac{2\pi}{\omega_{\mathrm{p}1}} \frac{\Lambda_\infty}{\ln\Lambda^\mathrm{c}},
\end{equation}
with $\Lambda_\infty = n\,[\epsilon_0T_\infty/(ne^2)]^{3/2}$ the plasma parameter based on the temperature $T_\infty$.
It also follows that the temperatures vary exponentially as $T_1 = T_\infty - 0.5[T_2(0)-T_1(0)]\exp\{-2t/t_\mathrm{th}\}$,
and similarly for $T_2$.

Here we do not find the temperature curves to be strictly exponential, mainly because there are four species.
The cold electrons interact with the hot electrons on a timescale $t_0$, 
but also with the hot ions on a timescale $m_\ionperso/m_\lec t_0 = 25t_0$.
The cold ions are heated by interactions with the hot ions on a timescale $(m_\ionperso/m_\lec)^{1/2} t_0 = 5t_0$,
and by interactions with the hot electrons on a timescale $m_\ionperso/m_\lec t_0 = 25t_0$.
Since the cold ions are heated more slowly than the cold electrons, a temperature difference 
between these two components appears and they also start heating or cooling each other.
Nevertheless, given the separation of scales we expect a measure of the slope around $t=0$ to reflect the electron-electron 
or ion-electron thermalization times when measured on the electron or ion temperature curves, respectively.

Results are shown in Fig.~\ref{fig:thermalization} (bottom). We see that the relation $t_\mathrm{th} \propto \rho_\mathrm{sp}$
is roughly correct for both electrons and ions. 
We also underline the difference with a real plasma, where $t_\mathrm{th}/T_\mathrm{pe} \sim \Lambda$ 
reaches $10^{10}$ or more, while it is on the order of $\Lambda_p \leq 10^4$ in PIC simulations.

%%%%%%%%%%%%%%%%%%%%%%%%%%%%%%%
\subsection{Field fluctuation level dependence: coarse-graining and finite superparticle size}
%%%%%%%%%%%%%%%%%%%%%%%%%%%%%%%
\label{Sec:noise_level}

\begin{figure}
 \centering
 \def\svgwidth{\columnwidth}
 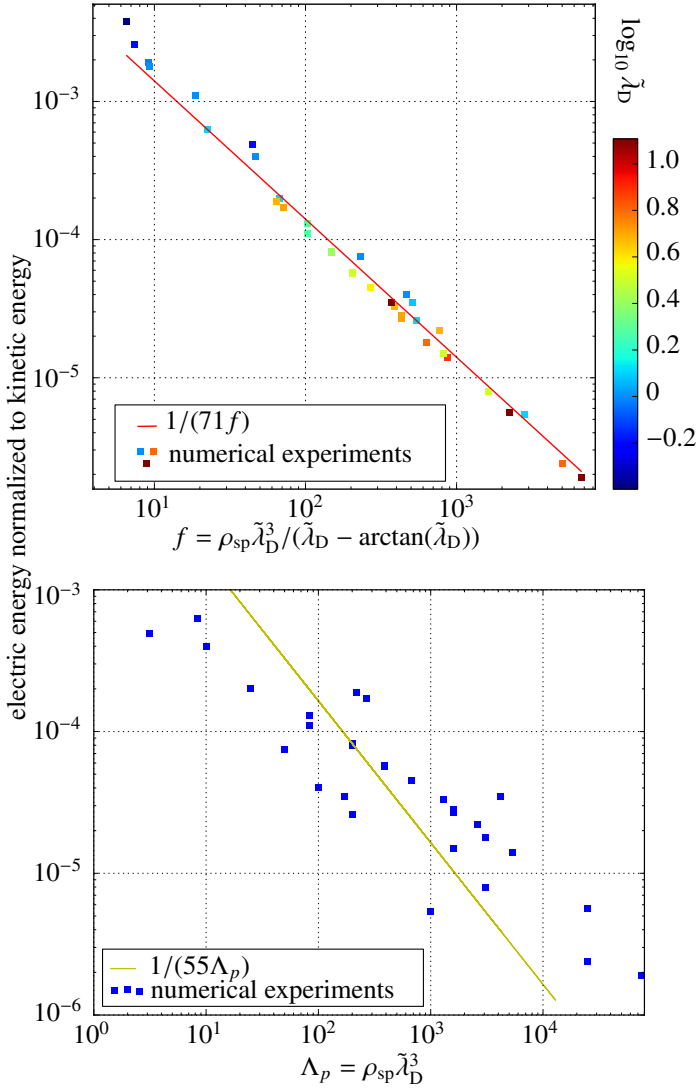
 \caption{\label{fig:fluctuations}%
                 \textbf{Top}:    Field energy levels as a function of 
                             $f = \rho_\mathrm{sp}\tilde{\lambda}_\mathrm{D}^3/(\tilde{\lambda}_\mathrm{D}-\arctan \tilde{\lambda}_\mathrm{D})$.
                             The colorbar is $\log_{10}(\tilde{\lambda}_\mathrm{D})$. Blue points at low $f$ have an 
                             under-resolved Debye length that could explain the mismatch with $1/f$.\newline
                 \textbf{Bottom}: Field energy levels versus $\Lambda_p=\rho_\mathrm{sp}\tilde{\lambda}_\mathrm{D}^3$. 
                             We clearly see the mismatch between $1/\Lambda_p$ and the results, 
                             even if the trend is correct. The large scatter is a hint that $\Lambda$ is not a 
                             relevant parameter to describe field fluctuations.\newline
                 \textbf{Top and bottom}:
                             Each point is the result from a simulation. 
                             The field energy levels are measured as the energy in the $x$ electric field, 
                             $\alpha^{-1}\,\int\!\dif V e_x^2/2$ (with $\alpha$ from Eq.~\ref{equ:alpha}), divided
                             by the kinetic energy of the superparticles, $\sum_\mathrm{sp} (\gamma_\mathrm{sp}-1)$.
 }
\end{figure}

The previous section showed that the behavior of PIC plasma quantities can be guessed 
by the substitution $\Lambda\rightarrow\Lambda_p$. 
While it is true for orders of magnitude estimates, this recipe is, however, not exact, 
and coarse-graining dependent quantities generally follow other relations than their real counterparts
with respect to physical parameters (temperature, Debye length, plasma parameter, etc.).
The main reason for this is that the finite volume of the superparticles 
implies a cutoff of the physical processes at smaller scales,
an effect that becomes even more important when the superparticle size is close to the Debye length:
$\lambda_\mathrm{D}/X_0 = \tilde{\lambda}_\mathrm{D}$ typically ranges between one (or less) and ten 
in simulations.

This section illustrates this double dependence ($\Lambda\rightarrow\Lambda_p$ and superparticle size)
with a detailed study of the level of electric field fluctuations $\varepsilon$ in a PIC thermal plasma.

In a real plasma in thermal equilibrium, it is given by \citep[Sect.~1.1]{Callen2006}
\begin{equation}\label{equ:electric_fluctuations_real_plasma}
 \varepsilon = \frac{\langle \epsilon_0 \textbf{E}^2/2\rangle}{3nT/2} \sim \frac{1}{\Lambda},
\end{equation}
where the symbol $\langle\cdot\rangle$ denotes an average over space.

\citet{Dieckmann2004} studied the spectrum of thermal fluctuations in a PIC plasma, but without investigating their levels.
\citet{Hockney1971} (see also \citet{Hockney1988}) measured ratios like $\varepsilon$ in a series of 2D simulations of thermal plasmas,
and found a good agreement with the empirical formula $\varepsilon \propto (\tilde{W}^2+\rho_\mathrm{sp}\tilde{\lambda}_\mathrm{D}^2)^{-1}$, where $\tilde{W}$ 
is the superparticle geometrical size in number of cells. Its algorithm was two dimensional, electrostatic,
and based on the integration of the Poisson equation.

We perform these simulations with our 3D electromagnetic code and {measure} the level of energy in the electric field. 
We use thermal velocities from 0.04$c$ to 0.10$c$, $\rho_\mathrm{sp}$ from 2 to 500, and $n_x$ from 10 to 128.
It results in $\tilde{\lambda}_\mathrm{D} = \lambda_\mathrm{D}/X_0$ from 0.4 to 12.8, and in 
$\Lambda_p = n_\mathrm{sp}\lambda_\mathrm{D}^3 = \rho_\mathrm{sp}\tilde{\lambda}_\mathrm{D}^3$ from 0.1 to 75000.
The fluctuation levels do not depend on the timestep 
(which varies from $n_t=2000$ down to close to the Courant limit $\Delta t\sim X_0/c$)
nor on box size (which is always bigger than $n_x$).
We {use} a pair plasma, but increasing the mass of the ions would only multiply 
the fluctuation levels by a constant factor.

The results are summarized in Fig.~\ref{fig:fluctuations}: $\varepsilon$ is found to be proportional to
$(\tilde{\lambda}_\mathrm{D}-\arctan \tilde{\lambda}_\mathrm{D})/\rho_\mathrm{sp}\tilde{\lambda}_\mathrm{D}^3$,
and not exactly to $1/\Lambda_p$.
To explain this, we generalize the computation of Hockney to three dimensions.

For a plasma in thermal equilibrium, the energy in the electric field at location $(\textbf{x},t)$
can be evaluated by adding the electric field produced at $\textbf{x}$  by charges at location $\textbf{x}_0$ and having a velocity $\textbf{v}_0$, 
$\textbf{E}_{\textbf{x}_0,\textbf{v}_0}(\textbf{x},t)$,
\begin{equation}\label{equ:field_energy_previous}
  \frac{\epsilon_0 \langle\textbf{E}^2(\textbf{x},t)\rangle}{2} = \int\!\dif^3\textbf{v}_0\dif^3\textbf{x}_0\,\frac{\epsilon_0\textbf{E}^2_{\textbf{x}_0,\textbf{v}_0}(\textbf{x},t)}{2} f_0(\textbf{x}_0,\textbf{v}_0),
\end{equation}
where $\langle\cdot\rangle$ means an ensemble average (which coincides with a spatial average);
$\textbf{E}_{\textbf{x}_0,\textbf{v}_0}(\textbf{x},t)$ is a generalization of the Debye 
electric field for moving particles \citep[chap.~9]{Nicholson1983};
and Eq.~\ref{equ:field_energy_previous} can be evaluated for a plasma of finite-sized particles
as \citep{Hockney1971,Birsdall1985}
\begin{equation}\label{equ:field_energy}
 \frac{\langle \epsilon_0 \textbf{E}^2/2\rangle}{3nT/2} = \frac{1}{3n}~\int \!\frac{4\pi k^2\dif k}{(2\pi)^3}\,\frac{1}{1+k^2\lambda_\mathrm{D}^2/S^2(ka)},
% \frac{1}{3n}~\iiint \!\frac{\dif^3k}{(2\pi)^3}W(\textbf{k}) = 
\end{equation}
where $S(ka)$ is the Fourier transform of the shape of the superparticles and $a$ the 
characteristic size of the superparticles (in our case $a\sim X_0$);
$S(ka)$ tends to 1 as $k^{-1}\gg a$.

Equation~\ref{equ:field_energy} cannot be used as such for a real plasma of point particles ($S(ka)=1$) because it includes
the electric field at arbitrarily small distances from the charge, which has an infinite energy. 
It leads to Eq.~\ref{equ:electric_fluctuations_real_plasma} only if a truncation at small distances is performed,
for example $k < (\alpha\lambda_\mathrm{D})^{-1}$ with $\alpha$ any constant: 
$\langle \epsilon_0 \textbf{E}^2/2\rangle$ in Eq.~\ref{equ:electric_fluctuations_real_plasma} is then the energy in the electric field for wavelengths 
larger than $\alpha\lambda_\mathrm{D}$. 
Alternatively and to avoid a cutting procedure, we note that the electric field in
Eq.~\ref{equ:electric_fluctuations_real_plasma} can be taken as
the total field produced by the particles to maintain the screening Debye clouds (the polarization electric field; see \citet[Sect.~1.1]{Callen2006}).
\note{$\varepsilon$ could also be evaluated as the ratio of the particles energy of interaction to their kinetic energies, 
respectively evaluated for each charge as $e^2/(2\epsilon_0 n^{-1/3})$ and $3T/2$.
The ratio of these two quantities then gives $1/(3\Lambda)^{2/3}$, and not $1/\Lambda$.
This is because the electric energy of interaction was taken as the unscreened potential at a distance $n^{-1/3}$.
If we take it as $e^2/(2\epsilon_0 \lambda_D)$, we indeed find $1/\Lambda$.}

In the case of a PIC plasma all processes at scales below the grid size $a=X_0$ are 
ignored, so that the upper bound of the integral is $k_\mathrm{max}=a^{-1}$ and there is no small scale divergence.
Since $S(ka)\sim 1$ for $k\ll a^{-1}$, and given that the integration stops at $k=a^{-1}$, we will assume that $S(ka) = 1$.

Changing to spherical coordinates, using $u=\lambda_\mathrm{D} k$ and $\langle E_x^2\rangle=\langle\textbf{E}^2\rangle/3$, we arrive at
\begin{equation}
 \frac{\langle \epsilon_0 E_x^2/2\rangle}{3nT/2} = \frac{1}{18\pi^2}\frac{1}{n\lambda_\mathrm{D}^3}~\int_{u_\mathrm{min}}^{u_\mathrm{max}}\!\dif u\, \frac{u^2}{1+u^2}.
\end{equation}
A primitive of the integral is $u-\arctan(u)$. We use 
$k_\mathrm{max}=1/X_0$ or $u_\mathrm{max}=\lambda_\mathrm{D}/X_0=\tilde{\lambda}_\mathrm{D}$,
while the largest wavelength is given by the simulation domain size and verifies $u_\mathrm{min}\ll u_\mathrm{max}$. 
Consequently, we obtain
\begin{equation}\label{equ:field_fluct_new}
 \frac{\langle \epsilon_0 E_x^2/2\rangle}{3nT/2} 
   = \frac{1}{18\pi^2}\frac{\tilde{\lambda}_\mathrm{D}-\arctan \tilde{\lambda}_\mathrm{D}}{\rho_\mathrm{sp}\tilde{\lambda}_\mathrm{D}^3},
\end{equation}
in good agreement with the simulations (Fig.~\ref{fig:fluctuations}, top panel),
except for the constant factors $18\pi^2 \sim 178$ (which can be attributed to an approximate choice of $u_\mathrm{max}$).

The two limits are interesting. For a very high resolution, $\tilde{\lambda}_\mathrm{D} \gg 1$, the field energy 
decreases as $1/(\rho_\mathrm{sp}\tilde{\lambda}_\mathrm{D}^2)$,
which is non-trivial and different from what is expected in a real plasma where 
it decreases as $1/\Lambda = 1/(n{\lambda}_\mathrm{D}^3)$. The empirical formula of \citet{Hockney1971}, generalized to 3D,
would also predict $\varepsilon\propto 1/(\rho_\mathrm{sp}\tilde{\lambda}_\mathrm{D}^3)$ in the high resolution limit.
However, our experiments clearly preclude this dependence, and are compatible with 
Eq.~\ref{equ:field_fluct_new} (see Fig. \ref{fig:fluctuations}).
The presence of a finite superparticle volume and of the grid is retained in our calculation only
in the upper bound of the integral: physically speaking, physical processes with $k^{-1}<X_0$ are smoothed out. 
This explains the difference between Eq.~\ref{equ:field_fluct_new} and that of a real plasma.

For low resolutions, $\tilde{\lambda}_\mathrm{D} \lesssim 1$, an expansion of the arctangent shows that the field energy 
behaves as $1/(54\pi\,\rho_\mathrm{sp})$, which is finite and independent of $\tilde{\lambda}_\mathrm{D}$.
In a real plasma, $\varepsilon$ would go to zero as the screening distance vanishes. 
That this is not the case here indicates that the screening distance does not vanish, because the finite size 
of the superparticles also plays the role of the screening mechanism.

%%%%%%%%%%%%%%%%%%%%%%%%%%%%%%%
\subsection{Comparing the PIC and Vlasov-Maxwell models}
%%%%%%%%%%%%%%%%%%%%%%%%%%%%%%%
\label{sec:from_real_to_PIC}

\begin{figure}
 \centering
 \includegraphics[width=\columnwidth]{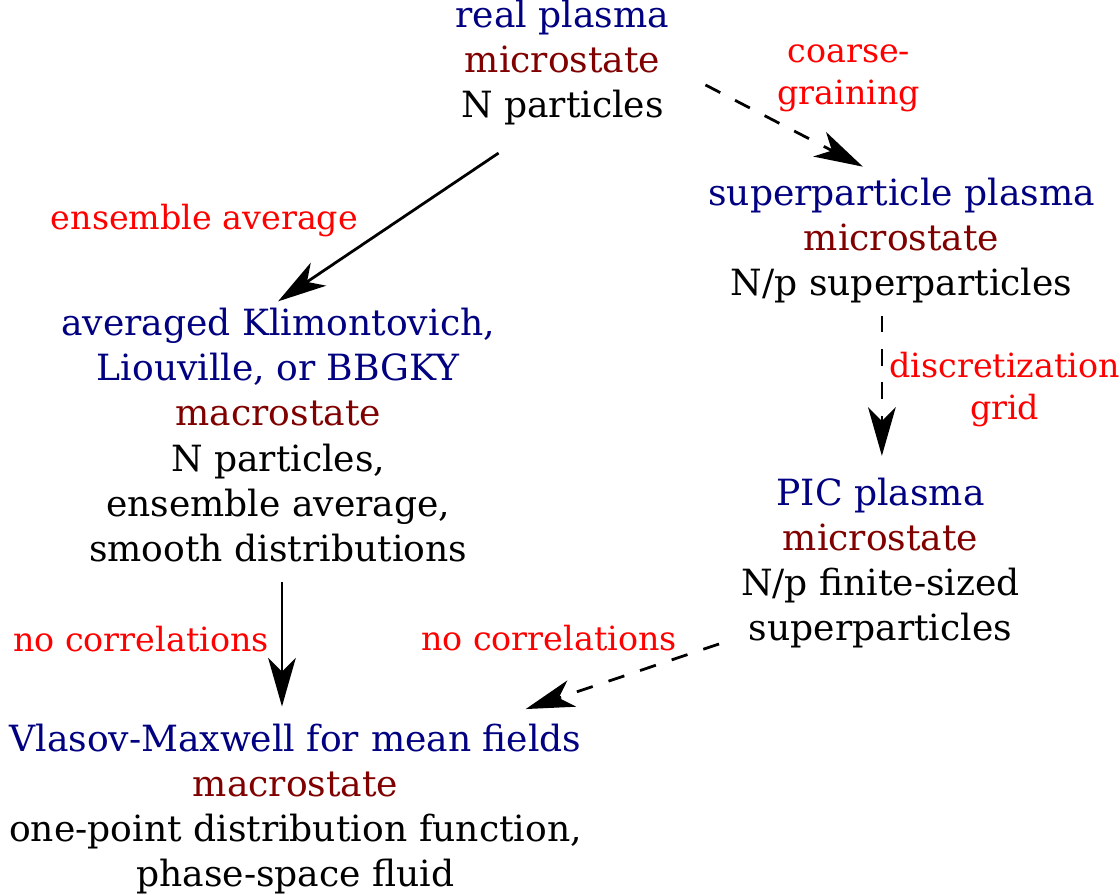}
 \caption{Different plasma models. Dashed arrows are transitions showing non-trivial effects. 
 Coarse-graining and discretization are discussed in Sects~\ref{Sec:plasma_behavior} to \ref{Sec:noise_level}.}
 \label{fig:plasma_models}
\end{figure}

We now highlight some differences between PIC models and kinetic models 
based on Liouville or Klimontovich equations. 
To do so, we recall how the Vlasov-Maxwell system is derived from these 
formalisms. 

A plasma is constituted of many charged particles in mutual electromagnetic interaction. 
Under relativistic conditions, a plasma microstate is fully characterized by 
the positions and velocities of the $N$ particles and by the value of the fields at all space points
(the fields must be treated as independent from the particles because of retarded interactions), 
plus the necessary boundary conditions.
Within the frame of classical electrodynamics, the time evolution of a microstate
is described by Maxwell equations
and by the equations of motion for the particles under the action of the Lorentz force:
\begin{equation}
 \begin{aligned}
  &\frac{\dif}{\dif t}(\gamma_j \textbf{v}_j) = \frac{q_j}{m_j} \left(\textbf{e}_m + \frac{\textbf{v}_j}{c}\wedge \textbf{b}_m\right), \\
  &\frac{\dif}{\dif t}\textbf{x}_j = \textbf{v}_j,  \\
  &\frac{\partial\textbf{b}_m}{\partial t} = -c\,\nabla\wedge\textbf{e}_m,  \\
  &\frac{\partial\textbf{e}_m}{\partial t} = c\,\nabla\wedge\textbf{b}_m - \frac{1}{\epsilon_0}\sum_{j=1}^N q_j \textbf{v}_j \delta(\textbf{x}-\textbf{x}_j),  \\
  &\nabla\cdot\textbf{e}_m = \frac{1}{\epsilon_0}\sum_{j=1}^N q_j \delta(\textbf{x}-\textbf{x}_j), \\
  &\nabla\cdot\textbf{b}_m = 0.
 \end{aligned}
\end{equation}
Here, $c$ is the speed of light, $\textbf{e}_m$ the microscopic electric field; 
$\textbf{b}_m=c\textbf{B}_m$ is $c$ times the microscopic magnetic field,
$q_j$, $m_j$, $\gamma_j$, $\textbf{v}_j$, and $\textbf{x}_j$ the charge, mass, Lorentz factor, 
velocity, and position of the particle number $j$. We also define its momentum $\textbf{p}_j=m_j\gamma_j\textbf{v}_j$.

At a given time $t$, a microstate is represented by a point 
$\{\textbf{x}_j,\textbf{p}_j\}_{j=1..N}$ in the $6N$-dimensional phase-space, and by the fields.
One can then consider the collection of microstates having the same macroscopic properties 
(which can depend on what one is looking for\note{Liouville equation for $f_N$ (Eq.~\ref{equ:Liouville}) is true whatever the 
distribution of microstates, be they compatible with a single macrostate, completely random, or arbitrarily
chosen. But the microstates must be compatible with a single macrostate if one wants to consider $f_N$
as the probability distribution of the microstates for this macrostate. Also, when one 
writes $f_1$ as the integral of $f_N$ over all other particles, it has a sense only if $f_N$ is the density 
of microstates not randomly chosen!}), 
place them as points in the $6N$-dimensional phase-space, 
and define the $N$-particle distribution function $f_N(t,\{\textbf{x}_j,\textbf{p}_j\}_{j=1..N})$ as the 
number density, at a given time, of these microstates in the $6N$-dimensional phase-space. 
Given that the number of microstates in phase-space is chosen as a continuum, 
$f_N$ is a smooth function \citep{Klimontovich1982,Nicholson1983}.
It defines a macrostate, i.e., an ensemble average of a collection of compatible microstates\note{Because of the preceding footnote.
See also the footnote with $\langle f^\mathrm{K}\rangle=f_1$.}.
The dynamic evolution is then obtained by the Liouville equation, which states that the number of 
microstates is conserved\note{That it conserves the number 
of microstates can be seen by writing Liouville equation in a conservation form:
\begin{equation}
 \frac{\partial f_N}{\partial t} + \sum_{i=1}^N \frac{\partial}{\partial \textbf{x}_i}(\textbf{v}_i f_N) + \sum_{i=1}^N \frac{\partial}{\partial \textbf{p}_i} (\dot{\textbf{p}}_i f_N) = 0.
\end{equation}
It is equivalent to Eq.~\ref{equ:Liouville} because $\nabla_\textbf{p}\cdot \dot{\textbf{p}} = (q/m)\nabla_\textbf{p}\cdot(\textbf{e}+\textbf{v}\wedge\textbf{B})=0$.
}
\begin{equation}\label{equ:Liouville}
 \frac{\partial f_N}{\partial t} + \sum_{i=1}^N \textbf{v}_i \cdot \frac{\partial f_N}{\partial \textbf{x}_i} + \sum_{i=1}^N \dot{\textbf{p}}_i \cdot \frac{\partial f_N}{\partial \textbf{p}_i} = 0,
\end{equation}
supplemented by Maxwell equations for the microscopic fields $\textbf{e}_m$ and $\textbf{b}_m$\note{They are still the microscopic fields
(compare with the non-relativistic case where $\textbf{b}_m=0$ and $\textbf{e}_m$ is 
computed from the exact potential $\propto\sum q_iq_j/|\textbf{r}_i-\textbf{r}_j|$). They are also microscopic in BBGKY hierarchy.
They become macroscopic only when we truncate BBGKY, neglecting correlations in their computation.}
with for sources the particles of the corresponding microstate.

When there is no background magnetic field, and when the particles are non-relativistic, 
the magnetic field remains negligible toward the electric field which
can be computed directly from the particle positions at time $t$ with the use of the electrostatic potential $V(r)=q^2/(4\pi\epsilon_0 r)$
(the Coulomb model). The Liouville equation can then be transformed to the infinite BBGKY hierarchy. 
The first BBGKY level involves the 
one-particle distribution function $f_1(t,\textbf{w})$ (with $\textbf{w}=(\textbf{x},\textbf{p})$) and the
two-point distribution function $f_2$ via the correlation function
$g_2(t,\textbf{w}_1,\textbf{w}_2) = f_2(t,\textbf{w}_1,\textbf{w}_2) - f_1(t,\textbf{w}_1)f_1(t,\textbf{w}_2)$:
\begin{equation}
\begin{aligned}
 &\frac{\partial f_1}{\partial t} + \textbf{v}_1\cdot\frac{\partial f_1}{\partial {\textbf{x}}} - \frac{\partial}{\partial {\textbf{x}_1}}\overline{V}_{t,\textbf{x}_1}[f_1]\cdot\frac{\partial f_1}{\partial \textbf{p}} = C[g_2], \label{equ:Vlasov_NR} \\
 &\overline{V}_{t,\textbf{x}_1}[f_1] = \int\!\dif^6\textbf{w}_2 V(|\textbf{x}_1-\textbf{x}_2|) f_1(t,\textbf{w}_2).
\end{aligned}
\end{equation}
Here $C$ is an integral operator, vanishing with $g_2$.
The quantity $\overline{V}_{t,\textbf{x}_1}[f_1]$ is the mean potential due to the particle distribution $f_1$.
Since $f_1(\textbf{w})$ is the probability of finding a particle near $\textbf{w}$ independently of the positions of all others, this potential 
does not include short-range correlations, but only long-range collective effects. It is a macroscopic quantity, just as the 
fields entering into the Vlasov-Maxwell system (Eq.~\ref{equ:Vlasov_Maxwell})\note{A note on the scales involved in these
macroscopic quantities : according to \citet{Nicholson1983}, the small element $\dif^6\textbf{w}$ in  
$f_1(\textbf{w})\dif^6\textbf{w} = \langle f^\mathrm{K}\rangle(\textbf{w})\dif^6\textbf{w}$ is to be taken as large toward interparticle spacing, 
so that a large number of particles contribute and the quantity do not fluctuate a lot, and small toward Debye length 
because at larger scales there are spatial variations (while $f(\textbf{w})$ is a local density). $n_0^{-1/3}\leq r \leq \lambda_D$ 
is of course possible in a plasma.}.

Approximations can then be made to truncate the BBGKY hierarchy. Evaluations of the right-hand side of 
Eq.~\ref{equ:Vlasov_NR} can lead, depending on the hypothesis made, to Boltzmann, Landau, Lenard-Balescu,
or more refined kinetic equations.
For a fully ionized plasma, the relevant approximation parameter is the number of particles per Debye sphere, 
or plasma parameter $\Lambda$.
For large $\Lambda$, collisions and correlations between small numbers of particles are negligible toward interactions 
of particles with the fields collectively generated by all others. Keeping only these collective interactions
is equivalent to a truncation of BBGKY hierarchy at its lowest level, i.e., $g_2=0$, and leads to the Vlasov-Maxwell system.
Generalized to relativistic plasmas\footnote{The relativistic Vlasov-Maxwell system 
is usually derived by using the Klimontovich formalism, not the Liouville formalism
\citep[see, e.g.,][]{Nicholson1983,Klimontovich1982}.}\note{
I don't think (but I might be wrong) that there
exists a BBGKY hierarchy from Liouville equation in the relativistic case. As explained in the text, in the relativistic case Liouville equation must 
be written along with the equations for the fields. But when one derive BBGKY from Liouville, one explicitly uses the fact that the 
force acting on the particles is the instantaneous Coulomb force. It seems very hard to derive BBGKY with a retarded force...

There is consequently no way to consistently derive Vlasov-Maxwell system in the relativistic case from Liouville equation.

It can be done with Klimontovich formalism (Vlasov and also BBGKY can be derived from Klimontovich, see \citet[p59]{Nicholson1983}; 
and see Callen (kinetics) where he shows that Kilontovich and BBGKY-from-Liouville both lead to the very same 
Lenard-Balescu equation), which is not an equation for the density of microstates but for the particles.
The points of the discussion remain the same: interesting properties are derived only after an ensemble average of 
Klimontovich distribution function $\langle f^\mathrm{K}\rangle$ and of the electric and magnetic fields. 
In the non-relativistic case, the two roads lead to the same result because of the identification $\langle f^\mathrm{K}\rangle=f_1$.
See \citet[chap.~3, p50, p54]{Nicholson1983}
for a simple explanation (ask me for the pdf of the book), or \citet{Klimontovich1982}.} and to two species, it reads
\begin{equation}\label{equ:Vlasov_Maxwell}
 \begin{aligned}
  &\frac{\partial f_{1,s}}{\partial t} + \textbf{v}\cdot\frac{\partial f_{1,s}}{\partial {\textbf{x}}} + q_s \left(\textbf{e}+\textbf{v}\wedge\textbf{b}\right)\cdot\frac{\partial f_{1,s}}{\partial \textbf{p}} = 0,~~~s=\mathrm{i},\,\mathrm{e},\\
  &\frac{\partial\textbf{b}}{\partial t} = -c\,\nabla\wedge\textbf{e},\\
  &\frac{\partial\textbf{e}}{\partial t} = c\,\nabla\wedge\textbf{b} - \frac{1}{\epsilon_0}\sum_{s=\mathrm{i},\mathrm{e}} \iiint\!\dif^3\textbf{p}\, q_s \textbf{v} f_{1,s}(t,\textbf{x},\textbf{p})\\
  &\nabla\cdot\textbf{e} = \frac{1}{\epsilon_0}\sum_{s=\mathrm{i},\mathrm{e}} \iiint\!\dif^3\textbf{p}\, q_s f_{1,s}(t,\textbf{x},\textbf{p}), \\
  &\nabla\cdot\textbf{b} = 0,
  \end{aligned}
\end{equation}
with $f_{1,s}(t,\textbf{x},\textbf{p})$ the one-particle distribution function, for electrons ($s=\mathrm{e}$) or ions ($s=\mathrm{i}$),
defined from $f_N$. It is also a smooth function, and 
the Vlasov-Maxwell system describes the evolution of a continuous fluid in the six-dimensional phase-space,
where information on the individual nature of the particles has been smoothed out.
In particular within this description, the plasma parameter $\Lambda$ is infinite, 
the {fluctuation- and collision-induced} thermalization time is infinite,
and the level of electric field fluctuations $\varepsilon$ is zero.
We note that these are analytical properties. Numerical solutions of the Vlasov-Maxwell-system will
also not strictly recover the collisionless behavior of the plasma. For instance, numerical diffusion arising necessarily from
the discretization of Eq.~\ref{equ:Vlasov_Maxwell} will also lead to a finite thermalization time. We are, however,
not aware of a comprehensive study of such effects for algorithms solving the discretized Vlasov-Maxwell system.

In contrast, the models underlying PIC simulations follow a different path, illustrated in Fig.~\ref{fig:plasma_models}.
It consists in following the time evolution of the microstate 
constituted by $N/p$ superparticles and the fields, Eqs~\ref{equ:dyn1}-\ref{equ:dyn5}, with $p$ reaching $10^{10}$ or more.
{One of the consequences is that} the PIC plasma has a plasma parameter $\Lambda_p = \Lambda/p$
(from Eq.~\ref{equ:Lambda_equals_p_Lambdap}) far smaller than that of the real plasma, so that it includes 
relatively large correlation and noise levels.

%%%%%%%%%%%%%%%%%%%%%%%%%%%%%%%
\subsection{Higher-order effects of coarse-graining}
%%%%%%%%%%%%%%%%%%%%%%%%%%%%%%%
\label{sec:coarse_graining}

We have highlighted that the coarse-graining step, the description of a real plasma of $N$ particles 
by a PIC plasma of $N/p$ superparticles, each containing $p$ real particles,
involves a reduction of the plasma parameter $\Lambda$ by a factor $p$. 
This is the main effect of coarse-graining.
Higher-order effects arise because after coarse-graining, the model ignores the internal dynamics and correlations 
of the particles contained within a superparticle.

This can be seen by writing explicitly the grouping:
we label the particles either by $w_n$, $n=1..N$ (with $w = (\textbf{x},\textbf{p})$), 
or by $w_{ij}$ with $i=1..N/p$ representing the group number, and $j=1..p$ the particle number within this group.
We denote by $\overline{w}_{\mathrm{sp},i}$ the position and velocity of the center-of-mass of the group number $i$.
The $N$-particle distribution function $f_N$ of the real plasma can then be written formally as
\begin{equation}\label{equ:decomp_fn_PIC}
\begin{aligned}
 f_N&(t,w_1,...,w_N) = g_\mathrm{corr}(t,w_1,...,w_N) \\
                      &+ f_{N/p}(t,\overline{w}_{\mathrm{sp},1},...,\overline{w}_{\mathrm{sp},N/p})
                      \times \prod_{i=1}^{N/p}f_{\mathrm{sp},i}(t,w_{i1},...,w_{ip}).
\end{aligned}
\end{equation}
This equation introduces $f_{N/p}(t,\overline{w}_{\mathrm{sp},1},...,\overline{w}_{\mathrm{sp},N/p})$, 
the analog of $f_N$ but for the center-of-mass of the particle groups (i.e., of the superparticles);
$f_{\mathrm{sp},i}(t,w_{i1},...,w_{ip})$, the distribution function of the particles contained within a group,
which represents the dynamics and correlations between particles of the same group;
and $g_\mathrm{corr}(t,w_1,...,w_N)$, the correlations ignored by writing $f_N = f_{N/p}\times f_{\mathrm{sp},1}..f_{\mathrm{sp},N/p}$,
i.e., the correlations between particles of different groups.
The PIC approximation then consists in setting $g_\mathrm{corr}=0$ 
and $f_{\mathrm{sp},i}(t,w_{i1},...,w_{ip})=\mathrm{constant}$.

A complete understanding of these approximations would require 
developing a BBGKY hierarchy from Eq.~\ref{equ:decomp_fn_PIC}
and making explicit the electric and magnetic field contributions from the particle groups.
This is a complex task. 
We can, however, stress important consequences of the assumption $f_{\mathrm{sp},i}(t,w_{i1},...,w_{ip})=\mathrm{constant}$:
\begin{itemize}
 \item The superparticles are assumed incompressible. The compressibility due to particle motion 
       within a particle group is thus absent from the coarse-grained plasma.
 \item The velocity dispersion of the particles within a group is ignored in the coarse-grained plasma.
       The kinetic pressure resulting from this dispersion is thus also absent.
 \item The electric fields present in the PIC plasma are computed from the superparticles.
       This is equivalent to saying that they are computed by taking into account only the monopole 
       distribution of charge created by the internal arrangement of the particles within a group,
       with higher-order multipole terms neglected.
       The same holds for the magnetic field.
\end{itemize}

%%%%%%%%%%%%%%%%%%%%%%%%%%%%%%%%%%%%%%%%%%%%%%%%%%%%%%%%%%%%%%%%%%%%%%%%%%%%%%%%%%%%%%%%%%%%%%%%%%%%%%%%%%%%%%%%%%%%%%%%%%%%%%%%%%%%%%%%%%%%%%%%
%%%%%%%%%%%%%%%%%%%%%%%%%%%%%%%%%%%%%%%%%%%%%%%%%%%%%%%%%%%%%%%%%%%%%%%%%%%%%%%%%%%%%%%%%%%%%%%%%%%%%%%%%%%%%%%%%%%%%%%%%%%%%%%%%%%%%%%%%%%%%%%%
\section{Discussion and conclusion}
%%%%%%%%%%%%%%%%%%%%%%%%%%%%%%%%%%%%%%%%%%%%%%%%%%%%%%%%%%%%%%%%%%%%%%%%%%%%%%%%%%%%%%%%%%%%%%%%%%%%%%%%%%%%%%%%%%%%%%%%%%%%%%%%%%%%%%%%%%%%%%%%
%%%%%%%%%%%%%%%%%%%%%%%%%%%%%%%%%%%%%%%%%%%%%%%%%%%%%%%%%%%%%%%%%%%%%%%%%%%%%%%%%%%%%%%%%%%%%%%%%%%%%%%%%%%%%%%%%%%%%%%%%%%%%%%%%%%%%%%%%%%%%%%%
\label{Sec:Discussion}

%%%%%%%%%%%%%%%%%%%%%%%%%%%%%%%
\subsection{Code validation}
%%%%%%%%%%%%%%%%%%%%%%%%%%%%%%%

We have presented our particle-in-cell code Apar-T 
(Sect.~\ref{Sec:Problem_solved} and Appendix~\ref{Sec:Numerical_implementation}),
and studied several validation tests.

Computation of the spectra from a magnetized or unmagnetized thermal plasma at rest 
(Sect.~\ref{Sec:cold_modes}) has proven very accurate, 
with the plasma pulsation and the right and left cutoff pulsations 
precisely recovered (Figs.~\ref{fig:PDF_B_zero} and \ref{fig:PDF_B}), even in cases where the Debye length
and the Larmor radius are not resolved.
This proves that the description of individual particle motions in a constant magnetic field and of collective
particle dynamics is accurate, and robust with respect to numerical resolution. 
In particular we showed that Larmor orbits in a constant magnetic field are well described 
provided that the cyclotron pulsation is well resolved, independently of the grid size.
We found, however, a numerical instability with abnormal behavior in the energy curves 
and high noise levels for under-resolved Debye length 
or under-resolved Larmor radius,
so that this parameter range should be avoided.

Simulations of the filamentation instability (Sect.~\ref{Sec:two-stream_linear}) showed a good agreement with linear cold theories, 
provided that the growth rates are computed from the temporal evolution of the Fourier modes. 
Discrepancies with theory then range from 5\% to 13\%, and can be explained by the wavenumber dependence of the growth rates.
On the other hand, the linear growth rates derived from the total energy curves, or equivalently from the sum of all Fourier modes,
present larger discrepancies with linear theory, ranging between 12\% to 61\%.
This is explained by the high level of fluctuations in PIC codes that prevent the fastest growing modes to dominate the total energy
before the end of the linear phase of the instability.

Simulations of the tearing instability in a relativistic pair plasma gave linear growth rates
within 8\% of those found by \citet{Petri2007} with an analytical linear Vlasov-Maxwell solution,
a result not varying significantly when changing the numerical resolution
(Sect.~\ref{Sec:tearing_instability}, Fig.~\ref{fig:tearing}, and Table~\ref{tab:results_tearing}).
This example, where the shape of the velocity distribution is a key feature, is thus in agreement 
with the Vlasov-Maxwell description. 
It also validates the new method used to load the relativistic (in both temperature and bulk velocity) Maxwell-J\"uttner
distribution that we present in Sect.~\ref{Sec:Load_particles} and Appendix~\ref{Sec:detail_load_Juttner}, 
as well as the general relations used for the relativistic Harris equilibrium (Appendix~\ref{Sec:Harris_details}).
We note that in this case the total energy curves can be used for evaluation of the linear growth rates because the linear phase
spans several orders of magnitude in field intensity, 
so that the fastest growing modes have enough time to dominate the energy.

All in all, these tests show that our code Apar-T is a sound basis for future explorations.
They also serve as a base to explore important questions
regarding the nature of a PIC plasma, that we summarize in the next section.

%%%%%%%%%%%%%%%%%%%%%%%%%%%%%%%
\subsection{PIC and real plasmas}
%%%%%%%%%%%%%%%%%%%%%%%%%%%%%%%
The widespread use of PIC codes for studying plasmas out of equilibrium 
calls for a deep understanding of the PIC model, and of its relation with a real plasma
and with the Vlasov-Maxwell description. Section~\ref{Sec:computer_vs_real_plasma}
attempted to provide some explanations.

We have seen that the PIC model lies on two building blocks.
The first stems from the capability of computers to handle only up to $\sim 10^{10}$ particles,
while real plasmas contain from $10^4$ to $10^{20}$ particles per Debye sphere. This means that 
a \textit{coarse-graining} step must be used, whereby of the order of $p \sim 10^{10}$ real particles are represented by
a single computer superparticle.
The second step is field storage on a grid with its subsequent finite superparticle size.

We have introduced the notion of coarse-graining dependent quantities, i.e., physical quantities depending on $p$.
The prototype of such quantities is the plasma parameter $\Lambda$, that behaves as $\Lambda_p \propto 1/p$.
This vast reduction of $\Lambda$ induces higher noise levels and correlations, 
but we have again seen that it does not threaten plasma and collisionless 
behavior as long as $\Lambda_p$ remains above unity. 
All coarse-graining dependent quantities can 
be expressed as the product of a fluid quantity (which is coarse-graining independent)
and a parameter expressing a number of particles per fluid volume.
Examples of such parameters include $\Lambda = n\lambda_\mathrm{D}^3$, $n(c/\omega_\mathrm{pe})^3$, $n(c/\omega_\mathrm{pi})^3$,~...
Their behavior in the PIC plasma can be guessed by taking into account the reduction by a factor $p$ of the number of particles,
leading for example to the substitution $\Lambda \rightarrow \Lambda_p = \Lambda/p$ in the relevant analytical expressions.
We checked this for the collision and fluctuation induced thermalization time (Sect.~\ref{Sec_plasma_behavior_thermalization_time}),
which is indeed proportional to the number of computer superparticles per cell; 
the lower the number the shorter the thermalization time.
\citet{Bret2013} similarly reduce the parameter $n(c/\omega_\mathrm{pe})^3$ by a factor $p$ when applying their theory for the 
magnetic fluctuation level in a drifting plasma to their PIC simulations.
However, the substitution $\Lambda \rightarrow \Lambda_p$ is strictly valid only for point-size particles,
and the large finite size of the superparticles, which reaches a fraction of a Debye length, 
suppresses interactions and fluctuations at shorter wavelengths and modifies these scalings.
We have detailed how this works for the electric field fluctuation level in a thermal plasma in Sect.~\ref{Sec:noise_level}.

We stress that the reduction of the collision and fluctuation induced thermalization time and 
of other related timescales (e.g., the slowing-down time of fast particles),
by 10 or more orders of magnitude, can have important consequences for the relevance of simulations:
one has to insure that collisionless kinetic processes remain more efficient than the artificially 
enhanced collisional and fluctuation induced PIC effects.
Similarly, we have seen in Sect~\ref{Sec:two-stream_linear} that the high level of fluctuations alter the linear 
spectrum of instabilities by preventing the fastest growing modes to dominate the total energy.

A more subtle effect of coarse-graining is due to the 
loss of the dynamics of the $p$ particles represented by each superparticle.
We intuitively expect 
that it will lead to the overall loss of
compressibility due to superparticle incompressibility, 
of the contribution to kinetic pressure of the particle velocity spreading within a superparticle,
and of the multipole contribution to the electric and magnetic fields created by the distribution of 
particles within a superparticle (see Sect.~\ref{sec:coarse_graining}). 
The relevance of these missing effects remains unclear.

%%%%%%%%%%%%%%%%%%%%%%%%%%%%%%%
\subsection{PIC and Vlasov-Maxwell plasmas}
%%%%%%%%%%%%%%%%%%%%%%%%%%%%%%%
\label{Sec:discussion_coarse_graining}

We have highlighted in Sect.~\ref{sec:from_real_to_PIC} that a PIC algorithm simulates
a plasma of finite-sized charges in their self-fields, and does not strictly solve the Vlasov-Maxwell system. 
Using the Vlasov equation assumes that the plasma is represented in phase space by a continuous fluid.
In this limit of an infinite number of particles, the plasma parameter $\Lambda$ and the thermalization time are infinite, and the
collision frequency and the thermal field fluctuation levels are zero.
Using a PIC algorithm amounts to dividing the continuous phase space fluid into discrete
elements, and to following their orbits. In this sense, one can say that we integrate the characteristics
of the Vlasov equation. However, the newly introduced graininess (which is far higher than that of the
original plasma) implies the presence of binary collisions and of correlations between superparticles that is not easy to evaluate,
in part because they are reduced by the finite size of the superparticles 
and the subsequent vanishing of the two-point force at short distances (see Sect.~\ref{Sec:plasma_behavior}).
The intricate dependence of Eq.~\ref{equ:field_fluct_new} is a hint to this complexity.
We note that \citet[chap.~12]{Birsdall1985} have derived a generalization of the Balescu-Guernsey-Lenard kinetic equation
that includes the use of a grid (and thus of finite sized superparticles), and of the discretization in space
and time of the equations. The correlations just mentioned are partly present in this equation, but difficult to extract.

These differences between PIC and Vlasov-Maxwell plasmas are especially enhanced in the 
linear phase of instabilities. We see two main points.
The first is that nonlinear effects absent from the linear theory, 
and possibly enhanced by the high noise level of the simulation (Sect.~\ref{Sec:noise_level}),
may have visible consequences (\citet[Sect.~13.6]{Birsdall1985}, \citet{Dieckmann2006}). 
This example is reported by \citet{Daughton2002} in the context of the drift kink instability of a current sheet: 
the instability is found to grow faster than predicted by the linear Vlasov theory because the early development of 
another instability quickly produces non-linear effects. 
\citet[Fig.\,23]{Bret2010b} also report significant early non-linear behavior in counter-streaming situations.
We have also reported the presence of field components due to non-linear effects in Fig.~\ref{fig:twostream2}.

The second point is that the high level of fluctuations delays the dominance of the fastest growing Fourier modes 
over the sum of other modes. 
The consequence is that effective linear growth rates measured from total energy curves 
appear slower than the growth rates of the fastest modes. 
This is even more important in instabilities where the linear phase is short, 
and explains the differences between the effective growth rates and the linear cold theory 
of the counter-streaming instability measured in Sect.~\ref{Sec:two-stream_linear},
with discrepancies reaching 60\% or more.
It may also explain the differences between theory and measured growth rates 
of \citet{Cottrill2008,Dieckmann2006,Haugboelle2012} for the counter-streaming instability.
On the other hand, the differences can be small if the linear phase lasts long enough
for the fastest mode to dominate the energy, as is the case for the relativistic tearing instability 
in Sect.~\ref{Sec:tearing_instability} or for the Weibel instability of \citet{Markidis2010}.

%%%%%%%%%%%%%%%%%%%%%%%%%%%%%%%
\subsection{Modeling astrophysical plasmas}
%%%%%%%%%%%%%%%%%%%%%%%%%%%%%%%

In the light of what has been said so far one may wonder what this all implies for the modeling of astrophysical plasmas.
We attempt to give some answers in the following.
We have shown that the PIC description of an astrophysical plasma bears some risk
because the plasma parameter $\Lambda$ is always underestimated, leading to systematic errors in
the evaluation of important parameters of the plasma such as the {collision and fluctuation induced} thermalization time. 
We stress that this does not lessen the important role of PIC algorithms for deepening our understanding of plasma physics.
In particular they also have their virtues, for instance the consideration 
of certain correlations and of direct particle encounters
(with the restrictions discussed in Sect.~\ref{Sec:computer_vs_real_plasma}).
They provide an accurate description of collisionless kinetic processes such as instabilities, 
and of the induced turbulence and eventual associated thermalization relevant to collisionless environments. 

The Vlasov-Maxwell equations perfectly
describe a plasma free of collisions and fluctuations.
However, their discretization will
again introduce different plasma characteristics. A thorough
discussion of these effects is still missing. On the other hand, a
collisionless description of astrophysical plasmas is not always
correct. On larger spatial and temporal scales, the description of
flows and the propagation of non-thermal particles must include
collisions to a certain degree. In this regime other models, and in
particular Fokker-Planck models, have been shown to
give a good description of the plasma and have provided significant
results. However, Fokker-Planck models have their own drawbacks,
notably that they are local and use dragging and diffusion
coefficients not self-consistently derived.

These discrepancies between real, PIC, and Vlasov-Maxwell plasmas are
complex, and it needs to be discussed in further details under what circumstances
which model and which numerical realization comes closest to a real
plasma. In the long term, it may be justified to use models
including correlations in a more systematic way, for example the
Landau or the Lenard-Balescu equation on the theoretical side, and
$\mathrm{P^3M}$ algorithms on the numerical side (that include short-range
particle-particle interactions \citep{Hockney1988}).  Both
approaches should then be faced with results from well-controlled
collisionless plasma experiments, which are presently in their
infancy \citep[see, e.g.,][]{Grosskopf2013}.

%%%%%%%%%%%%%%%%%%%%%%%%%%%%%%%%%%%%%%%%%%%%%%%%%%%%%%%%%%%%%%%%%%%%%%%%
%%%%%%%%%%%%%%%%%%%%%%%%%%%%%%%%%%%%%%%%%%%%%%%%%%%%%%%%%%%%%%%%%%%%%%%%
\begin{acknowledgements} 
  We would like to thank J\'er\^ome P\'etri for information and discussions about linear growth rates.
  We also thank the anonymous referee for comments that allowed to significantly improve the manuscript.
  We acknowledge support from the French Stellar Astrophysics Program PNPS. 
  Some of the simulations have been performed at p\^ole scientifique de mod\'elisation num\'erique,
  PSMN, at ENS-Lyon, whose staff we thank for their steady technical support.
  Larger simulations were performed using HPC ressources from GENCI (Grand \'Equipement National de Calcul Intensif)
  at CINES and CCRT, under the allocation x2012046960.
\end{acknowledgements}
%
%
%
%
%
%
%%%%%%%%%%%%%%%%%%%%%%%%%%%%%%%%%%%%%%%%%%%%%%%%%%%%%%%%%%%%%%%%%%%%%%%%%%%%%%%%%%%%%%%%%%%%%%%%%%%%%%%%%%%%%%%%%%%%%%%%%%%%%%%%%%%%%%%%%%%%%%%%
%%%%%%%%%%%%%%%%%%%%%%%%%%%%%  Bibliography  %%%%%%%%%%%%%%%%%%%%%%%%%%%
%%%%%%%%%%%%%%%%%%%%%%%%%%%%%%%%%%%%%%%%%%%%%%%%%%%%%%%%%%%%%%%%%%%%%%%%%%%%%%%%%%%%%%%%%%%%%%%%%%%%%%%%%%%%%%%%%%%%%%%%%%%%%%%%%%%%%%%%%%%%%%%%
%
\bibliographystyle{apj} 
\bibliography{biblio_particles.bib}

\addcontentsline{toc}{section}{References}
\appendix

%%%%%%%%%%%%%%%%%%%%%%%%%%%%%%%%%%%%%%%%%%%%%%%%%%%%%%%%%%%%%%%%%%%%%%%%%%%%%%%%%%%%%%%%%%%%%%%%%%%%%%%%%%%%%%%%%%%%%%%%%%%%%%%%%%%%%%%%%%%%%%%%
%%%%%%%%%%%%%%%%%%%%%%%%%%%%%%%%%%%%%%%%%%%%%%%%%%%%%%%%%%%%%%%%%%%%%%%%%%%%%%%%%%%%%%%%%%%%%%%%%%%%%%%%%%%%%%%%%%%%%%%%%%%%%%%%%%%%%%%%%%%%%%%%
\section{Numerical implementation}
%%%%%%%%%%%%%%%%%%%%%%%%%%%%%%%%%%%%%%%%%%%%%%%%%%%%%%%%%%%%%%%%%%%%%%%%%%%%%%%%%%%%%%%%%%%%%%%%%%%%%%%%%%%%%%%%%%%%%%%%%%%%%%%%%%%%%%%%%%%%%%%%
%%%%%%%%%%%%%%%%%%%%%%%%%%%%%%%%%%%%%%%%%%%%%%%%%%%%%%%%%%%%%%%%%%%%%%%%%%%%%%%%%%%%%%%%%%%%%%%%%%%%%%%%%%%%%%%%%%%%%%%%%%%%%%%%%%%%%%%%%%%%%%%%
\label{Sec:Numerical_implementation}
This Appendix is the direct continuation of Sect.~\ref{Sec:Problem_solved}.

%%%%%%%%%%%%%%%%%%%%%%%%%%%%%%%
\subsection{Computation of the current}
%%%%%%%%%%%%%%%%%%%%%%%%%%%%%%%
\label{sec:current_simple}
Before going through the normalization of Eqs.~\ref{equ:dyn1}-\ref{equ:dyn5}, we have to understand how the 
current is computed. As said in Sect. \ref{sec:volume}, interpolation of particle quantities to grid nodes is 
done by attributing to the superparticles a finite volume $V_\mathrm{sp} = X_0^3$. 
We consider a superparticle,
and the cell that contains its center. At time $t$, the superparticle occupies a volume $V_t$ of the cell.
The charge in the cell is given by $Q_\mathrm{cell}(t) = q_\mathrm{sp}\times(V_t/V_\mathrm{sp})$.
The charge continuity equation then gives
\begin{equation}
 \dintcl\!\textbf{j}\cdot \mathrm{d}\textbf{S} = - \frac{dQ_\mathrm{cell}}{dt} = \frac{V_{t+dt}-V_t}{V_\mathrm{sp}}\frac{q_\mathrm{sp}}{dt}.
\end{equation}

The superparticle volume necessarily intersects three faces of the cell that contains its center:
one of perpendicular along $\textbf{x}$, one along $\textbf{y}$, and one along $\textbf{z}$.
Consequently, the motion of this superparticle will create a current through these three faces, and we can write
\begin{equation}
 \dintcl\!\textbf{j}\cdot \mathrm{d}\textbf{S} = j_xX_0^2 + j_yX_0^2 + j_zX_0^2.
\end{equation}

We have to know which part of the volume variation $V_{t+dt}-V_t$ is attributed to each part of the current.
The displacement of the superparticle between $t$ and $t+dt$ is denoted $(\Delta x,\Delta y,\Delta z)$.
The volume variation depends on this displacement, and the part of it proportional to $\Delta x$
is attributed to $j_x$, and similarly for the $y$ and $z$ components.
More specifically, we can write $V_{t+dt}-V_t = A_x\Delta x + A_y\Delta y + A_z\Delta z$.
The areas $A_i$ can be evaluated with some geometry (see Sect.~\ref{sec:current_details}). Then (and similarly for $y$ and $z$):
\begin{equation}
 j_x = -\frac{1}{X_0^2} \frac{q_\mathrm{sp}}{dt} \frac{A_x\Delta x}{V_\mathrm{sp}}.
\end{equation}

This way of computing the current ensures that the discrete charge conservation equation is fulfilled, and justifies
the advection of the divergence of the fields.

%%%%%%%%%%%%%%%%%%%%%%%%%%%%%%%
\subsection{Normalization}
%%%%%%%%%%%%%%%%%%%%%%%%%%%%%%%
The problem is formulated with as many equations as variables (Eqs.~\ref{equ:dyn1}-\ref{equ:dyn5}, variables
$\textbf{e}$, $\textbf{b}$, $\textbf{j}$, $\textbf{x}_\mathrm{sp}$, and $\textbf{v}_\mathrm{sp}$ for each superparticle),
and it is possible to normalize the equations in a way independent of any physical quantity.
We denote normalized quantities with a tilde.

We choose to normalize lengths by $X_0$. Consequently, the normalized step-size is unity.
Times are normalized by $T_0 = X_0/c$, and 
velocities are then naturally normalized to $X_0/T_0 = c$.
For the fields $\textbf{E}$ and $\textbf{B}$, we use $\textbf{e}=\textbf{E}$ and $\textbf{b}=c\textbf{B}$.
These last two quantities are normalized by $e_0 = b_0 = m_\lec c^2/(eX_0)$.
With this, Eqs.~\ref{equ:dyn1}, \ref{equ:dyn2}, and \ref{equ:dyn3} transform into
\begin{subequations}
 \label{equ:dynamicnorm}
 \begin{align}
 \dfrac{\dif\tilde{\textbf{x}}}{\dif\tilde{t}} &= \tilde{\textbf{v}}, \label{equ:dyn1norm}\\
 \dfrac{\dif\gamma \tilde{\textbf{v}}}{\dif\tilde{t}} &= \left[\dfrac{m_\lec}{m_s} \dfrac{q_s}{e}\right] (\tilde{\textbf{e}} + \tilde{\textbf{v}}\wedge \tilde{\textbf{b}}), \label{equ:dyn2norm}\\
 \dfrac{\partial\tilde{\textbf{b}}}{\partial\tilde{t}} &= -\tilde{\nabla}\wedge\tilde{\textbf{e}},
 \end{align}
\end{subequations}
with $m_s=m_\lec$ or $m_\ionperso$ and $q_s=-e$ or $e$ ($e$ is positive).

For the current $\textbf{j}$, the algorithm computes the quantity $A_x\Delta x/V_\mathrm{sp}$. Writing for example the $x$ component of
Eq.~\ref{equ:dyn4} gives 
\begin{equation}\label{equ:max2bis}
 \dfrac{\partial\tilde{e_x}}{\partial \tilde{t}} = (\tilde{\nabla}\wedge\tilde{\textbf{b}})_x + \underbrace{\left[\dfrac{1}{\epsilon_0} \dfrac{e^2}{m_\lec c^2X_0}p \dfrac{q_s}{e}\right]}_{\alpha}\,\left[\dfrac{\tilde{A}_x\Delta \tilde{x}}{\tilde{V}_{sp}} \right]\dfrac{1}{\widetilde{dt}}.
\end{equation}
Using Eq.~\ref{equ:neandrho}, we can write
\begin{equation}\label{equ:alpha}
 \alpha = \dfrac{2\,\mathrm{sgn}(q_s)}{n_x^2\rho^0_\mathrm{sp}},
\end{equation}
where $\mathrm{sgn}$ is the sign of the charge.

With this, all the equations are completely independent of any physical quantity related to the simulated problem,
and depend only on space discretization ($n_x$) and particle coarse-graining ($\rho^0_\mathrm{sp}$).
Time discretization ($n_t$) will play a role in time integration. 

As a final comment, we note that $\rho_\mathrm{sp}^0$ is {a priori} unrelated to the actual
number of superparticles per cell during the simulation. One should, however, make these two values not too far apart,
because the size and timesteps are a fraction $n_x$ and $n_t$
of the skin depth and plasma period of a $n_\lec^0$-density plasma. 
If, for example, the superparticle density in the simulation is
twice $\rho_\mathrm{sp}^0$, then $n_x$ cells will now represent two skin depths of the 
$2n_\lec^0$-density plasma, and the resolution
will decrease. This must be kept in mind in simulations where high density contrasts appear.

%%%%%%%%%%%%%%%%%%%%%%%%%%%%%%%
\subsection{Discrete version of the equations}
%%%%%%%%%%%%%%%%%%%%%%%%%%%%%%%
In this section we drop the tilde over normalized quantities. We denote the time at which they are considered by a superscript
and their spatial location on the grid by a subscript.

%%%%%%%%%%%%%%%%%%%%%%%%%%%%%%%
\subsubsection{The main loop}

The strategy is to use a leap-frog scheme. It has the advantages of being time centered and reversible, and second order
in time and space.

Before the loop, $\textbf{b}$ and $\textbf{v}$ are known at time $t-dt/2$, and $\textbf{e}$ and $\textbf{x}$ are known at time $t$.
This should also be true for the initial conditions, so that initially we integrate 
backward the velocities and the $\textbf{b}$-field by $-dt/2$.
Injected particles (if any) should also be correctly staggered.

The structure of the main loop is the following:
\begin{enumerate}%[itemsep=0pt]
  \item Half advance of $\textbf{b}^{t-dt/2}$ with $\nabla\wedge\textbf{e}^t$; $\textbf{b}$ is now at time\,$t$.
  \item Update of $\textbf{v}^{t-dt/2}$ with $\textbf{b}^t$ and $\textbf{e}^t$; $\textbf{v}$ is now at time $t+dt/2$.
  \item Update of $\textbf{x}^t$ with $\textbf{v}^{t+dt/2}$; $\textbf{x}$ is now at time $t+dt$.
  \item Half advance of $\textbf{b}^t$ with $\nabla\wedge\textbf{e}^t$; $\textbf{b}$ is now at time $t+dt/2$.
  \item Boundary for $\textbf{b}$.
  \item Full advance of $\textbf{e}^t$ with $\nabla\wedge\textbf{b}^{t+dt/2}$; $\textbf{e}$ is now at time $t+dt$.
  \item Boundary for $\textbf{e}$.
  \item Boundary for particles.
  \item Computation of the currents from $\textbf{v}^{t+dt}$ and $\textbf{x}^{t+dt/2}$.
  \item Filtering of the currents.
  \item Boundary for the currents.
  \item Add currents to $\textbf{e}^{t+dt}$.
\end{enumerate}

%%%%%%%%%%%%%%%%%%%%%%%%%%%%%%%
\subsubsection{Integration of the fields}

\begin{figure}
 \centering
 \includegraphics[width=\columnwidth]{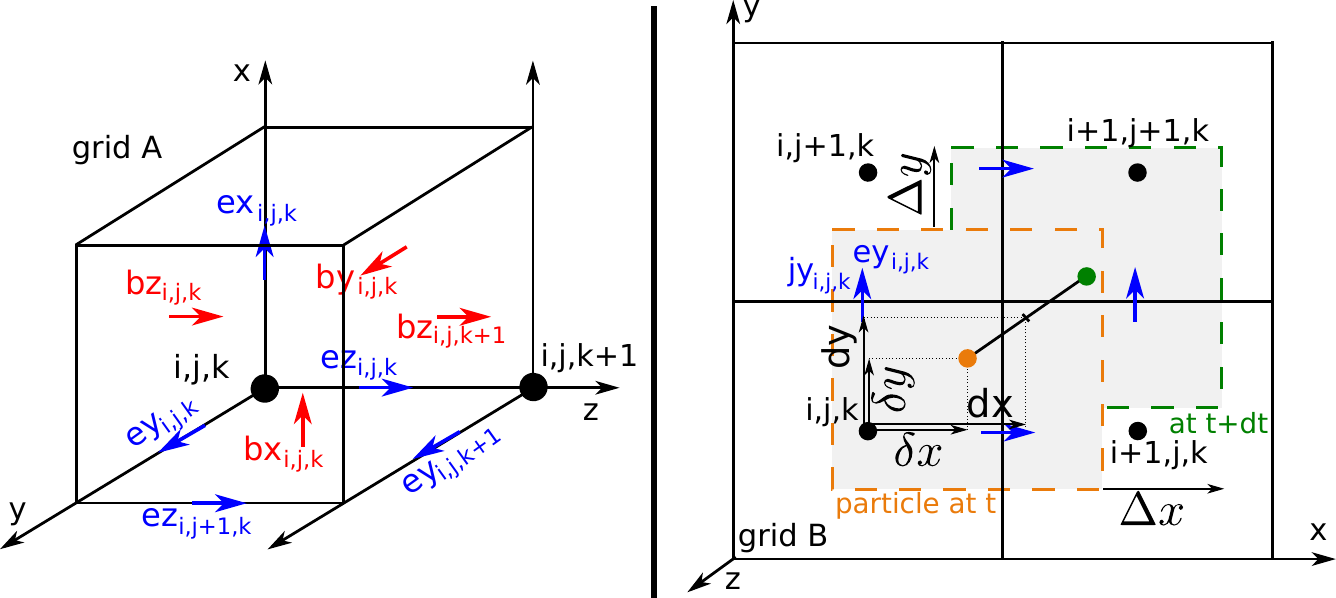}
 \caption{\label{fig:Yee}Grid and fields locations.}
\end{figure}

The fields are stored on the grid in a staggered way, with $\textbf{e}$ at the center of the grid edges 
and $\textbf{b}$ at the center of the grid faces (Fig. \ref{fig:Yee}). This is the so-called Yee lattice.
It allows an easy integration of the fields, and is second-order accurate in time and space \citep[Sect.~15]{Birsdall1985},
\begin{equation}
\f{b}{x}{i,j,k}{t+dt/2} = \f{b}{x}{i,j,k}{t-dt/2} + dt\,\left(\f{e}{y}{i,j,k+1}{t} - \f{e}{y}{i,j,k}{t} - \f{e}{z}{i,j+1,k}{t} + \f{e}{z}{i,j,k}{t}\right), 
\end{equation}
and similarly for the other components of $\textbf{b}$ and for $\textbf{e}$.

To reduce the effects of \v{C}erenkov emission (see Appendix~\ref{sec:localeffects}), we have also implemented 
a fourth order solver \citep{Greenwood2004}.

%%%%%%%%%%%%%%%%%%%%%%%%%%%%%%%
\subsubsection{Moving the particles}

Integration of the equation of motion for the superparticles is done with the algorithm
described by \citet[Sect.~15.4]{Birsdall1985}. It is a relativistic generalization of 
the leap-frog scheme, time centered, time reversible, and second order accurate.
We note however that, as pointed out by \citet{Vay2008}, it can have shortcomings for ultrarelativistic particles.
In short, Eqs.~\ref{equ:dyn1norm} and \ref{equ:dyn2norm} are discretized as
\begin{subequations}\label{equ:mover}
 \begin{align}
    \begin{split}
    \left(\textbf{u}^{t+dt/2} - \textbf{u}^{t-dt/2}\right)/{dt} &= \\
     &\!\!\!\!\!\!\!\!\!\!\!\!\!\!\!\!\!\!\!\!\!\!\!\!\!\!\!\!\!\!\!\!\!\!\!\frac{m_\lec}{m}\frac{q}{e} \left\{ \textbf{e}^t + \frac{\textbf{u}^{t+dt/2}+\textbf{u}^{t-dt/2}}{2\gamma^t} \wedge \textbf{b}^t \right\},
    \end{split} \label{equ:mover1} \\ 
 \frac{\textbf{x}^{t+dt} - \textbf{u}^{t}}{dt} &= \frac{\textbf{u}^{t+dt/2}}{\sqrt{1+(\textbf{u}^{t+dt/2})^2}}, \label{equ:mover2}
 \end{align}
\end{subequations}
with $\textbf{u}=\gamma\textbf{v}$ and $\gamma$ the Lorentz factor.
Defining $\textbf{u}^- = \textbf{u}^{t-dt/2} + qm_\lec/(em) \textbf{e}^t dt/2$ and
$\textbf{u}^+ = \textbf{u}^{t+dt/2} - qm_\lec/(em) \textbf{e}^t dt/2$ and substituting into \ref{equ:mover1} leads to
\begin{equation}
 \frac{\textbf{u}^{+} - \textbf{u}^{-}}{dt} = \frac{qm_\lec/(em)}{2\gamma^t} (\textbf{u}^{+}+\textbf{u}^{-}) \wedge \textbf{b}^t.
\end{equation}
This equation is the classical rotation around a $\textbf{b}$ field (\citet[Sect.~4.4]{Birsdall1985}, \citet[Sect.~4.7.1]{Hockney1988}), 
and is solved via
\begin{equation}
 \textbf{u}^+ = \textbf{u}^- + \frac{2}{1+ \left( \frac{\Omega dt}{2} \right)^2} \left( \textbf{u}^- + \textbf{u}^- \frac{\Omega dt}{2} \right) \wedge {\Omega},
\end{equation}
with the rotation vector $\Omega = qm_\lec/(2em\gamma^t)\textbf{b}^t$.
Finally, we use $\gamma^t = \sqrt{1+(u^-)^2} = \sqrt{1+(u^+)^2}$.
In Eq.~\ref{equ:mover1}, the fields have to be known at time $t$. It explains the need for
the half advances of $\textbf{b}$ in the general scheme. 

The interpolation of the fields at particle positions
is done via a trilinear interpolation. We denote by $(i,j,k)$ the nodes of the main grid $A$ (Fig. \ref{fig:Yee}),
and we introduce a second grid $B$ whose cell centers are on $(i,j,k)$.
Consider a superparticle at position $x=i+\delta x$, $y=j+\delta y$, $z=k+\delta z$.
The superparticle is actually a charge cloud of volume equal to a cell, and this volume intersects
the cell of the second grid with center $(i,j,k)$ in a volume $V_{i,j,k}=(1-\delta x)(1-\delta y)(1-\delta z)$,
the cell of the second grid with center $(i+1,j,k)$ in a volume $V_{i+1,j,k}=\delta x(1-\delta y)(1-\delta z)$, and so on.
For a quantity $f$ defined at grid points $(i,j,k)$, the weight associated to $f_{i,j,k}$ is $V_{i,j,k}$, the one
associated to $f_{i+1,j,k}$ is $V_{i+1,j,k}$, and so on for a total of 8 points.

However, neither $\textbf{e}$ nor $\textbf{b}$ are defined at grid points $(i,j,k)$ (Fig. \ref{fig:Yee}), and they must be first
interpolated at grid points before applying the above procedure. This is done for example
with $f_{i,j,k} = 0.5(e_{x,i-1,j,k}+e_{x,i,j,k})$ or $f_{i,j,k} = 0.25(b_{z,i,j,k}+b_{z,i-1,j,k}+b_{z,i-1,j-1,k}+b_{z,i,j-1,k})$.
Details can be found in \citet{Matsumoto1993, Messmer1998}.

We note that the superparticle shape used for interpolation of fields to particle position and for interpolation of the current to grid nodes
is the same. This is required to avoid the existence of a self-force on the superparticles and to conserve 
the total momentum \citep[Sect.~8.6]{Birsdall1985}.

%%%%%%%%%%%%%%%%%%%%%%%%%%%%%%%
\subsubsection{Computation of the current}
\label{sec:current_details}

The current $\textbf{j}_{i,j,k}$ is defined at the same locations as $\textbf{e}_{i,j,k}$.
For current deposition, we again consider the volumes occupied by the superparticle
in the grid $B$ cells. As the superparticle moves, these volumes vary. 
We denote by $(i+\delta x,j+\delta y,k+\delta z)$ the position of the superparticle
at $t-dt$, and we assume that it moves from $(\Delta x,\Delta y,\Delta z)$ between $t-dt$ and $t$.

Consider, for example, the volume of the superparticle in the cell of center $(i,j,k)$. 
Its variation is given by $dV = (1-\delta x-\Delta x)(1-\delta y-\Delta y)(1-\delta z-\Delta z) - (1-\delta x)(1-\delta y)(1-\delta z)$.
Defining $d_x = \delta x + \Delta x/2$, $c_x = 1-d_x$, and similarly for $y$ and $z$, one finds
\begin{equation}\label{equ:current_dep}
 \begin{split}
  dV &= \Delta x \left[ -c_yc_z - \Delta y \Delta z /12 \right] ~~~\rightarrow j_{x|i,j,k} \\
     &+ \Delta y \left[ -c_zc_x - \Delta z \Delta x /12 \right] ~~~\rightarrow j_{y|i,j,k}\\
     &+ \Delta z \left[ -c_xc_y - \Delta x \Delta y /12 \right] ~~~\rightarrow j_{z|i,j,k}.
 \end{split}
\end{equation}
As explained in Appendix~\ref{sec:current_simple}, the part of \ref{equ:current_dep} proportional to the displacement along $x$ 
is attributed to $j_{x|i,j,k}$, and so on.

A similar treatment is done with the cells that intersect the superparticle volume. These are cells centered in $(i+\epsilon,j+\eta,k+\xi)$, with $\epsilon$, $\eta$ and
$\xi$ equal either to 0 or 1 (8 cells). For each of these cells, only the faces intersecting the 
superparticle volume are concerned, so that in total there are only 12 currents to update.

Currents can be smoothed before being added to $\textbf{e}$. This has the effect of 
reducing electromagnetic noise (see Appendix~\ref{sec:globaleffects}).
This is done in the following way for the current of cell $(i,j,k)$: attribute 
a weight of 1 to cell $(i,j,k)$;
of 0.5 to cells $(i\pm1,j,k)$, $(i,j\pm1,k)$, and $(i,j,k\pm1)$; 
of 0.25 to cells $(i\pm1,j\pm1,k)$,  $(i\pm1,j,k\pm1)$, and $(i,j\pm1,k\pm1)$;
of 0.125 to cells $(i\pm1,j\pm1,k\pm1)$; 
and normalize the sum of the weights to 1.

%%%%%%%%%%%%%%%%%%%%%%%%%%%%%%%
\subsection{Boundaries}
%%%%%%%%%%%%%%%%%%%%%%%%%%%%%%%

Periodic and reflective boundaries are available. 
The latter simulate a perfect conductor at the domain boundary 
by imposing the correct values for the electric and magnetic fields
($\textbf{b}=\textbf{e}=0$ inside the conductor, 
$\textbf{b}_\mathrm{normal}=\textbf{e}_\mathrm{tangential}=0$ at the conductor surface),
and by reflecting the particles.

%%%%%%%%%%%%%%%%%%%%%%%%%%%%%%%
\subsection{Parallel efficiency}

The code parallelization was performed and tested by \citet[chap.~4]{Messmer1998}. It uses Fortran 90 and MPI.
The simulation domain is decomposed in sub-domains of equal length along the $z$ direction,
and all the cells and particles of each sub-domain are assigned to a processor. To minimize 
communications between processors, ghost cells for the fields are added to each sub-domain. 
Communication between neighboring processors occurs at each step involving the boundaries: for
particles leaving or entering the domain, for the fields, and for the currents. 

The domain is currently decomposed along one direction only. This is relevant for 
simulations of collisionless shocks where the domain is elongated along the flow direction, 
or for 2D magnetic reconnection simulations where the presence of the over-dense current sheet 
at the domain center would lead to load balancing issues if a 2D domain decomposition were used.

\begin{figure}
 \centering
 \def\svgwidth{\columnwidth}
 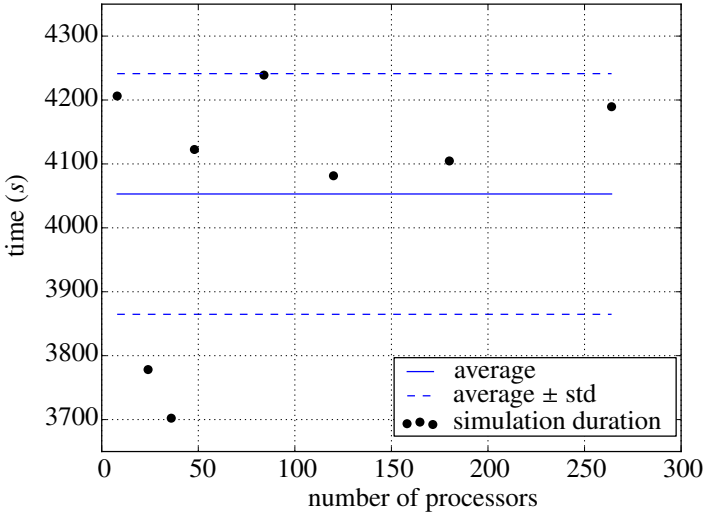
 \caption{\label{fig:scaling}Simulation time duration versus number of cores, the spatial domain extending in proportion to the number of cores.
                             The standard deviation corresponds to 4\% of the mean value.}
\end{figure}

We have tested the efficiency of this implementation with simulations using 16 superparticles per cell and
a domain size of $60\times60\times(16n_\mathrm{proc})$, where the number of cores varied from $n_\mathrm{proc}=16$ to $256$. 
The corresponding (weak) scaling results, shown in Fig.~\ref{fig:scaling}, are satisfactory.
Simulation times scatter with a standard deviation of $4\%$ around a constant value. 
The scatter is probably due to the uncontrolled node geometry.

%%%%%%%%%%%%%%%%%%%%%%%%%%%%%%%%%%%%%%%%%%%%%%%%%%%%%%%%%%%%%%%%%%%%%%%%%%%%%%%%%%%%%%%%%%%%%%%%%%%%%%%%%%%%%%%%%%%%%%%%%%%%%%%%%%%%%%%%%%%%%%%%
%%%%%%%%%%%%%%%%%%%%%%%%%%%%%%%%%%%%%%%%%%%%%%%%%%%%%%%%%%%%%%%%%%%%%%%%%%%%%%%%%%%%%%%%%%%%%%%%%%%%%%%%%%%%%%%%%%%%%%%%%%%%%%%%%%%%%%%%%%%%%%%%
\section{Numerical effects}
%%%%%%%%%%%%%%%%%%%%%%%%%%%%%%%%%%%%%%%%%%%%%%%%%%%%%%%%%%%%%%%%%%%%%%%%%%%%%%%%%%%%%%%%%%%%%%%%%%%%%%%%%%%%%%%%%%%%%%%%%%%%%%%%%%%%%%%%%%%%%%%%
%%%%%%%%%%%%%%%%%%%%%%%%%%%%%%%%%%%%%%%%%%%%%%%%%%%%%%%%%%%%%%%%%%%%%%%%%%%%%%%%%%%%%%%%%%%%%%%%%%%%%%%%%%%%%%%%%%%%%%%%%%%%%%%%%%%%%%%%%%%%%%%%
\label{Sec:numerical_stability}
We have said in Sect.~\ref{Sec:computer_vs_real_plasma} that passing from a real plasma to a PIC model implies a discretization of the equations.
This step comes with numerical issues that have been largely studied by 
\citet{Birsdall1985} and \citet{Hockney1988}. We highlight part of their work here.

%%%%%%%%%%%%%%%%
\subsection{Local numerical effects}
%%%%%%%%%%%%%%%%
\label{sec:localeffects}
\begin{itemize}
 \item Stability of the electric part of the superparticle motion integrator used here requires 
 that $\Omega \Delta t < 2$, with $\Omega$ the pulsation of oscillation of the superparticles (usually the plasma pulsation). 
 The magnetic part is unconditionally stable.
 \item Courant condition for the stability of the field integrator in vacuum is $c\,\Delta t < X_0/\sqrt{2}$. 
 \item The dispersion relation of electromagnetic waves in vacuum is modified by the grid.
This modification depends on the angle of propagation with respect to the grid, and waves can have a phase velocity smaller 
than $c$ \citep{Greenwood2004}. If superparticles with velocity close to $c$ are present, 
they can overtake light waves and emit \v{C}erenkov radiation. This results in the production of non-physical fields. 
The situation can be improved with a higher order interpolation scheme for the fields.
\end{itemize}

%%%%%%%%%%%%%%%%
\subsection{Global numerical effects}
%%%%%%%%%%%%%%%%
\label{sec:globaleffects}
By considering the algorithm as a whole, Birsdall and Langdon were able to identify numerical effects
not predicted by the consideration of subparts alone.

For example, 
the discrete space representation of the continuous quantities introduces a periodicity in Fourier space
of period $k_0 = 2\pi/X_0$ (with $X_0$ the grid spacing). 
A physical mode of wavenumber $k=2\pi/\lambda$ will then have,
in the numerical plasma, aliases of wavenumbers $k + nk_0$, and $-k + nk_0$, with $n$ an integer. 
Instabilities can arise if the physical mode couples resonantly with one of the aliases. 
This coupling cannot occur if $k < -k + k_0$, i.e., if $\lambda > 2X_0$
(we note that it is Nyquist-Shannon criterion to avoid spectral aliasing).
Just as in signal processing, the strength of the aliases can be reduced
by low-pass filtering the time-series, and this is what is done by attributing a cloud shape to the superparticles.
Aliases are even more reduced when the superparticle shapes have a fast decaying Fourier transform, that is, when they are smoother.

We mention in particular the following effects due to grid aliasing:
\begin{itemize}
  \item A cold beam of velocity $v_\mathrm{beam}$ becomes unstable if the Doppler shifted frequency of Langmuir oscillations 
  is near the grid-crossing frequency $k_\mathrm{grid}\, v_\mathrm{beam}$. The beam is then heated. 
  It is not the case if $\lambda_\mathrm{D} / X_0 > 0.046$.
  \item $\lambda_\mathrm{D} / X_0 > 1/\pi$ is needed to avoid an artificial numerical heating of a Maxwellian plasma. 
  Otherwise, the plasma is heated up to the point where $\lambda_\mathrm{D}$ reaches $X_0/\pi$. 
 \item The rate of passage of the superparticles through the cell faces produces a high-frequency noise;
 the rougher the superparticle shapes, the more important is the noise.
\end{itemize}

Similarly to grid effects, a finite timestep implies that harmonics differing from a multiple 
of $2\pi/\Delta t$ are not differentiated by the algorithm, and there are time aliases as well. 
\begin{itemize}
  \item This implies no other instabilities in the case of a non-magnetized Maxwellian plasma.
  \item In a magnetized plasma, artificial coupling of cyclotron harmonics can lead to instabilities.
\end{itemize}

A last point is that the effects of the grid, as well as other errors, act as a random force $F(t)$ on the superparticles. 
Consequently, the velocity of a superparticle undergoes a random walk, $\mathrm{d}v/\mathrm{d}t\propto F(t)$, 
and the kinetic energy $\langle v^2\rangle$ increases linearly with time.
\citet[Sect.~9.2]{Hockney1988} shows that this is indeed the cause of plasma self-heating in superparticle simulations.
This is also what we find in our thermal simulations (see Sect.~\ref{sec:energy_conservation}).

%%%%%%%%%%%%%%%%
\subsection{Qualitative constraints on timestep and sizestep}
%%%%%%%%%%%%%%%%
\begin{itemize}
 \item The step-size $X_0$ of the grid (which is also roughly the superparticle size), 
       and the time-step $\Delta t$, must be smaller than the
       scales of the phenomena studied. This scale can be an instability wavelength or growth rate, 
       the cyclotron radius or pulsation, gradient scales, etc.
 \item The same is true for the mean distance between superparticles: $n_\mathrm{sp}^{-1/3} < \lambda_\mathrm{relevant}$.
       This is equivalent to having a high enough number of superparticles per volume $\lambda_\mathrm{relevant}^3$.
       The case $\lambda_\mathrm{relevant} = \lambda_\mathrm{D}$ applies to the description of plasma behavior.
 \item If thermal effects are important, then one should insure that the distribution function $g(\textbf{p})$ is
       well represented on scales where these effects are important. It requires a high enough number of superparticles 
       per relevant volume.
       \citet[Sect.~15.19]{Birsdall1985} mention that it is sufficient to have a good representation of the relevant projections of $g$.
 \item The plasma should remain collisionless: the collision time should be greater than relevant timescales 
       (instability growth rates, etc.).
\end{itemize}

%%%%%%%%%%%%%%%%
\subsection{Limitations for the computation of photon spectra}
%%%%%%%%%%%%%%%%
\begin{itemize}
 \item The highest frequency represented is $2\pi c/X_0 = 2\pi n_x \omega_\mathrm{pe}$, so that high energy radiation is 
    absent from the code and must be computed separately to extract photon spectra.
    This is done for example by \citet{Hededal2005,Frederiksen2010,Nishikawa2011}; and \citet{Cerutti2012b} from superparticle motions, with 
    the inclusion of radiative energy losses. 
    However, even in these cases, effects such as plasma frequency cutoff, Raizin effect, or transition radiation
    are not described because they are due to the back-reaction of the plasma particles 
    on the electromagnetic waves, waves that are absent from the PIC code and are only computed afterward.
\end{itemize}

%%%%%%%%%%%%%%%%%%%%%%%%%%%%%%%%%%%%%%%%%%%%%%%%%%%%%%%%%%%%%%%%%%%%%%%%%%%%%%%%%%%%%%%%%%%%%%%%%%%%%%%%%%%%%%%%%%%%%%%%%%%%%%%%%%%%%%%%%%%%%%%%
%%%%%%%%%%%%%%%%%%%%%%%%%%%%%%%%%%%%%%%%%%%%%%%%%%%%%%%%%%%%%%%%%%%%%%%%%%%%%%%%%%%%%%%%%%%%%%%%%%%%%%%%%%%%%%%%%%%%%%%%%%%%%%%%%%%%%%%%%%%%%%%%
\section{Relativistic Harris current sheet}
%%%%%%%%%%%%%%%%%%%%%%%%%%%%%%%%%%%%%%%%%%%%%%%%%%%%%%%%%%%%%%%%%%%%%%%%%%%%%%%%%%%%%%%%%%%%%%%%%%%%%%%%%%%%%%%%%%%%%%%%%%%%%%%%%%%%%%%%%%%%%%%%
%%%%%%%%%%%%%%%%%%%%%%%%%%%%%%%%%%%%%%%%%%%%%%%%%%%%%%%%%%%%%%%%%%%%%%%%%%%%%%%%%%%%%%%%%%%%%%%%%%%%%%%%%%%%%%%%%%%%%%%%%%%%%%%%%%%%%%%%%%%%%%%%
\label{Sec:Harris_details}
Harris configuration is one of the rare fully consistent solutions of the Vlasov-Maxwell system in a non-homogeneous case.
Its generalization to the relativistic case is not difficult, and was partly done by \citet{Hoh1966} for a Maxwell-J\"uttner
distribution and non-relativistic current speeds. The fully relativistic case appeared later for pair plasmas 
with the same temperature for both species,
for example in \citet{Kirk2003} or \citet{Petri2007}.
Here, we propose a formulation for the relativistic case with an arbitrary temperature 
and mass ratio for ions (singly ionized) and electrons.

We define a frame $\mathcal{R}_0$, where the magnetic field is assumed to have the dependence given by Eq.~\ref{B_field_Harris},
and the particles sustaining this field are assumed to have a Maxwell-J\"uttner distribution function given in $\mathcal{R}_0$ by
\begin{equation}\label{equ:Jutt_Harris}
 f(\textbf{x},\textbf{p}) = \frac{\mu_s\,n'_s(x)}{4\pi K_2(\mu_s)} \exp\left\{ -\mu_s\Gamma_s\left(\sqrt{1+p^2}  - \beta_s p_y\right) \right\},
\end{equation}
with $s=\ionperso$ for ions or $\lec$ for electrons, $\mu_s=1/\Theta_s=m_sc^2/T_s$, $\textbf{p}=\gamma\textbf{v}/c$, 
and $K_2$ the modified Bessel function of the second kind;
$U_s$ is the bulk velocity of species $s$, and $\Gamma_s$ the associated Lorentz factor.
We note that $f_s$ is normalized with respect to $\textbf{p}$ to $\Gamma_s n'_s(x)$, so that $n'_s$ is the density of species $s$
in its comobile frame (noted with a prime), and $\Gamma_sn'_s$ its density in $\mathcal{R}_0$.

Inserting Eq.~\ref{equ:Jutt_Harris} into the Vlasov equation expressed in $\mathcal{R}_0$,
\begin{equation}\label{equ:Vlasov_Harris}
  \textbf{v}\cdot\frac{\partial f_s}{\partial\textbf{x}} + \frac{q_s}{m_sc}\, \textbf{v}\wedge\textbf{B}(\textbf{x})\cdot\frac{\partial f_s}{\partial\textbf{p}} = 0,
\end{equation}
leads to the relation for the comobile number density
\begin{equation}
 n'_s(x) = \frac{n'_{0,s}}{\cosh^2(x/L)}
\end{equation}
with
\begin{equation}\label{equ:velocity_Harris}
   \frac{\Gamma_sU_s}{c} = -\frac{2T_s}{q_sB_0Lc}
    = -2\Theta_s \frac{d'_\lec}{L} \frac{\omega'_\mathrm{pe}}{\omega_{\mathrm{c}s}}\mathrm{sgn}(q_s),
\end{equation}
with $\mathrm{sgn}(q_s)$ the sign of the charge, $\omega_{\mathrm{c}s}=|q_s|B_0/m_s$, 
$\omega'_\mathrm{pe}=\sqrt{n'_0e^2/(\epsilon_0m_\lec)}$, and $d'_\lec=c/\omega'_\mathrm{pe}$.

We note that the absence of electric field in Eq.~\ref{equ:Vlasov_Harris} implicitly assumes that the plasma
is quasi neutral in $\mathcal{R}_0$, which is true only if the overall charge density in this frame vanishes:
\begin{equation}
 \Gamma_i n'_{0,\ionperso} = \Gamma_\lec n'_{0,\lec}.
\end{equation}

We now use the Maxwell-Amp\`ere equation in $\mathcal{R}_0$: $\nabla\wedge\textbf{B} = \mu_0 e [\Gamma_\ionperso n'_\ionperso(x) U_\ionperso - \Gamma_\lec n'_\lec(x) U_\lec]$. Insertion of $n'(x)$ 
and $B(x)$ leads, not surprisingly, to a pressure balance between the unmagnetized center of the sheet and the 
magnetically dominated outer domain:
\begin{equation}\label{equ:pressure_balance}
 \frac{B_0^2}{2\mu_0} = n'_{0,\lec}T_\lec + n'_{0,\ionperso}T_\ionperso.
\end{equation}

Manipulating Eq.~\ref{equ:pressure_balance} and defining $\chi = (1+T_\ionperso/T_\lec)/2$ we obtain
\begin{equation}\label{equ:pressure_ad}
 \Theta_\lec = \frac{1}{4\chi} \left( \frac{\omega_\mathrm{ce}}{\omega'_\mathrm{pe}} \right)^2.
\end{equation}

Given a temperature ratio $\chi$, the four variables $L/d'_\lec$, $\Theta_\lec$, $\omega_\mathrm{ce}/\omega'_\mathrm{pe}$ and $\Gamma_\lec U_\lec/c$ 
are constrained by the two equations \ref{equ:velocity_Harris} (for $s=\lec$) and \ref{equ:pressure_ad}. 
Consequently, one needs to specify two of them. Then, the ion velocity and temperature are easily deduced 
with Eq.~\ref{equ:velocity_Harris} for $s=\ionperso$ and with $\chi$;
$L$ and $\omega_\mathrm{ce}$ expressed in units of $d_\lec$ and $\omega_\mathrm{pe}$, that are useful for a setup in a simulation,
are then deduced with $\Gamma_\lec$.
A special case is when the temperatures are equal: then $\Gamma_\ionperso U_\ionperso = - \Gamma_\lec U_\lec$, $n'_\lec=n'_\ionperso$, 
and $\Theta_\lec = (1/4)(\omega_\mathrm{ce}/\omega'_\mathrm{pe})^2$.

%%%%%%%%%%%%%%%%%%%%%%%%%%%%%%%%%%%%%%%%%%%%%%%%%%%%%%%%%%%%%%%%%%%%%%%%%%%%%%%%%%%%%%%%%%%%%%%%%%%%%%%%%%%%%%%%%%%%%%%%%%%%%%%%%%%%%%%%%%%%%%%%
%%%%%%%%%%%%%%%%%%%%%%%%%%%%%%%%%%%%%%%%%%%%%%%%%%%%%%%%%%%%%%%%%%%%%%%%%%%%%%%%%%%%%%%%%%%%%%%%%%%%%%%%%%%%%%%%%%%%%%%%%%%%%%%%%%%%%%%%%%%%%%%%
\section{A method for loading a drifting Maxwell-J\"{u}ttner distribution}
%%%%%%%%%%%%%%%%%%%%%%%%%%%%%%%%%%%%%%%%%%%%%%%%%%%%%%%%%%%%%%%%%%%%%%%%%%%%%%%%%%%%%%%%%%%%%%%%%%%%%%%%%%%%%%%%%%%%%%%%%%%%%%%%%%%%%%%%%%%%%%%%
%%%%%%%%%%%%%%%%%%%%%%%%%%%%%%%%%%%%%%%%%%%%%%%%%%%%%%%%%%%%%%%%%%%%%%%%%%%%%%%%%%%%%%%%%%%%%%%%%%%%%%%%%%%%%%%%%%%%%%%%%%%%%%%%%%%%%%%%%%%%%%%%
\label{Sec:detail_load_Juttner}
This Appendix directly follows the notations of Sect.~\ref{Sec:Load_particles}, and presents a concrete way to
load a drifting Maxwell-J\"{u}ttner distribution in a PIC code.

Starting from the distribution in Eq.~\ref{equ:Jutt}, we make a first change of variables 
$(p_x,p_y,p_z)\rightarrow(x,y,z)=(p_x/\gamma_y,p_y,p_z/\gamma_y)$, where $\gamma_y = \sqrt{1+p_y^2}$, and we then change 
to cylindrical coordinates $(r,\theta,y)$ with the axis along $y$. We integrate along $\theta$. Finally, a last change of variables 
$(r,y)\rightarrow(u,y)=(\sqrt{1+r^2},y)$ leads to the distribution for the random variables $(U,Y)\in [1,+\infty[\,\times\,]-\infty,+\infty[$\,:
\begin{equation}
\begin{aligned}
 j_{U,Y}(u,y) = \frac{\mu}{2 K_2(\mu) \Gamma_0} &\gamma_y \exp\left\{\mu\Gamma_0\beta_0\,y\right\} \\
     &\times  u \, \exp\left\{ -\mu\Gamma_0\gamma_y\,u \right\}.
\end{aligned}
\end{equation}

It is then easy to obtain the marginal distribution for $Y$:
\begin{equation}\label{equ:distr_y}
\begin{aligned}
 j_Y(y) &= \int_1^\infty \!\dif u\, j_{U,Y}(u,y) \\
	&= \frac{1+\mu\Gamma_0\gamma_y}{2\mu\Gamma_0^3K_2(\mu)}\,\exp\left\{ -\mu\Gamma_0\left(\gamma_y-\beta_0 y\right) \right\}.
\end{aligned}
\end{equation}
From this, we deduce the conditional probability distribution of $U$ given the value $y$ of $Y$:
\begin{equation}\label{distr_u}
 j_{U|y}(u)  = \frac{j_{U,Y}(u,y)}{j_Y(y)}= \frac{a_y^2e^{a_y}}{1+a_y}\,u\exp\left\{-a_y u\right\},
\end{equation}
with $a_y = \mu\Gamma_0\sqrt{1+y^2}$.

Then, for each particle, one has to generate $y=p_y$ according to distribution \ref{equ:distr_y}, 
compute $a_y$, and generate $u$ according to distribution \ref{distr_u}.

For the first step, we use the method of the inversion of the cumulative distribution. 
This method is based on the fact that if $W$ is a uniform random variable 
on $[0,1]$, if $F(s) = \int_{-\infty}^s\!f(x)\,\dif x$ is the cumulative distribution 
of the distribution $f$ and $F^{-1}$ its inverse, then $F^{-1}(W)$ follows 
the distribution $f$. In practice, one has to choose random numbers $w_i$ in $[0,1]$, 
and the $y_i=F^{-1}(w_i)$ will be distributed according to $f$. 

There is, however, no analytic expression for the cumulative distribution $J(y) = \int_{-\infty}^y\!j_Y(z)\,\dif\,z$. 
We compute numerically $J^{-1}(t)$ on a grid of points $t_i = i/N$, $i=1..(N-1)$.
{For each index $i$, we want to find $y_i$ such that $\int_{-\infty}^{y_i}\!j_Y(z)\,\dif\,z = t_i$.
We thus compute numerically the integral $\int_{-\infty}^{y}\!j_Y(z)\,\dif\,z$ up to the point where it reaches $t_i$,
and then attribute the value of $y$ to $y_i$. We use the following algorithm:}
\begin{enumerate}
\item Choose a maximal integration step $\delta y_\mathrm{max}$, and set $\delta y = \delta y_\mathrm{max}$. Choose a tolerance $\mathrm{tol}$.
 \item Start from a low enough value $y_0$ such that $j_Y(y_0)\ll 1$, and set $y=y_0$. Also set $i=1$.
 \item Set $J = 0$, or if possible $J = \int_{-\infty}^{y_0}\!j_Y(z)\dif z$.
 \item Compute $J = J + \delta y \, j_Y(y)$.
 \item $J$ is now an estimation of $\int_{-\infty}^y\!j_Y(z)\,\dif z$.
 
  If $|J - t_i| < \mathrm{tol}$, {then the desired $y_i=J^{-1}(t_i)$ is $y$}. Set $i=i+1$, and go back to step 4.
  
  If $J<t_i$, {set $y=y+\delta y$}, and go back to step 4.
  
  If $J>t_i$, set $\delta y = \delta y/2$ and go back to step 4.
\end{enumerate}
We run this algorithm once for the needed $\mu$ and $\Gamma_0\beta_0$ and store the $y_i$ in a file. 
Then, during the particle initialization, we choose random integers $i$ between $1$ and $N-1$ and set $p_y = y_i$. 

Once $y$ is known, we have to pick a $u$ according to Eq.~\ref{distr_u}. 
Since $a_y$ can be anything between $\mu\Gamma_0$ and $+\infty$, we cannot generate a 
file before the program run, and we have to invert $F$ on the flight. 
It turns out that we can integrate $j_{U|y}$. After some basic manipulations, we arrive at 
\begin{equation}\label{equ:finding_u}
  v = \int_1^u\!\dif t\, j_{U|y}(t) ~~~~ \Leftrightarrow ~~~~ l(x) = w\,l(a_y),
\end{equation}
where $x=a_yu$, $l(x)=(1+x)\exp(-x)$, and $v$ and $w$ are two random numbers between 0 and 1. 
Inversion of the right side of Eq.~\ref{equ:finding_u} is easily done with a Newton method because of the smoothness 
of the function $l$. Starting from $x=a_y$ is a good idea, and one must enforce a minimum number of iterations.

In Fig.~\ref{fig:testJuttner}, we show that the method is accurate.

%%%%%%%%%%%%%%%%%%%%%%%%%%%%%%%%%%%%%%%%%%%%%%%%%%%%%%%%%%%%%%%%%%%%%%%%%%%%%%%%%%%%%%%%%%%%%%%%%%%%%%%%%%%%%%%%%%%%%%%%%%%%%%%%%%%%%%%%%%%%%%%%
%%%%%%%%%%%%%%%%%%%%%%%%%%%%%%%%%%%%%%%%%%%%%%%%%%%%%%%%%%%%%%%%%%%%%%%%%%%%%%%%%%%%%%%%%%%%%%%%%%%%%%%%%%%%%%%%%%%%%%%%%%%%%%%%%%%%%%%%%%%%%%%%
\section{Analytical expressions for quantities averaged over a Maxwell-J\"uttner distribution}
%%%%%%%%%%%%%%%%%%%%%%%%%%%%%%%%%%%%%%%%%%%%%%%%%%%%%%%%%%%%%%%%%%%%%%%%%%%%%%%%%%%%%%%%%%%%%%%%%%%%%%%%%%%%%%%%%%%%%%%%%%%%%%%%%%%%%%%%%%%%%%%%
%%%%%%%%%%%%%%%%%%%%%%%%%%%%%%%%%%%%%%%%%%%%%%%%%%%%%%%%%%%%%%%%%%%%%%%%%%%%%%%%%%%%%%%%%%%%%%%%%%%%%%%%%%%%%%%%%%%%%%%%%%%%%%%%%%%%%%%%%%%%%%%%
\label{sec:detail_calc_averages}

This Appendix details a method used to obtain the analytical expressions of Table~\ref{tab:averages_Juttner}
for the averaged quantities weighted by a Maxwell-J\"uttner distribution. 
The notations are those in Sect.~\ref{Sec:Load_particles}.

\subsection{In the comoving frame}

We start with averages over the distribution with zero drift velocity (i.e., the distribution in the frame comoving with the plasma).
 The calculations are easier when the distribution function is expressed in terms of Lorentz factors $\gamma_0$:
\begin{equation}
 g_{0}(\gamma_0) = \frac{\mu}{K_2(\mu)}\, \gamma_0\sqrt{\gamma_0^2-1}\,\exp\left\{-\mu\gamma_0\right\},~~~\gamma_0\in[1,+\infty[.
\end{equation}
We introduce two relations\footnote{To evaluate the right-hand side of Eq.~\ref{equ:I_of_mu_evaluated} we use \citep[Eq.~10.32.8]{NIST:DLMF}
\begin{equation}\label{equ:calc_of_I}
 K_\nu(z) = \frac{\sqrt{\pi}}{\Gamma(\nu+1/2)}(z/2)^\nu  \int_1^{+\infty}\!\dif\gamma (\gamma^2-1)^{\nu-1/2} \exp\left\{-z\gamma\right\},
\end{equation}
with Euler's gamma function: $\Gamma(1/2+1)=\sqrt{\pi}/2$.}:
\begin{equation}\label{equ:I_of_mu_evaluated}
 I(\mu) \equiv \int_1^{+\infty}\!\dif\gamma_0 \sqrt{\gamma_0^2-1}\,\exp\left\{-\mu\gamma_0\right\} = \mu^{-1}K_1(\mu),
\end{equation}
and \citep[Eq.~10.29.4]{NIST:DLMF}
\begin{equation}\label{equ:calc_of_der_of_Kmu}
 \frac{\dif}{\dif \mu}\left[\mu^{-\nu}K_\nu(\mu)\right] = -\mu^{-\nu}K_{\nu+1}(\mu).
\end{equation}

The method is then to derive Eq.~\ref{equ:I_of_mu_evaluated} as many times as needed with respect to $\mu$,
with the help of Eq.~\ref{equ:calc_of_der_of_Kmu}.  
For example the integral of $g_{0}$ is easily deduced from $\dif I/\dif\mu$.

\subsection{Drifting distribution}

We now consider the distribution with a drift velocity $\beta_0\hat{\textbf{y}}$.
We assume that we want the average of a quantity $M$,
\begin{equation}
 \langle M \rangle = \iiint_{\mathbb{R}^3}\!\dif^3\textbf{p}\, g(\textbf{p}) M(\textbf{p}),
\end{equation}
with $g$ given by Eq.~\ref{equ:Jutt}.
We use the change of variables $p_{0,x} = p_x$, $p_{0,y} = \Gamma_0(p_y-\beta_0\sqrt{1+p^2})$, $p_{0,z} = p_z$, which amounts to 
passing back into the comoving frame.
%It involves $\sqrt{1+\tilde{p}^2_0} = \gamma_0 = \Gamma_0(\sqrt{1+\tilde{p}^2} -\tilde{U}_0 \tilde{p}_y)$. 
Using $\dif^3\textbf{p}/\gamma = \dif^3\textbf{p}_0/\gamma_0$, we arrive at
\begin{equation}\label{equ:m_avg_from_lab_to_0}
 \langle M\rangle = \Gamma_0^{-1} \iiint_{\mathbb{R}^3}\!\frac{\dif^3\textbf{p}_0}{\gamma_0}\, g_0(\textbf{p}_0) \gamma M(\textbf{p})
                   = \Gamma_0^{-1} \left\langle \frac{\gamma M(\textbf{p})}{\gamma_0} \right\rangle_0,
\end{equation}
where $\langle\cdot\rangle_0$ means that the average is taken with $g_0$, 
and where $\gamma$ and $\textbf{p}$ are to be expressed with comoving quantities (subscript 0).
With this last formula, one is left with averages in the comoving frame and can use the previous method.

\end{document}